\documentclass[prd, twocolumn, superscriptaddress, nofootinbib, 
floatfix]{revtex4-2}

\usepackage{lipsum}
\usepackage{xcolor}
\usepackage{etoolbox}
\usepackage{graphicx}
\usepackage{wrapfig}
\usepackage{amsmath}
\usepackage{amsfonts}
\usepackage{amssymb}
\usepackage{multirow}
\usepackage{slashed}
\usepackage{physics}
\usepackage{ulem}
\usepackage{diagbox}
\usepackage{float}
\usepackage{placeins}

\usepackage[colorlinks=true, linkcolor=blue, citecolor=blue, urlcolor=blue]{hyperref}

\DeclareRobustCommand{\fig}[1]{Fig.~\ref{fig:#1}}

\DeclareRobustCommand{\sec}[1]{Sec.~\ref{sec:#1}}

\DeclareRobustCommand{\tb}[1]{Table~\ref{tb:#1}}

\DeclareRobustCommand{\refcite}[1]{Ref.~\cite{#1}}

\newcommand\bets{\begin{table*}}
\newcommand\eets[1]{\label{tb:#1}\end{table*}}

\begin{document}

\title{Transverse-momentum-dependent pion structures from lattice QCD: \\ Collins-Soper kernel, soft factor, TMDWF, and TMDPDF}

\author{Dennis Bollweg}
\affiliation{Computing and Data Sciences Directorate, Brookhaven National Laboratory, Upton, New York 11973, USA}

\author{Xiang Gao}
\affiliation{Physics Department, Brookhaven National Laboratory, Building. 510A, Upton, NY 11973, USA}

\author{Jinchen He}
\email{jinchen@umd.edu}
\affiliation{Maryland Center for Fundamental Physics, University of Maryland, College Park, MD 20742, USA}
\affiliation{Physics Division, Argonne National Laboratory, Lemont, IL 60439, USA}

\author{Swagato Mukherjee}
\affiliation{Physics Department, Brookhaven National Laboratory, Building. 510A, Upton, NY 11973, USA}

\author{Yong Zhao}
\affiliation{Physics Division, Argonne National Laboratory, Lemont, IL 60439, USA}

\date{\today}

\begin{abstract}
We present the first lattice quantum chromodynamics (QCD) calculation of the pion valence-quark transverse-momentum-dependent parton distribution function (TMDPDF) within the framework of large-momentum effective theory (LaMET). Using correlators fixed in the Coulomb gauge (CG), we computed the quasi-TMD beam function for a pion with a mass of 300 MeV, a fine lattice spacing of $a = 0.06$ fm and multiple large momenta up to 3 GeV. The intrinsic soft functions in the CG approach are extracted from form factors with large momentum transfer, and as a byproduct, we also obtain the corresponding Collins-Soper (CS) kernel. Our determinations of both the soft function and the CS kernel agree with perturbation theory at small transverse separations ($b_\perp$) between the quarks. At larger $b_\perp$, the CS kernel remains consistent with recent results obtained using both CG and GI TMD correlators in the literature. By combining next-to-leading logarithmic (NLL) factorization of the quasi-TMD beam function and the soft function, we obtain $x$-dependent pion valence-quark TMDPDF for transverse separations $b_\perp \gtrsim 1$~fm. Interestingly, we find that the $b_\perp$ dependence of the phenomenological parameterizations of TMDPDF for moderate values of $x$ are in reasonable agreement with our QCD determinations. In addition, we present results for the transverse-momentum-dependent wave function (TMDWF) for a heavier pion with 670~MeV mass.
\end{abstract}

\maketitle

\section{Introduction}\label{sec:intro}
In high-energy scatterings, transverse-momentum-dependent parton distribution functions (TMDPDFs) provide a fundamental description of the transverse momentum and polarization degrees of freedom of quarks and gluons within hadrons~\cite{Boussarie:2023izj}. These distributions play a crucial role in understanding the intricate dynamics of quark-gluon interactions and the phenomenon of color confinement. Knowledge of TMDPDFs is essential for predicting observables in multi-scale, non-inclusive high-energy processes, such as semi-inclusive deep-inelastic scattering (SIDIS) and Drell-Yan (DY), based on QCD factorization. Experiments at facilities including the Jefferson Lab 12 GeV Program~\cite{Dudek:2012vr}, the Electron-Ion Collider (EIC)~\cite{Accardi:2012qut,AbdulKhalek:2021gbh}, and the Large Hadron Collider (LHC)~\cite{Amoroso:2022eow} rely heavily on accurate knowledge of the TMDPDFs.

Significant progress has been made in the phenomenological parameterizations of TMDPDFs, particularly for the nucleon, through global analyses of experimental data~\cite{Davies:1984sp,Ladinsky:1993zn,Landry:2002ix,Konychev:2005iy,Sun:2014dqm,DAlesio:2014mrz,Bacchetta:2017gcc,Scimemi:2017etj,Bertone:2019nxa,Scimemi:2019cmh,Bacchetta:2019sam,Hautmann:2020cyp,Bury:2022czx,Bacchetta:2022awv,Isaacson:2023iui,Aslan:2024nqg,Bacchetta:2024yzl,Moos:2023yfa,Yang:2024drd,Bacchetta:2024qre,Moos:2025sal}. While these studies have significantly improved our understanding of the transverse-momentum structure of quarks and gluons within the nucleon, they remain in their early stages due to the limited availability of experimental data sensitive to the non-perturbative region. Compared to nucleons, much less is known about the transverse momentum structure of the pion~\cite{Vladimirov:2019bfa,Cerutti:2022lmb,Barry:2023qqh}. As the lightest QCD bound state, the pion plays a fundamental role in hadron structure and non-perturbative QCD. Its TMDPDFs provide crucial insights into the internal dynamics of quarks and gluons in a strongly interacting relativistic system, with direct implications for hadronization processes and the underlying mechanisms of non-perturbative QCD. However, the scarcity of experimental data on pion TMDPDFs, combined with the inherent nature of the parameterization dependence of global analyses, leaves significant gaps in our non-perturbative first-principles understanding. Lattice QCD calculations provide a natural approach to gain insight into non-perturbative TMD structures of the pion starting from first-principles QCD. 

Early lattice QCD studies primarily focused on computing the moments of TMDPDFs~\cite{Hagler:2009mb, Musch:2011er, Yoon:2015ocs, Yoon:2017qzo}. More recently, large-momentum effective theory (LaMET)~\cite{Ji:2013dva,Ji:2014gla,Ji:2020ect,Ji:2024oka} has provided a framework for directly calculating the $x$-dependence of parton distributions~\cite{Ji:2022ezo}, including TMDPDFs~\cite{Ji:2014hxa,Ji:2018hvs,Ebert:2018gzl,Ebert:2019okf,Ebert:2019tvc,Ji:2019sxk,Ji:2019ewn,Vladimirov:2020ofp,Ebert:2020gxr,Ji:2020jeb,Ji:2021znw,Ebert:2022fmh,Rodini:2022wic,Schindler:2022eva,Deng:2022gzi,Zhu:2022bja,delRio:2023pse,Ji:2023pba,Ji:2024hit}. In the large-momentum limit, quasi-TMDs, defined as equal-time correlators in highly boosted hadron states, can be related to their light-cone counterparts through perturbative factorization, making first-principles calculations of TMDPDFs possible within lattice QCD. 

This framework has driven substantial progress in recent years. One of its key achievements is the determination of the Collins-Soper (CS) kernel, which governs the scale dependence of TMDPDFs~\cite{Shanahan:2020zxr,LatticeParton:2020uhz,Li:2021wvl,Schlemmer:2021aij,Shanahan:2021tst,LatticePartonLPC:2022eev,Shu:2023cot,LatticePartonLPC:2023pdv,Avkhadiev:2023poz,Avkhadiev:2024mgd,Bollweg:2024zet}. Another crucial development is the proposal to extract the intrinsic soft function~\cite{Ji:2019sxk,Deng:2022gzi}, which accounts for the soft gluon radiation and plays a vital role in TMD factorization, from the meson form factors with large momentum transfer~\cite{LatticeParton:2020uhz,Li:2021wvl,LatticePartonLPC:2023pdv}. This step completes the factorization framework that connects quasi-TMDs to light-cone TMDs in the large-momentum limit. These advances have significantly enhanced first-principles lattice calculations of TMDPDFs. Recent achievements include studies of the pion TMD wave function (TMDWF)~\cite{LatticeParton:2023xdl}, the extraction of the unpolarized nucleon TMDPDF~\cite{LatticePartonCollaborationLPC:2022myp}, and investigations of the Boer-Mulders functions of the pion and nucleon~\cite{LPC:2024voq,LPC:2025spt}.

Despite these advancements, a major challenge in lattice calculations of gauge-invariant (GI) quasi-TMDPDFs arises from the structure of the quark-bilinear operators, which are separated in both longitudinal and transverse directions and connected by a staple-shaped Wilson link to maintain gauge invariance. The presence of Wilson lines introduces linear divergences, requiring careful renormalization~\cite{Constantinou:2019vyb,Shanahan:2019zcq,Ebert:2019tvc,Green:2020xco,Zhang:2022xuw, Avkhadiev:2023poz,Ji:2021uvr,Alexandrou:2023ucc}, and leads to an exponential suppression of the signal-to-noise ratio (SNR) as the quark field separation increases~\cite{Bollweg:2024zet}. This suppression presents a significant obstacle to precise lattice determinations of TMDPDFs, particularly in the low transverse momentum region or at large spatial separations, where high precision is crucial. Furthermore, controlling the discretization effects and power corrections at finite momentum remains a challenge in ensuring reliable extrapolations to the continuum and physical limits.

To address this issue, a new approach based on LaMET has been proposed~\cite{Gao:2023lny, Zhao:2023ptv} to extract TMDPDFs from Coulomb gauge (CG) fixed quark correlators. Unlike traditional GI correlators, CG correlators do not require Wilson lines. Despite this difference, they fall within the same universality class~\cite{Hatta:2013gta,Ji:2020ect} as GI correlators with staple-shaped Wilson lines, since both reduce to light-cone TMD correlators in the infinite boost limit. Consequently, CG quasi-TMDPDFs can be matched to light-cone TMDPDFs through a perturbative factorization~\cite{Zhao:2023ptv}. The absence of Wilson lines simplifies renormalization and significantly enhances the SNR of boosted correlators, particularly at large transverse separations.

In this work, we present the first lattice QCD calculation of the unpolarized valence-quark pion TMDPDF using the CG method. We compute the quasi-TMDPDF and quasi-TMDWF matrix elements of the pion in CG. These matrix elements enable the extraction of the CS kernel, the intrinsic soft function, the TMDWF, and ultimately the unpolarized valence TMDPDF of the pion. Furthermore, we improve the perturbative accuracy of the matching procedure, particularly in the extraction of the intrinsic soft function, by resumming logarithmic terms in the Sudakov kernel, an aspect that has not been accounted for in previous lattice QCD studies. Finally, the comparison of our results with global fits of the experimental data shows encouraging consistency, demonstrating the bright potential of the CG approach for TMD physics from lattice QCD.

The paper is organized as follows: we first introduce the theoretical framework for calculating the TMDPDF from the CG quasi-TMDPDFs in Sec.~\ref{sec:framework}, which also includes the CS kernel and the intrinsic soft factor; the lattice setup is presented in Sec.~\ref{sec:lattice_setup}; then we present the detailed analysis of the quasi-TMD in Sec.~\ref{sec:quasi-tmd}; the CS kernel, the intrinsic soft factor, the full pion TMDWF and TMDPDF are analyzed in Sec.~\ref{sec:tmdpdf} to be compared with phenomenological results; finally, we conclude in Sec.~\ref{sec:conclusion}.

\section{Theoretical framework}
\label{sec:framework}

\subsection{Light-cone TMDPDF from CG quasi-TMDPDF}

As proposed in Ref.~\cite{Zhao:2023ptv}, the light-cone TMDPDF can be derived from the CG quasi-TMD beam function defined as
\begin{align}
\tilde{f}_{\Gamma}(x, b_\perp, P^z; \mu) = P^z \int \frac{d z}{2 \pi} e^{i z (x P^z)} \tilde{h}_{\Gamma}(z, b_\perp, P^z; \mu) ~,
\label{eq:qtmdpdf_def_1}
\end{align}  
where the matrix elements are given by
\begin{align}
\tilde{h}_{\Gamma}(z, b_\perp, P^z; \mu) = \frac{1}{2 P_\Gamma} \bra{P} \left. \bar{q}(z, b_\perp) \Gamma q (0) \right|_{\vec{\nabla} \cdot \vec{A} = 0} \ket{P} ~,
\label{eq:qtmdpdf_def_2}
\end{align}  
where $P_\Gamma$ is the kinetic factor for normalization. In this work, we choose $ \Gamma = \gamma^t $ for quasi-TMD beam function and $P_\Gamma = P^t$. The parameter $\mu$ represents the $\overline{\text{MS}}$ renormalization scale, and the space-like separation between the quark and the antiquark is denoted as $\vec{b} \equiv (z, \vec{b}_\perp)$. The hadron state carries momentum $P = (P^t, 0, 0, P^z)$ and satisfies the relativistic normalization condition $\langle P | P \rangle = 2 P^t (2\pi)^3 \delta^{(3)}(\vec{0})$. When the hadron moves with a large momentum $P^z$, the quasi-TMD beam function factorizes as
\begin{align}
\begin{aligned}
    \tilde{f}_{\Gamma}(x, b_\perp, P^z; \mu) = H_f&\left(x, P^z; \mu\right) B (x, b_\perp, xP^+; \mu, \nu)\\
    &\times S^0_C(b_\perp;\mu, \nu) ~,
\end{aligned}
\label{eq:qtmdpdf_sep}
\end{align}  
where $B (x, b_\perp, xP^+; \mu, \nu)$ is the light-cone beam function with zero-bin subtraction, following the procedure outlined in Refs.~\cite{Manohar:2006nz,Stewart:2009yx}. The term $S^0_C(b_\perp;\mu, \nu)$ represents the zero-bin contribution, and the operator definitions of both functions are detailed in Ref.~\cite{Zhao:2023ptv}. The parameters $\mu$ and $\nu$ correspond to the renormalization scales associated with the ultraviolet (UV) and rapidity divergences, respectively. The function $H_f\left(x, P^z; \mu\right) = |C_{\rm TMD} (xP^z; \mu)|^2$ represents the hard kernel that matches the QCD quark field operator to Soft-Collinear Effective Theory (SCET)~\cite{Vladimirov:2020ofp}.  

The light-cone TMDPDF can be defined from the light-cone beam function $B $ and the soft function $S$ as
\begin{align}
f (x, b_\perp; \mu, \zeta) = B (x, b_\perp, xP^+; \mu, \nu) S (b_\perp, y_n;\mu, \nu) ~,
\label{eq:tmdpdf_sep}
\end{align}  
where the dependence on the rapidity renormalization scale $\nu$ cancels out on the right-hand side, leaving only a dependence on the Collins-Soper scale $\zeta \equiv 2 (x P^+)^2 e^{-2 y_n}$. Combining Eqs.~\eqref{eq:qtmdpdf_sep} and \eqref{eq:tmdpdf_sep}, the factorization formula connecting the quasi-TMDPDF and the light-cone TMDPDF is given by~\cite{Ebert:2018gzl,Ebert:2019okf,Ji:2019sxk,Ji:2019ewn,Zhao:2023ptv}  
\begin{align}
\begin{aligned}
&\sqrt{ S_I\left(b_{\perp}, y_n ; \mu\right)} \cdot \tilde{f}_{\Gamma}(x, b_\perp, P^z; \mu) \\
&\quad \quad \quad \quad \quad =f (x, b_\perp; \mu, \zeta) H_f\left(x, P^z; \mu\right) + \mathrm{p. c.} ~,
\label{eq:factorization_tmdpdf_short}
\end{aligned}
\end{align}  
where p.c. indicates the power corrections, and the intrinsic soft function $S_I$ is defined as  
\begin{align}
S_I\left(b_{\perp}, y_n ; \mu\right) \equiv \left( \frac{S (b_\perp, y_n;\mu, \nu)}{S^0_C(b_\perp;\mu, \nu)} \right)^2 ~\,.
\end{align}
Here $S^0_C$ is the CG quasi-soft function, as defined in Ref.~\cite{Zhao:2023ptv}, which exactly cancels the rapidity divergences in $S$. The above factorization formula is of the same form as that for the GI quasi-TMDPDFs~\cite{Ebert:2018gzl,Ebert:2019okf,Ji:2019sxk,Ji:2019ewn}, while the method to calculate $S_I$ was first proposed in Ref.~\cite{Ji:2019sxk}, which enables a complete determination of the TMDs from the lattice.

The intrinsic soft function satisfies the evolution equation  
\begin{align}
\frac{d}{d y_n} \ln S_I\left(b_{\perp}, y_n ; \mu\right) = -2 \gamma^{\overline{\rm MS}}(b_\perp; \mu) ~,
\end{align}  
where $\gamma^{\overline{\rm MS}}(b_\perp; \mu)$ is the Collins-Soper kernel. Therefore, the intrinsic soft function at any $y_n$ satisfies 
\begin{align}
\begin{aligned}
    &S_I\left(b_{\perp}, y_n ; \mu\right) \\
    & \quad = S_I\left(b_{\perp}, y_n=0 ; \mu\right) \cdot \exp(-2 y_n \cdot  \gamma^{\overline{\rm MS}}(b_\perp; \mu)) ~.
\end{aligned}
\end{align}
One can redefine the intrinsic soft function as $S_I\left(b_{\perp}; \mu\right) \equiv S_I\left(b_{\perp}, y_n = 0; \mu\right)$, allowing us to rewrite Eq.~\eqref{eq:factorization_tmdpdf_short} in a more explicit form,  
\begin{align}
\begin{aligned}
    &\sqrt{S_I\left(b_{\perp} ; \mu\right) } \cdot \tilde{f}_{\Gamma}\left(x, b_{\perp}, P^z; \mu\right) = f\left(x, b_{\perp}; \mu, \zeta \right) \\ 
    & \quad \quad \times  H_f \left(x, P^z; \mu\right)  \exp \left[\frac{1}{2}\ln \frac{\left(2 x P^z\right)^2}{\zeta} \gamma^{\overline{\mathrm{MS}}}\left(b_{\perp}; \mu\right)\right] \\
    & \quad \quad +\mathcal{O}\left(\frac{\Lambda_{\mathrm{QCD}}}{x P^z}, \frac{1}{b_{\perp}\left(x P^z\right)}\right) ~.
    \label{eq:factorization_tmdpdf}
\end{aligned}
\end{align}  
This factorization framework provides a systematic approach to relating quasi-TMDPDFs extracted from lattice QCD to their light-cone counterparts, ensuring a well-defined separation between perturbative and non-perturbative contributions. However, power corrections of order 
$\mathcal{O}\left(\frac{\Lambda_{\mathrm{QCD}}}{x P^z}, \frac{1}{b_{\perp} x P^z}\right)$~\cite{Liu:2023onm} 
introduce systematic uncertainties, particularly in the small-$x$ region. Despite these limitations, the framework offers valuable first-principles constraints in the large-$x$ regime, where power corrections are expected to be relatively small.

\subsection{Quasi-TMDWF and CS kernel}

Similar to the quasi-TMD beam function defined in Eqs.~\eqref{eq:qtmdpdf_def_1} and \eqref{eq:qtmdpdf_def_2}, the light-cone TMDWF can be derived from the CG quasi-TMDWF, which is defined as  
\begin{align}
    \tilde{\phi}_\Gamma (x, b_\perp, P^z; \mu) = P^z \int \frac{d z}{2 \pi} e^{i z (x P^z)} \tilde{\varphi}_{\Gamma}(z, b_\perp, P^z; \mu)~,
    \label{eq:qtmdwf_def_1}
\end{align}  
with the quasi-TMDWF matrix elements given by  
\begin{align}
\begin{aligned}
    &\tilde{\varphi}_{\Gamma}(z, b_\perp, P^z; \mu) = \frac{e^{-i z P^z / 2}}{f_\pi P_\Gamma} \\
    &\quad \times \bra{\Omega} \left. \bar{q}(z/2, b_\perp) \Gamma q (-z/2, 0) \right|_{\vec{\nabla} \cdot \vec{A} = 0} \ket{P} ~,
    \label{eq:qtmdwf_def_2}
\end{aligned}
\end{align}  
which corresponds to a pion-to-vacuum matrix element, the $\ket{\Omega}$ represents the QCD vacuum state, the $f_\pi$ is the pion decay constant, and $P_\Gamma$ is the kinetic factor for normalization. In this work, we choose $ \Gamma = \gamma^z \gamma^5 $ for quasi-TMDWFs with nonzero momenta and $P_\Gamma = i P^z$, while for zero-momentum quasi-TMDWFs, we set $ \Gamma = \gamma^t \gamma^5 $ and $P_\Gamma = P^t$.   

In the large-momentum limit, the quasi-TMDWF can be matched to the light-cone TMDWF via the factorization formula
\begin{align}
\begin{aligned}
&\sqrt{ S_I\left(b_{\perp}, y_n ; \mu\right)} \cdot \tilde{\phi}_{\Gamma}\left(x, b_{\perp}, P^z ; \mu\right) \\
&\quad \quad =\phi (x, b_\perp, y_n; \mu, \zeta, \bar{\zeta}) H_\phi\left(x, \bar{x}, P^z; \mu\right) + \mathrm{p. c.} ~,
\label{eq:factorization_tmdwf_short}
\end{aligned}
\end{align}  
with $\bar{x} = 1 - x$, $\bar{\zeta} \equiv 2 (\bar{x} P^+)^2 e^{-2 y_n}$, and the hard kernel 
\begin{align}
    H_\phi (x, \bar{x}, P^z; \mu) = C_{\rm TMD} (xP^z; \mu) \cdot C_{\rm TMD} (\bar{x} P^z; \mu)~,
\end{align}
calculated up to NLO in \refcite{Zhao:2023ptv}.

An essential component of quasi-TMD factorization is the CS kernel $ \gamma^{\overline{\rm MS}}(b_\perp; \mu) $, which governs the rapidity evolution of TMDs and is crucial to achieve consistent matching between quasi-TMDPDFs and light-cone TMDPDFs. The CS kernel can be extracted from the ratios of quasi-TMD matrix elements computed at different hadron momenta~\cite{Ji:2014hxa, Ebert:2018gzl,Ji:2019sxk}. In this work, we employ the ratio of CG quasi-TMDWFs~\cite{Ji:2019sxk,Ji:2021znw} as
\begin{align}
\begin{aligned}
    &\gamma^{\overline{\rm MS}}(b_\perp; \mu)\\
    &\quad=\gamma^{\overline{\rm MS}}(b_\perp, P_1, P_2; \mu) + \text{p.c.} \\
    &\quad=\frac{1}{\ln \left(P_2 / P_1\right)} \ln \frac{H_\phi \left(x, \bar{x}, P_1; \mu\right) \tilde{\phi}_{\gamma^z \gamma^5}\left(x, b_{\perp}, P_2;  \mu\right)}{H_\phi \left(x, \bar{x}, P_2; \mu\right) \tilde{\phi}_{\gamma^z \gamma^5}\left(x, b_{\perp}, P_1; \mu \right)} \\
    &\quad\quad + \mathcal{O}\left(\frac{\Lambda_{\mathrm{QCD}}}{x P^z}, \frac{1}{b_{\perp}\left(x P^z\right)}, \frac{\Lambda_{\mathrm{QCD}}}{\bar{x} P^z}, \frac{1}{b_{\perp}\left(\bar{x} P^z\right)}\right) ~.
    \label{eq:cs_kernel_calc}
\end{aligned}
\end{align}
This method offers a key advantage over direct extractions from quasi-TMDPDFs, as quasi-TMDWFs naturally peak around $ x = 0.5 $, a region where power corrections are significantly suppressed. On the other hand, the quasi-TMDPDFs usually decrease quickly at moderate to large $x$, making the extraction of the CS kernel more susceptible to power corrections.

\subsection{Intrinsic soft function}\label{sec:softfunction}

Another essential component of the factorization of quasi-TMDs is the intrinsic soft function, which accounts for soft gluon radiation in the process. Its calculation is particularly challenging for two reasons. First, as a non-perturbative quantity, it cannot be computed using standard perturbative techniques. Second, because the soft function involves correlations along two light-cone directions, it cannot be directly simulated on the lattice, even with a large-momentum boost.

A practical approach to determining the intrinsic soft function, as proposed in Refs.~\cite{Ji:2019sxk}, is to use the TMD factorization of a pseudoscalar light-meson form factor, which is defined as
\begin{align}
\begin{aligned}
    &F\left(b_{\perp}, P_1, P_2, \Gamma, \Gamma^\prime \right) \\
    &\quad \equiv -4N_c \frac{\left\langle P_2 \right| \bar{q}\left(b_{\perp}\right) \Gamma q\left(b_{\perp}\right) \bar{q}(0) \Gamma^{\prime} q(0)\left| P_1 \right\rangle}{f_\pi^2 (P_1\cdot P_2)} ~,
    \label{eq:ff_def}
\end{aligned}
\end{align}
where $\ket{P_1} $ and $ \ket{P_2} $ denoting the pion states with momentum $P_1$ and $P_2$. The coefficient $-4 N_c = -12$ is a normalization factor~\cite{Deng:2022gzi}. This form factor can be extracted using the following ratio
\begin{align}
\begin{aligned}
    &R_F\left(b_{\perp}, P_1, P_2, \Gamma, \Gamma^{\prime}\right) \\
    &\quad \equiv -4N_c \frac{\left\langle P_2\right| \bar{q}\left(b_{\perp}\right) \Gamma q\left(b_{\perp}\right) \bar{q}(0) \Gamma^{\prime} q(0)\left|P_1\right\rangle}{\langle 0| \bar{q}(0) \gamma^\mu \gamma^5 q(0)\left|P_1\right\rangle\left\langle P_2\right| \bar{q}(0) \gamma_\mu \gamma^5 q(0)|0\rangle} ~.
    \label{eq:ff_ratio}
\end{aligned}
\end{align}
To extract the leading-twist contribution, one can choose the Dirac matrices as $ \Gamma = \Gamma^\prime \in \{ I, \gamma^5, \gamma^\perp, \gamma^\perp \gamma^5 \} $~\cite{Deng:2022gzi, LatticePartonLPC:2023pdv}. In this work, we choose $\Gamma = \Gamma^\prime \in \{ \gamma^\perp, \gamma^\perp \gamma^5 \}$. The denominator of the ratio $R_F$ consists of two-point functions used to normalize the form factor, and the index $\mu$ should be summed. The momenta in the external pion states are taken as $ P_1 = (P^t, 0, 0, P^z) $ and $ P_2 = (P^t, 0, 0, -P^z) $, moving back-to-back to achieve a large momentum transfer $ Q^2 = (2P^z)^2 $. In this kinematic regime, the form factor can be factorized in terms of the TMDWF $ \phi (x, b_\perp, y_n; \mu, \zeta) $~\cite{Ji:2019sxk, Ji:2021znw, Deng:2022gzi} as
\begin{align}
\begin{aligned}
    &F (b_\perp, P^z) = \int dx_1 dx_2 H_F (x_1, x_2, P^z; \mu) \\
    & \quad \quad \times \phi^\dagger (x_1, b_\perp, y_n; \mu, \zeta_1, \bar{\zeta}_1) \phi (x_2, b_\perp, -y_n; \mu, \zeta_2, \bar{\zeta}_2) ~.
    \label{eq:factorization_ff}
\end{aligned}
\end{align}
where $ H_F $ is the hard kernel encapsulating the short-distance dynamics of the process. The $y_n$-dependence cancels between $\phi$ and $\phi^\dagger$, and the hard kernel can be expressed as the product of two Sudakov kernels~\cite{Deng:2022gzi}:  
\begin{align}
\begin{aligned}
    &H_F (x_1, x_2, P^z; \mu) \\
    &\quad= C_{\rm Sud} (x_1, x_2, P^z; \mu) \cdot C_{\rm Sud} (\bar{x}_1, \bar{x}_2, P^z; \mu) ~,
\end{aligned}
\end{align}
where $ C_{\rm Sud} $ represents the Sudakov kernel, which has been computed at one-loop order in the literature~\cite{Manohar:2003vb, Deng:2022gzi}. The renormalization group (RG) resummed results for $ C_{\rm Sud} $ at next-to-leading logarithmic (NLL) accuracy are provided in App.~\ref{app:rg_resum}.  

By combining the factorization of the quasi-TMDWF in Eq.~\eqref{eq:factorization_tmdwf_short} with the form factor factorization in Eq.~\eqref{eq:factorization_ff}, the intrinsic soft function at $ y_n = 0 $ can be extracted as  
\begin{align}
    S_I (b_\perp; \mu) = \frac{F(b_\perp, P^z)}{\int dx_1 dx_2 H_F(x_1, x_2, P^z; \mu) \tilde{\Phi}^\dagger (x_1) \tilde{\Phi} (x_2) } ~,
    \label{eq:soft_func_calc}
\end{align}  
where the reduced quasi-TMDWF $ \tilde{\Phi} (x) $, is defined as  
\begin{align}
    \tilde{\Phi} (x) \equiv \frac{\tilde{\phi}_{\Gamma}\left(x, b_{\perp}, P^z ; \mu\right)}{H_\phi \left(x, \bar{x}, P^z; \mu\right)} ~.
    \label{eq:reduced_qtmdwf_def}
\end{align} 
Using different renormalization schemes for quasi-TMDWF, one can get different intrinsic soft functions, which can be perturbatively converted to each other at small $b_\perp$. However, the scheme dependence will eventually cancel between the renormalized quasi-beam and intrinsic soft functions. The details of scheme conversion can be found in App.~\ref{app:scheme_conversion}.

\section{Lattice setup}\label{sec:lattice_setup}

We perform a numerical lattice QCD calculation using a $2+1$-flavor gauge ensemble generated by the HotQCD Collaboration~\cite{HotQCD:2014kol}. The ensemble employs the Highly Improved Staggered Quark (HISQ) action~\cite{Follana:2006rc} with a lattice spacing of $ a = 0.06 $ fm and a volume of $ L_s^3 \times L_t = 48^3 \times 64 $. The valence sector is treated using tadpole-improved clover Wilson fermions on a hypercubic (HYP) smeared~\cite{Hasenfratz:2001hp} gauge background. The clover coefficient is set to $ c_{\rm sw} = u_0^{-3/4} $, where $ u_0 $ is the average plaquette after HYP smearing. For this ensemble, we use $ c_{\rm sw} = 1.0336 $ in both the time and spatial directions. The valence quark masses are tuned to yield a valence pion mass of 300 MeV, with a corresponding hopping parameter of $ \kappa \approx 0.12623 $.

The key requirement for the factorization of quasi-TMDs is a sufficiently large hadron momentum. To achieve a higher boost factor, we take advantage of the three-dimensional rotational symmetry of the CG approach and adopt off-axis momenta for the quasi-TMD, choosing momentum directions along $\vec{n} = (n^x, n^y, 0)$. The hadron momenta on the lattice are given by $P^z = \frac{2\pi |\vec{n}|}{L_s a}$. In this study, we consider $ n^x = n^y \in \{3, 4, 5\} $, which allows us to reach a maximum hadron momentum of $ P^z = 3.04 $ GeV for the quasi-TMD calculations. To optimize the signal-to-noise ratio and enhance overlap with large-momentum hadron ground states, we employ boosted Gaussian smearing~\cite{Bali:2016lva}, using the same setup as in Ref.~\cite{Gao:2023lny}. To extract the ground-state contribution, we compute quasi-TMD three-point functions for multiple source-sink separations, choosing $ t_{\rm sep} / a = 6, 8, 10, 12 $. Calculations are performed on 553 gauge configurations and we apply the All-Mode Averaging (AMA) technique~\cite{Shintani:2014vja} to further improve the signal. The stopping criteria for the sloppy and exact inversions are set to $10^{-4}$ and $10^{-10}$, respectively, aligning with the settings in Ref.~\cite{Gao:2020ito}. For the smaller separation $ t_{\rm sep} / a = 6 $, we compute 1 exact source and 32 sloppy sources per configuration, while for the larger separations $ t_{\rm sep} / a = 8, 10,12 $, we compute 4 exact sources and 128 sloppy sources. 

In addition, we compute the large-momentum pion form factor following the strategy outlined in Ref.~\cite{LatticeParton:2020uhz}. Specifically, we evaluate the two-current three-point function  
\begin{align}
\begin{aligned}
    &C_{\mathrm{F}}(t_{\rm sep},\tau) \\
    &\quad\quad = \left\langle \pi_{\rm w}(-\vec{P},t_{\rm sep})\, j_u (b_\perp, \tau)\, j_d(0, \tau)\, \pi_{\rm w}^\dagger(\vec{P},0)\right\rangle,
\end{aligned}
\end{align}
where the two currents $j_u (b_\perp, \tau) \equiv \bar{u}(b_\perp, \tau) \Gamma u(b_\perp, \tau)$ and $j_d(0, \tau) \equiv \bar{d}(0, \tau) \Gamma d(0, \tau)$ are separated by a transverse distance $ b_\perp $, inserted at the same time slice $ \tau $. The Dirac structures are chosen as $\Gamma \in \{ \gamma^\perp, \gamma^\perp \gamma^5 \}$ to extract the leading-twist contribution~\cite{Deng:2022gzi, LatticePartonLPC:2023pdv}. Unlike the two-point and quasi-TMD correlator calculations, where Gaussian-smeared sources are used, here we adopt wall sources $ \pi_{\rm w} $ for the pions. Given the significant computational demands of this calculation and the universality of the soft factor, insensitive to the choice of hadron species, we choose to employ a heavier valence pion mass of 670 MeV. Consequently, our analysis of the pion form factor is limited to a subset of 100 gauge configurations using wall sources in $L_t=64$ time slices. We compute the form factor at a single momentum $ \vec{n} = (0, 0, 6) $, corresponding to $ P^z = 2.58 $ GeV. To ensure consistency in the extraction of the intrinsic soft function, we also compute the quasi-TMDWF using the same valence pion mass. For quasi-TMDWF, we use momenta in the $ z $-direction with $ \vec{n} = (0, 0, n^z) $ and $ n^z \in \{8, 9, 10\} $, reaching a maximum momentum of $ P^z = 4.30 $ GeV. This part of the calculation is performed on the same 553 HotQCD gauge configurations used in the main quasi-TMD analysis, with 1 exact source and 32 sloppy sources per configuration.

\section{quasi-TMDPDF matrix elements}
\label{sec:quasi-tmd}

\subsection{Two-point function and dispersion relation}

\begin{figure}[th!]
    \centering
    \includegraphics[width=.9\linewidth]{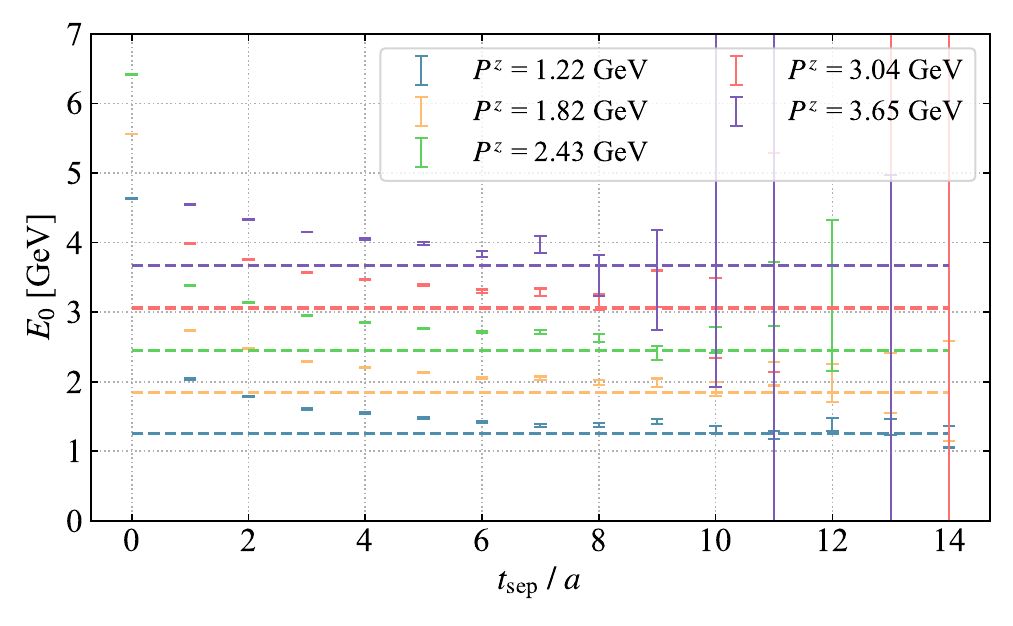}
    \includegraphics[width=.9\linewidth]{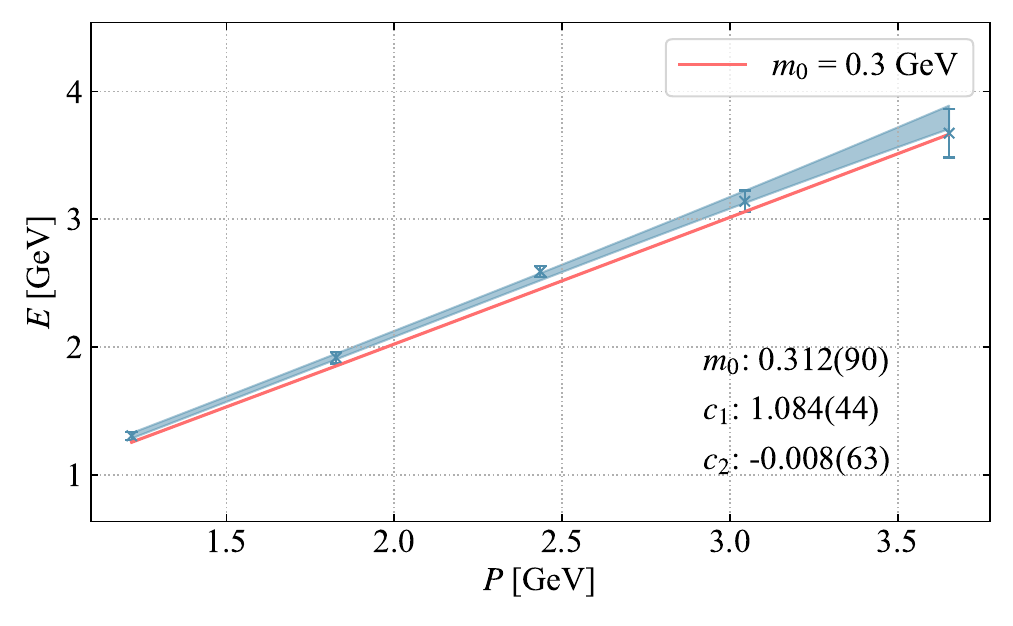}
    \caption{Upper panel: effective mass plot of $C^{\rm SS}_{\text{2pt}}$ with various momenta. The dashed lines are the ground state energies of pion calculated using the dispersion relation $E_0 = \sqrt{m_0^2 + (P^z)^2}$, where the valence pion mass is $m_0 = 300$ MeV. Lower panel: the ground-state energies $E_0$ extracted from two-state fit of the two-point functions. The red line represents the exact dispersion relations with $m_0 = 300$ MeV.
    \label{fig:2pt}}
\end{figure}

To extract the ground-state quasi-TMD matrix elements, it is essential to first determine the energy spectrum associated with the pion interpolator $ \pi^\dagger $. The two-point correlator is defined as  
\begin{align}
    C^{s s'}_{\text{2pt}} \left(t_{\rm sep}, P^z \right) = \langle \pi_{s'} (P^z, t_{\rm sep}) \pi^\dagger_{s} (P^z, 0) \rangle ~,
\end{align}
where $ \pi^\dagger_s $ and $ \pi_{s'} $ represent the pion source and sink defined by
\begin{align}
\begin{aligned}
&\pi_s(\vec{x},t_{\rm sep})=\overline{d}_s(\vec{x},t_{\rm sep})\gamma_5 u_s(\vec{x},t_{\rm sep}),\\ 
& \pi_s(\vec{P},t_{\rm sep})= \sum_{\vec{x}} \pi_s(\vec{x},t_{\rm sep})  e^{-i \vec{P}\cdot\vec{x}}.
\label{eq:pisource}
\end{aligned}
\end{align}
In this work, we use a Gaussian-smeared source and sink ($ s = s' = S $), following the setup in Ref.~\cite{Izubuchi:2019lyk}. The superscript $SS$ will be removed in the following text for simplicity.

By inserting a complete set of states, the two-point correlator can be expressed as a sum over energy eigenstates:
\begin{align}
    C_{\text{2pt}}\left(t_{\rm sep}\right) = \sum_{n=0}^{N_{\mathrm{s}}-1} \frac{\left|z_n\right|^2}{2 E_n} \left(e^{-E_n t_{\rm sep}}+e^{-E_n\left(L_t-t_{\rm sep}\right)}\right) ~.
    \label{eq:2pt}
\end{align}  
The overlap amplitude $ z_n = \langle n | \pi^\dagger | \Omega \rangle $ quantifies the projection of the pion interpolator onto the $ n $th energy eigenstate, and $ N_s $ denotes the number of excited states with the same quantum numbers as the pion, which are considered within the fit function.

To investigate the asymptotic behavior of $ C_{\text{2pt}} $ at large $ t_{\rm sep} $, we define the effective mass as  
\begin{align}
    m_{\rm eff} (t_{\rm sep} ) = \ln\left( \frac{C_{\text{2pt}}\left(t_{\rm sep}\right)}{C_{\text{2pt}}\left(t_{\rm sep} + a\right)} \right) ~.
\end{align}  
The effective masses for different momenta are shown in the upper panel of \fig{2pt}. As $ t_{\rm sep} $ increases, the effective masses approach plateaus, which align with the dashed lines computed from the relativistic dispersion relation $ E_0 = \sqrt{m_0^2 + (P^z)^2}$. This agreement indicates that the ground state is effectively isolated from the excited-state tower for $ t_{\rm sep} \gtrsim 8a $.

In practical analysis, the upper bound $ N_s $ must be truncated, as higher excited-state contributions decay rapidly with increasing $ t_{\rm sep} $. In this work, we perform a two-state fit by setting $ N_s = 2 $, which allows us to efficiently extract the ground-state contribution while accounting for the leading excited-state contamination. The lower panel of \fig{2pt} presents the extracted ground-state energies as a function of the hadron momentum. To test the validity of the relativistic dispersion relation, we fit the data points using the functional form  
\begin{align}
    E = \sqrt{m_0^2 + c_1 P^2 + c_2 a^2 P^4} ~,
\end{align}  
where $ E $ is the ground-state energy, $ P $ is the hadron momentum, and $ a $ is the lattice spacing. The coefficients $ c_1 $ and $ c_2 $ parameterize the possible discretization effects. As shown by the fit bands in \fig{2pt}, the extracted ground-state energies exhibit excellent agreement with the expected relativistic dispersion relation up to $ P \approx 3.6 $ GeV. This agreement shows that discretization effects remain small in the dispersion relation within the momentum range studied in this work.

\subsection{Bare quasi-TMD beam function matrix elements of pion}

\begin{figure*}
    \centering
    \includegraphics[width=.32\linewidth]{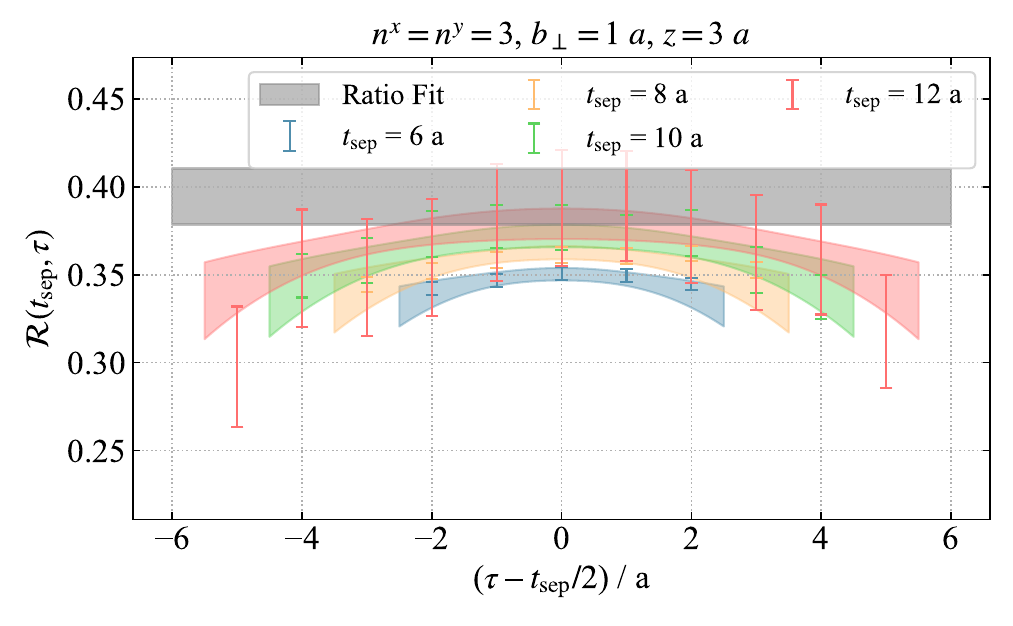}
    \includegraphics[width=.32\linewidth]{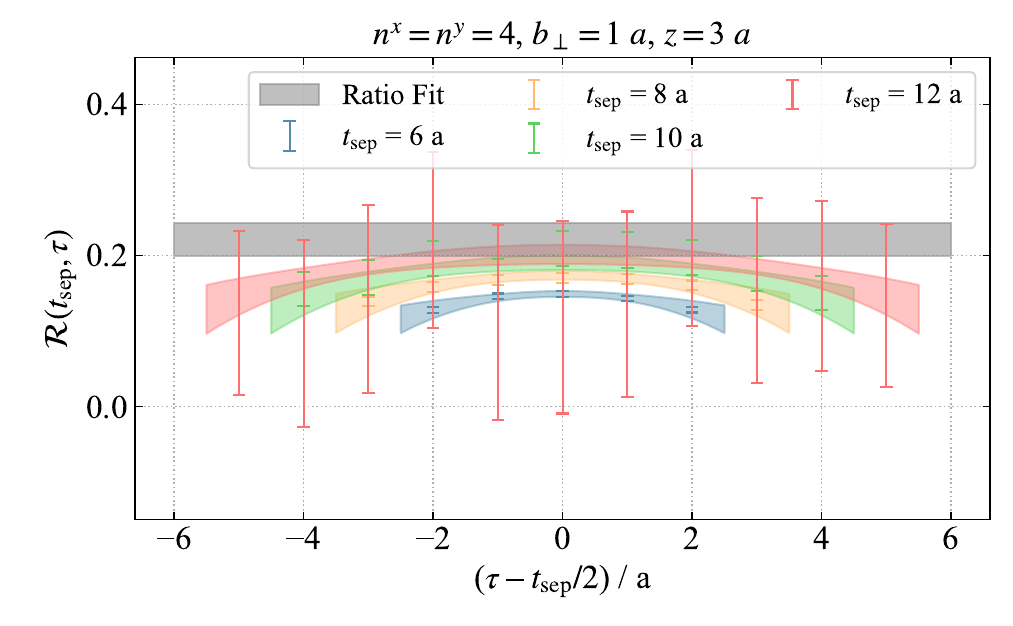}
    \includegraphics[width=.32\linewidth]{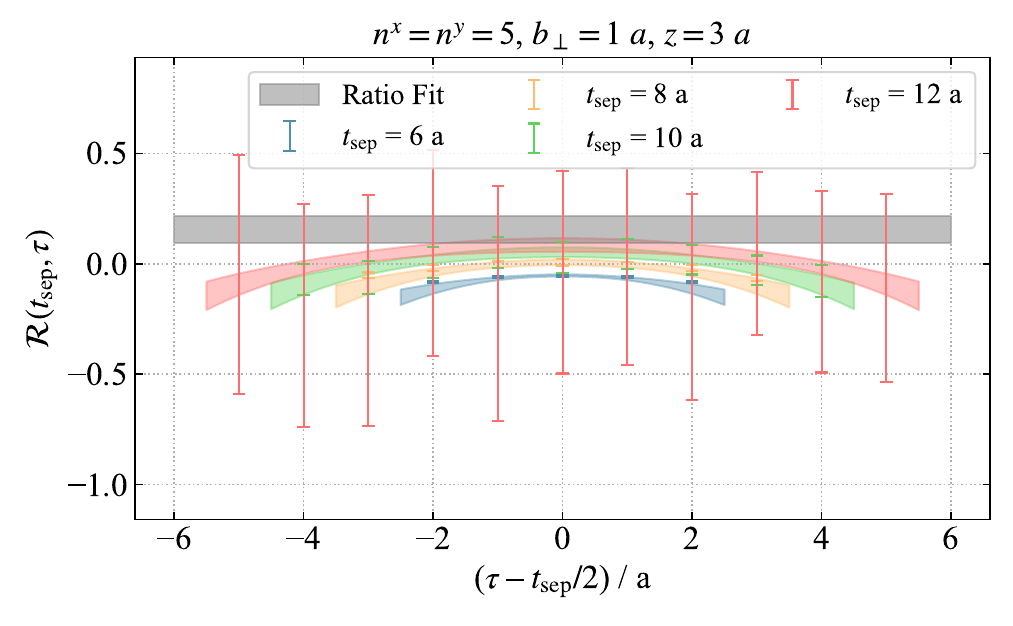}
    \includegraphics[width=.32\linewidth]{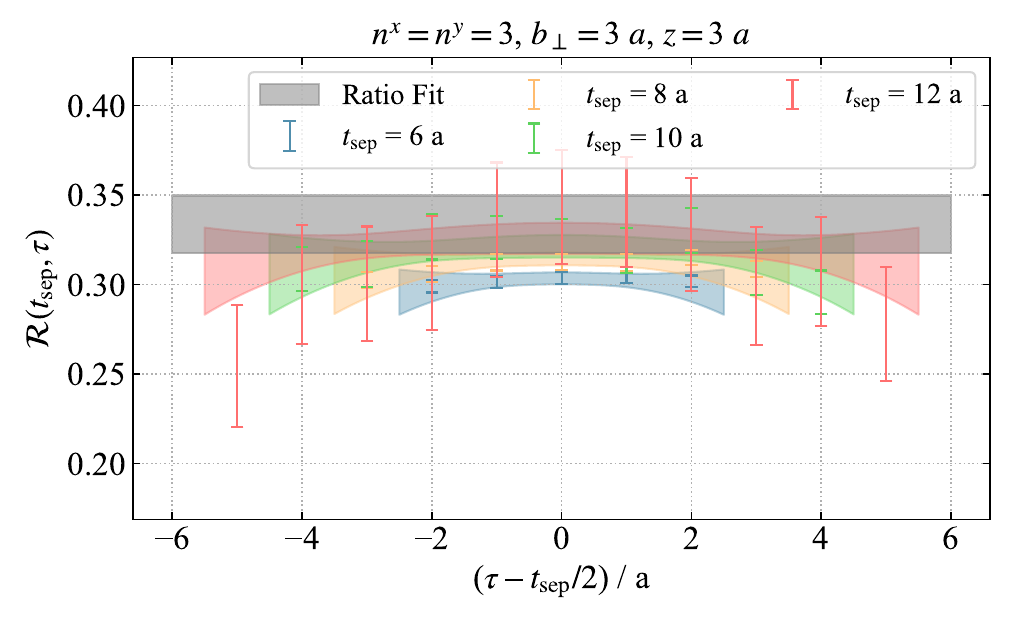}
    \includegraphics[width=.32\linewidth]{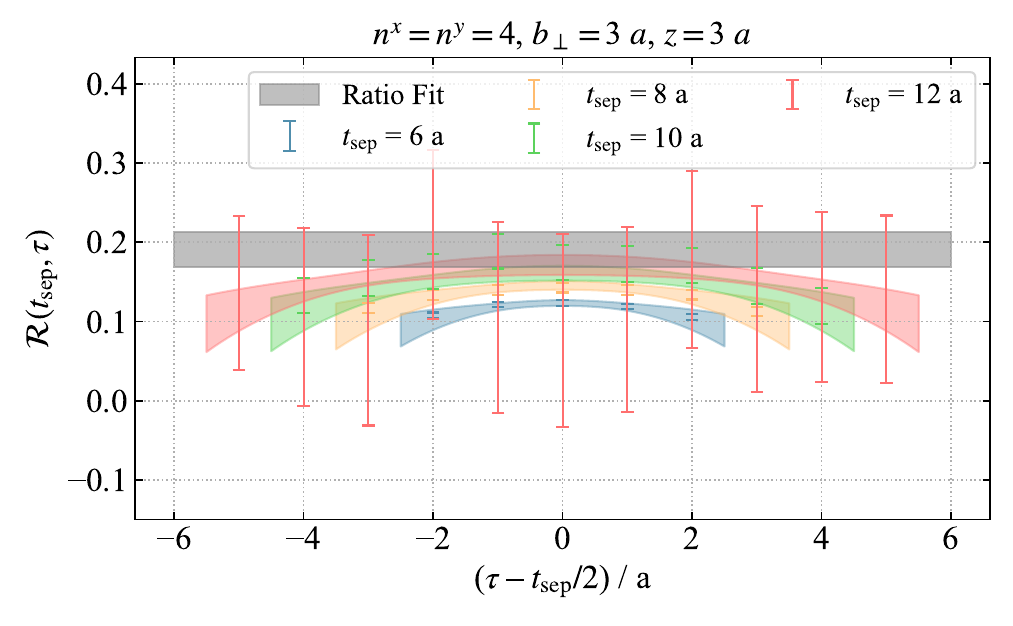}
    \includegraphics[width=.32\linewidth]{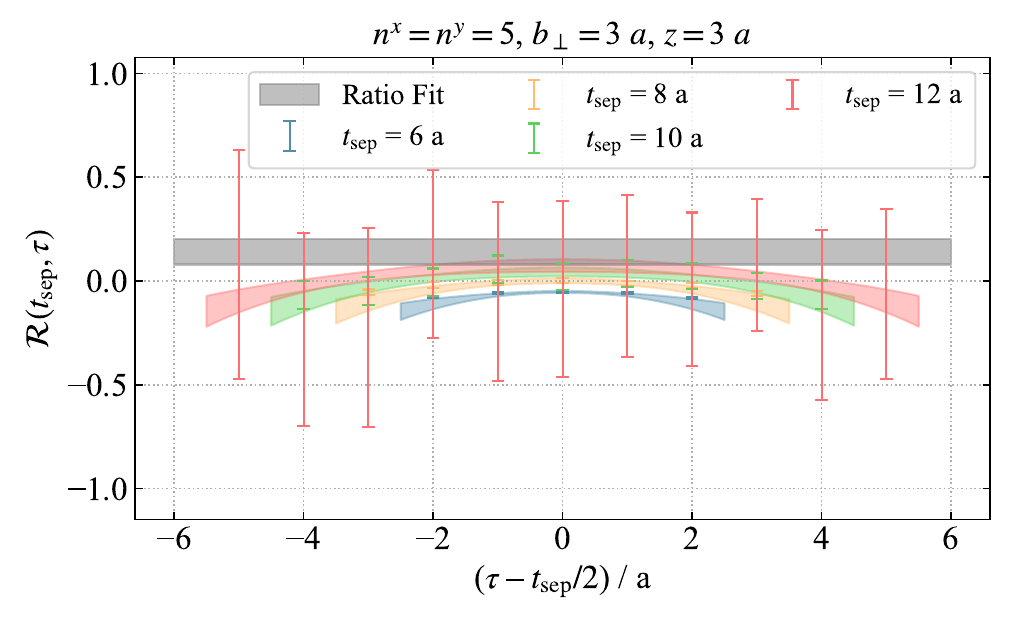}
    \caption{From left to right, the real part of the ratios $R_{\tilde{h}}$ between three-point and two-point functions for the pion quasi-TMD, corresponding to hadron momenta $P^z=1.83$, 2.43 and 3.04 GeV, are shown as functions of $t_{\rm sep}$ and $\tau$. The upper and lower panels are for the cases with $(b_\perp, z) = (1, 3)~a$ and $(b_\perp, z) = (3, 3)~a$, respectively. The colored bands are two-state fit results while the gray band is the estimated ground-state matrix element.
    \label{fig:ratio_fit_qtmdpdf}}
\end{figure*}

\begin{figure*}[th!]
    \centering
    \includegraphics[width=.32\linewidth]{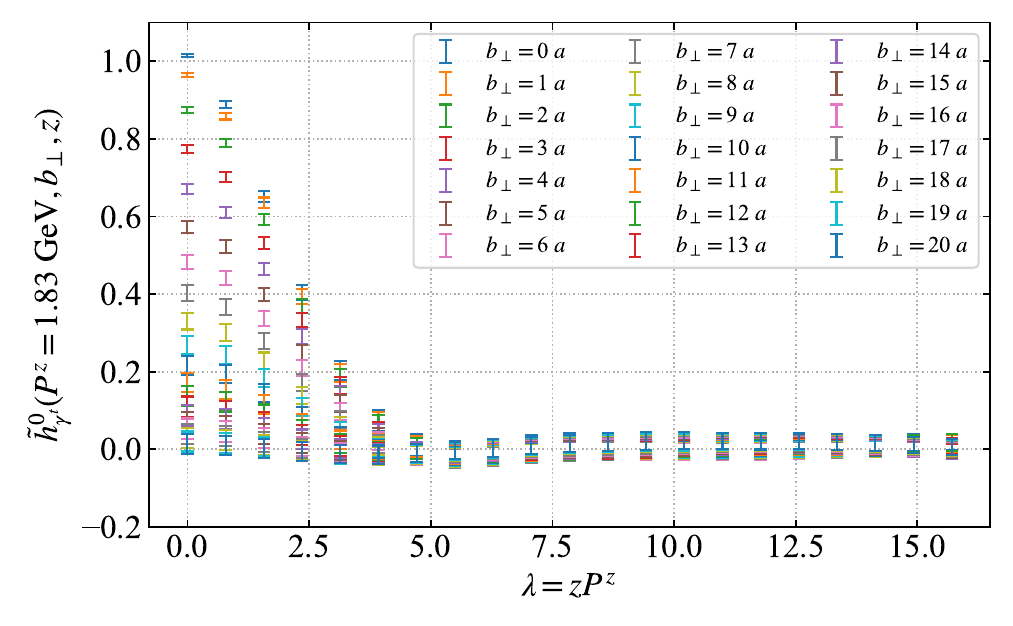}
    \includegraphics[width=.32\linewidth]{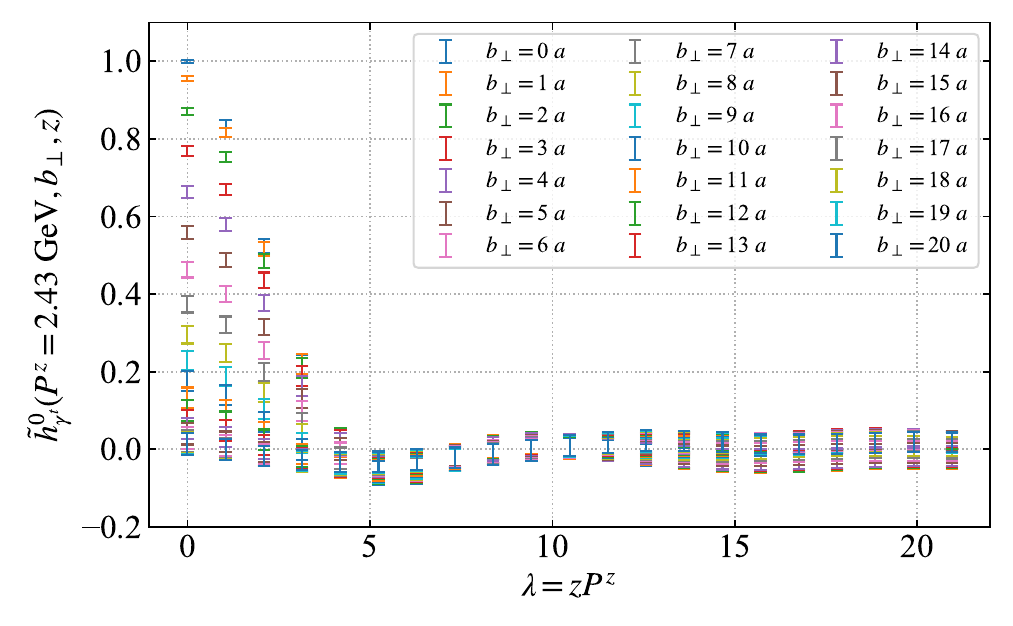}
    \includegraphics[width=.32\linewidth]{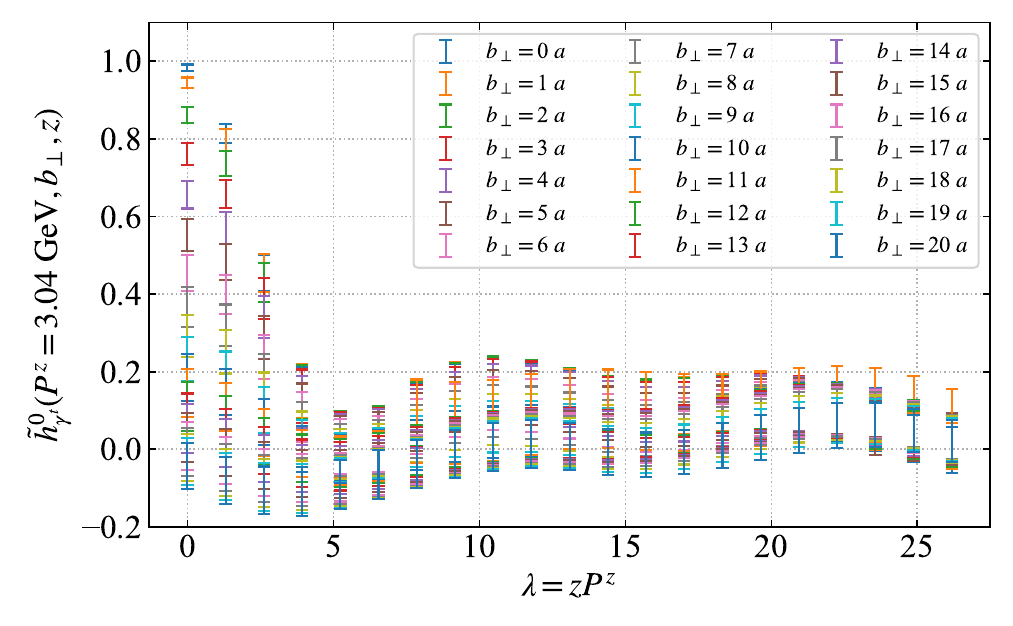}
    \caption{The bare matrix elements of quasi-TMD in the coordinate space are plotted as the function of $b_\perp$ and $\lambda = z P^z$. These matrix elements are extracted from the two-state chained fit of $C_{\rm 2pt}$ and $R_{\tilde{h}}$. From left to right, the three panels correspond to hadron momenta of $P^z=1.83$, 2.43 and 3.04 GeV, respectively. It is observed that for all three hadron momenta, the quasi-TMD decays as the transverse separation $b_\perp$ increases and asymptotically approaches zero in the large $\lambda$ regime.
    \label{fig:bare_zdep_qtmdpdf}}
\end{figure*}

The quasi-TMD beam function is extracted from the three-point correlator 
\begin{align}
C_{\tilde{h}}\left(t_{\mathrm{sep}}, \tau \right)=\left\langle \pi_s(\vec{P},t_{\rm sep}) \bar{q}(z, b_\perp, \tau) \gamma^t q (\vec{0}, \tau) \pi_s^\dagger(\vec{P},0)\right\rangle~,
\label{eq:c3pt_tmdpdf}
\end{align}  
which can be expressed in terms of a spectral decomposition: 
\begin{align}
\begin{aligned}
    C_{\tilde{h}}\left(t_{\mathrm{sep}}, \tau \right) = \sum^{N_s -1}_{n, m = 0} \frac{z_n^\dagger O_{n m} z_m}{4 E_n E_m} e^{-E_n\left(t_{\mathrm{sep}}- \tau \right)} e^{-E_m \tau} ~,
    \label{eq:qtmdpdf_3pt}
\end{aligned}
\end{align}  
where $ O_{n m}=\langle n|\bar{q}(z, b_\perp) \gamma^t q (0)|m \rangle$ is proportional to the matrix elements of the quasi-TMD operator. To extract the ground-state matrix element $ O_{00} = 2 E_0 \tilde{h}_{\gamma^t}$, we employ a two-state chained fit using the two-point correlator $ C_{\rm 2pt} $ and the ratio  
\begin{align}
    R_{\tilde{h}}\left(t_{\mathrm{sep}}, \tau \right) = \frac{C_{\tilde{h}}\left(t_{\mathrm{sep}}, \tau \right)}{C_{\rm 2pt} \left(t_{\mathrm{sep}} \right)} ~.
    \label{eq:qtmdpdf_ratio}
\end{align}  
In the limit $ t_{\rm sep} \to \infty $, the ratio $ R_{\tilde{h}} $ asymptotically converges to the bare quasi-TMD matrix element $ \tilde{h}_{\gamma^t}$. The chained fit procedure first fits the two-point correlator $ C_{\rm 2pt} $, then uses the posterior distributions of $ E_0 $, $ E_1 $, $ z_0 $, and $ z_1 $ as priors in the subsequent fit of $ R_{\tilde{h}} $.  

For illustration, six examples of the chained fit applied to the quasi-TMD matrix elements are shown in Fig.~\ref{fig:ratio_fit_qtmdpdf}. The error bars represent the data points of the ratio $R_{\tilde{h}}$, the colored bands depict the results of the two-state fit, and the gray band indicates the extracted ground-state matrix element. Additional details on the ground-state fit can be found in App.~\ref{app:gsfit}.  

The extracted bare matrix elements of the quasi-TMD function in the coordinate space are presented in Fig.~\ref{fig:bare_zdep_qtmdpdf}. The transverse separation $ b_\perp $ is plotted up to $ 1.2 $ fm for three different hadron momenta: $ P^z = 1.83 $, $ 2.43 $, and $ 3.04 $ GeV. In all cases, the quasi-TMD function decreases as $ b_\perp $ increases and asymptotically approaches zero in the large-$z$ regime, consistent with expected physical behavior.

\subsection{Renormalization and extrapolation} 

\begin{figure*}[th!]
    \centering
    \includegraphics[width=.32\linewidth]{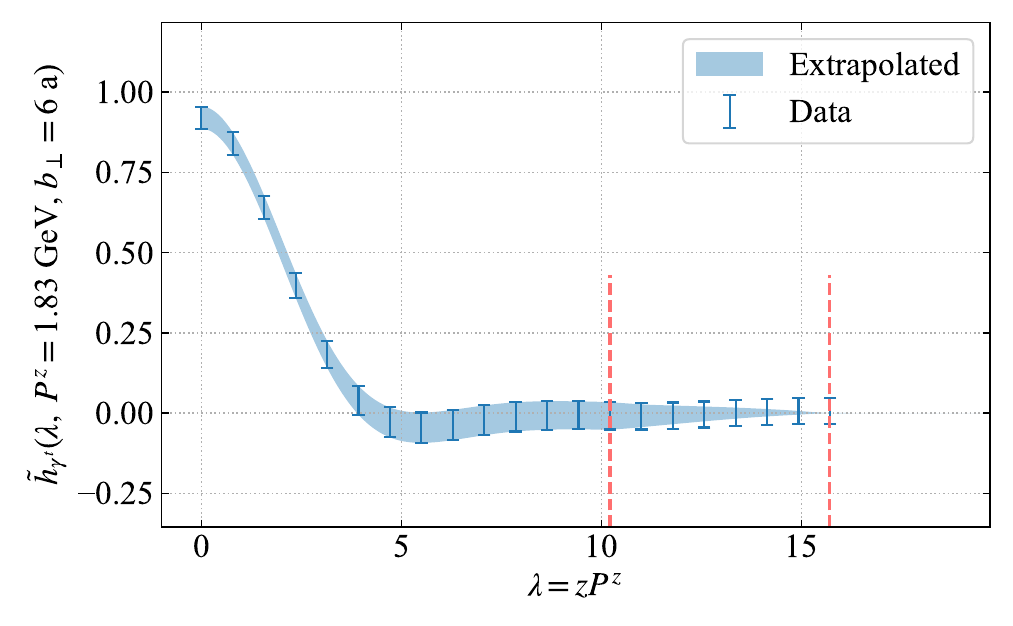}
    \includegraphics[width=.32\linewidth]{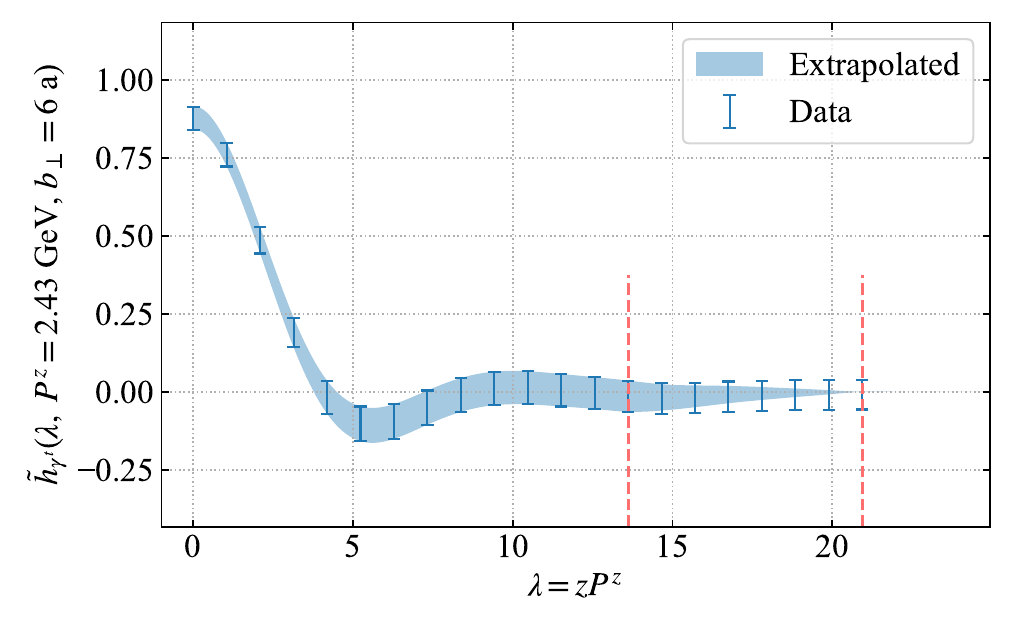}
    \includegraphics[width=.32\linewidth]{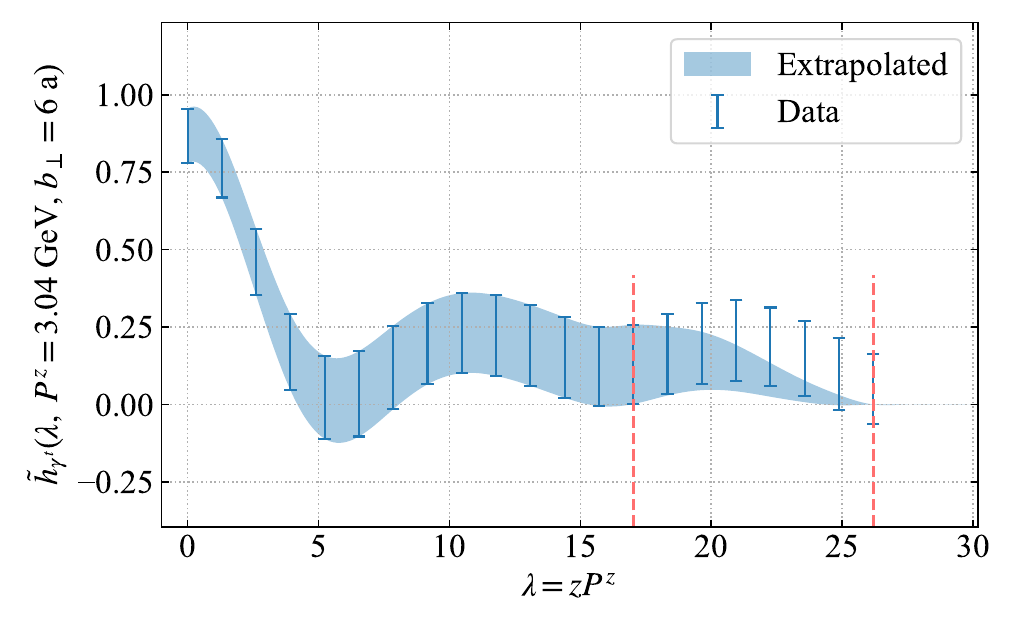}
    \caption{The extrapolation of the renormalized quasi-TMD in coordinate space for various hadron momenta is shown in three panels. From left to right, they correspond to hadron momenta $P^z=1.83$, 2.43 and 3.04 GeV, respectively. The regions between the two red dashed lines indicate the extrapolation range using Eq.~\eqref{eq:extrapolation_form}. For different hadron momenta, the same starting point of $z = 0.78$ fm is chosen for the extrapolation.
    \label{fig:extra_qtmdpdf}}
\end{figure*}

As discussed in Ref.~\cite{Gao:2023lny}, the absence of Wilson lines in the CG correlator eliminates linear divergences, allowing the renormalized operator to be defined as  
\begin{align}
    \bar{q}_0 (z,b_\perp) \Gamma q_0 (0) = Z_q (a) \left[ \bar{q} (z,b_\perp) \Gamma q(0) \right] ~,
\end{align}  
where $Z_q$ is the CG quark wave function renormalization factor. This renormalization is independent of both the external hadron states and the spatial separation of the quark bilinear operator. Consequently, an appropriate ratio can be constructed to cancel out the renormalization factor.

Following the approach in Ref.~\cite{Bollweg:2024zet}, we renormalize the quasi-TMD matrix elements using the ratio  
\begin{align}
    \tilde{h}_{\gamma^t}(z, b_\perp, P^z; \mu) = \frac{\tilde{h}^0_{\gamma^t}(z, b_\perp, P^z; a)}{\tilde{\varphi}^0_{\gamma^t \gamma^5}(z=0, b_\perp, P^z=0; a)} ~,
    \label{eq:renorm_qtmdpdf}
\end{align}  
and similarly for the quasi-TMDWF, it is renormalized as
\begin{align}
    \tilde{\varphi}_{\gamma^z \gamma^5}(z, b_\perp, P^z; \mu) = \frac{\tilde{\varphi}^0_{\gamma^z \gamma^5}(z, b_\perp, P^z; a)}{\tilde{\varphi}^0_{\gamma^t \gamma^5}(z=0, b_\perp, P^z=0; a)} ~.
    \label{eq:renorm_qtmdwf}
\end{align}  
Here, $\tilde{\varphi}^0$ denotes the bare quasi-TMDWF matrix elements defined in Eq.~\eqref{eq:qtmdwf_def_2} and the details of extraction can be found in App.~\ref{app:tmdwf}. 

The renormalized matrix elements in the coordinate space are presented in \fig{extra_qtmdpdf}. As shown, the quasi-TMD matrix elements decay rapidly as a function of $\lambda = zP^z$, reaching approximately zero for $\lambda \gtrsim 5$. However, at large distances, while the values remain statistically consistent with zero, the statistical uncertainties persist at a constant level, which will lead to non-physical fluctuations in the direct Fourier transform. Due to the finite correlation length of spatial correlators in QCD~\cite{Gao:2021dbh}, the quasi-TMD matrix elements in coordinate space are expected to exhibit exponential decay when the coordinate separation $z$ is large. Moreover, as demonstrated in Ref.~\cite{Gao:2021dbh}, the extracted quasi-distributions in momentum space within the moderate $x$ region remain largely insensitive to the choice of extrapolation strategy. Therefore, we apply a non-fit extrapolation method to smooth the uncertainties of the renormalized quasi-TMD matrix elements at long distances. The extrapolation is performed using the following form:  
\begin{align}
    \tilde{h}^{\rm ext} = w \cdot \tilde{h} + (1 - w) \cdot 0 ~,
    \label{eq:extrapolation_form}
\end{align}  
where $w$ is a weight function that transitions linearly from 1 to 0 within the range $\lambda \in [z_{\rm ext}P^z, z_{\rm max} P^z]$, the $z_{\rm ext} = 0.78$ fm is the starting point of extrapolation and $z_{\rm max} = 1.2$ fm is the largest longitudinal separation of quasi-TMD. It is expected that $z > z_{\rm ext}$ is large enough to see the exponential decay behavior of quasi-TMD. In addition, the comparison of TMDPDF using different $z_{\rm ext}$ can be found in App.~\ref{app:extrapolation}. The extrapolation range is indicated by the two red dashed lines in Fig.~\ref{fig:extra_qtmdpdf}. After applying this extrapolation, the uncertainty bands smoothly converge to zero, mitigating non-physical fluctuations.

\section{Pion valence quark TMDPDF}
\label{sec:tmdpdf}

As shown in the factorization formula Eq.~\eqref{eq:factorization_tmdpdf}, the computation of the unpolarized pion valence-quark TMDPDF relies on three key inputs: the Collins-Soper (CS) kernel, the intrinsic soft function, and the quasi-TMD beam function matrix elements discussed in Sec.~\ref{sec:quasi-tmd}. In this section, we present the numerical results for each of these components and, ultimately, the extracted unpolarized valence TMDWF and TMDPDF of the pion.

\begin{figure}[th!]
    \centering
    \includegraphics[width=0.9\linewidth]{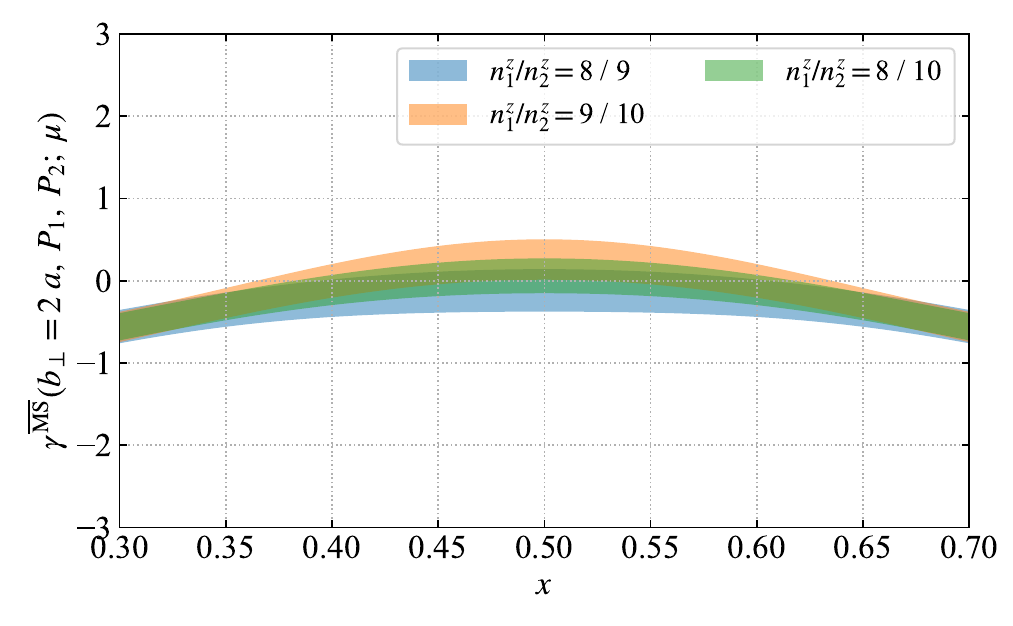}
    \includegraphics[width=0.9\linewidth]{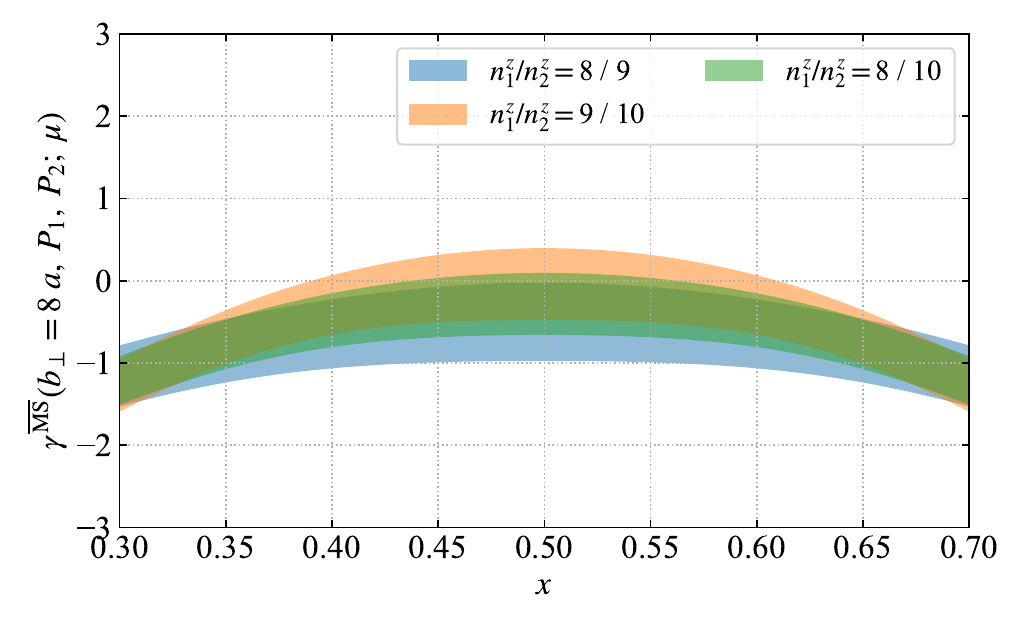}
    \caption{The ratio $\gamma^{\overline{\rm MS}}(b_\perp, P_1, P_2; \mu)$, as defined in Eq.~\eqref{eq:cs_kernel_calc}, of combinations of momentum $ n_1^z/n_2^z $. The results at $\mu=2$~GeV for $b_\perp = 2a$ (upper panel) and $b_\perp = 8a$ (lower panel) are presented.
    }
    \label{fig:cs_kernel_different_p}
\end{figure}

\subsection{The Collins-Soper kernel}

The CS kernel can be extracted from the ratio of quasi-TMDWFs with different momenta, as described in Eq.~\eqref{eq:cs_kernel_calc}. The bare matrix elements of the quasi-TMDWFs are extracted in App.~\ref{app:tmdwf}, which follows the same strategy as in Ref.~\cite{Bollweg:2024zet}. For the matching procedure, we use the fixed-order one-loop results for the CG matching coefficient $ C_{\text{TMD}} $ and the corresponding hard kernel $ H_\phi $, as provided in Ref.~\cite{Zhao:2023ptv}. Furthermore, we account for large logarithmic terms in the hard kernel by applying renormalization group evolution to improve the accuracy of the perturbative matching up to NLL; details on this resummation can be found in App.~\ref{app:rg_resum}.

\fig{cs_kernel_different_p} shows the ratio $ \gamma^{\overline{\rm MS}}(b_\perp, P_1, P_2; \mu) $, as defined in Eq.~\eqref{eq:cs_kernel_calc}, which is extracted from different combinations of momentum $ n_1^z/n_2^z $. The results are consistent across different momentum ratios within the uncertainty bands. However, a slight $ x $ dependence and variations between different momentum ratios suggest the presence of power corrections, since the pion mass $m_\pi=670$ MeV is quite heavy, contributing to systematic uncertainties. To estimate these uncertainties, we select two sets of closely spaced momentum values: $ n^z_1 / n^z_2 = 8 / 9 $ and $ n^z_1 / n^z_2 = 9 / 10 $. Within each Jackknife sample, we collect two momentum pairs and include all data points over the interval $ x \in [0.34, 0.66] $. The mean value $ \langle \gamma \rangle_i $ and the standard deviation $ \sigma_i $ are then calculated for each Jackknife sample $ i $. The final mean value, along with the corresponding statistical and systematic uncertainties, is estimated as follows:
\begin{align}
\begin{aligned}
    \text{Mean} &= \text{Avg}[ \langle \gamma \rangle_i ] \\
    \text{Stat} &= \text{Std}[ \langle \gamma \rangle_i ] \cdot \sqrt{N_s - 1} \\
    \text{Sys} &= \text{Avg}[ \sigma_i ] ~,
\end{aligned}
\end{align}
where ``Avg" means taking the average and ``Std" means taking the standard deviation, the factor $\sqrt{N_s - 1}$ arises from the Jackknife resampling procedure.
\begin{figure}[th!]
    \centering
    \includegraphics[width=.9\linewidth]{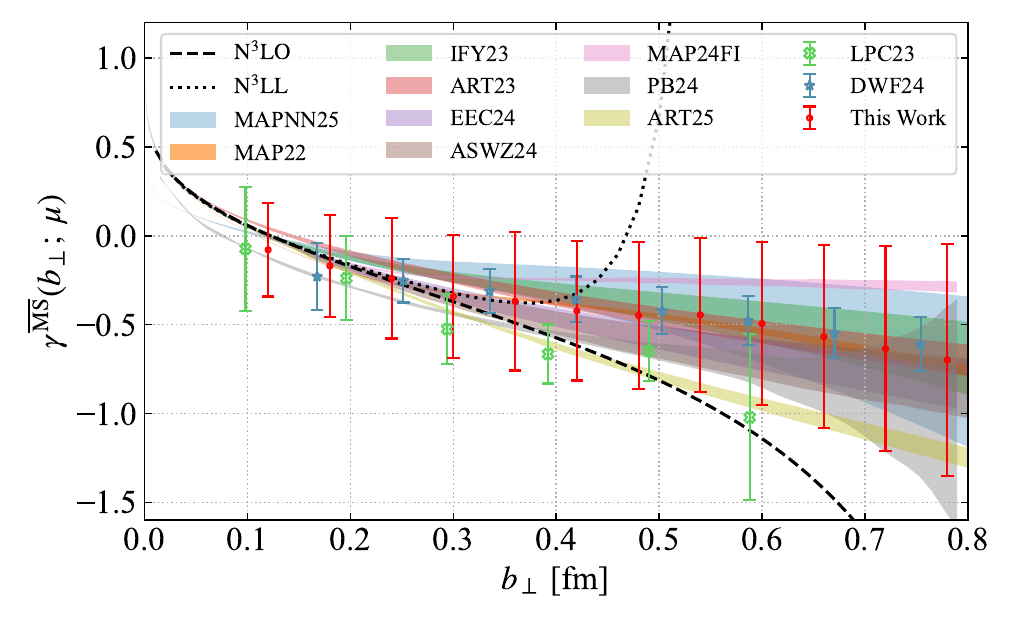}
    \caption{The CS kernel at $\mu = 2$ GeV in this work is presented using red points with error bars. The 3-loop perturbative results~\cite{Li:2016ctv, Vladimirov:2016dll} for the CS kernel are denoted as $\rm N^3LO$ and $\rm N^3LL$ in the figure. The CS kernels from recent phenomenological parameterizations of experimental data are shown from MAP22~\cite{Bacchetta:2022awv}, IFY23~\cite{Isaacson:2023iui}, ART23~\cite{Moos:2023yfa},
    MAP24FI~\cite{Bacchetta:2024qre},
    EEC24~\cite{Kang:2024dja},
    PB24~\cite{Martinez:2024mou},
    MAPNN25~\cite{Bacchetta:2025ara} and
    ART25~\cite{Moos:2025sal}. 
    In addition, some recent lattice calculations are presented from LPC23~\cite{LatticePartonLPC:2023pdv}, ASWZ24~\cite{Avkhadiev:2024mgd} and DWF24~\cite{Bollweg:2024zet}.
    \label{fig:cs_kernel}}
\end{figure}

After incorporating both statistical and systematic uncertainties, Fig.~\ref{fig:cs_kernel} presents our final results for the CS kernel at $\mu = 2$ GeV. In the small $ b_\perp $ region, our results align well with the three-loop perturbative calculations from Refs.~\cite{Li:2016ctv, Vladimirov:2016dll}, labeled $ \rm N^3LO $ and $ \rm N^3LL $. Moreover, our calculation remains reliable in the large $ b_\perp $ region, where perturbative methods break down.

To further contextualize our findings, we compare them with CS kernels extracted from recent global fits of experimental data, including MAP22~\cite{Bacchetta:2022awv}, IFY23~\cite{Isaacson:2023iui}, ART23~\cite{Moos:2023yfa},
MAP24FI~\cite{Bacchetta:2024qre},
EEC24~\cite{Kang:2024dja},
PB24~\cite{Martinez:2024mou},
MAPNN25~\cite{Bacchetta:2025ara}
and ART25~\cite{Moos:2025sal}. Additionally, we include recent lattice QCD results from LPC23~\cite{LatticePartonLPC:2023pdv}, ASWZ24~\cite{Avkhadiev:2024mgd}, and DWF24~\cite{Bollweg:2024zet}. 

A particularly important observation is the agreement between our calculation and the DWF24 result from chirally symmetric domain-wall fermion configurations, both of which employ the CG framework despite using different lattice actions. Moreover, it is observable that the two most recent global analysis results (MAP24FI and ART25) exhibit a deviation from their prior results in different directions. Nevertheless, both remain consistent with this work due to the large uncertainty in our CS kernel, primarily stemming from power corrections at such a heavy pion mass, which resulted in non-flat curves in \fig{cs_kernel_different_p}. In our future study, the precision of our CS kernel could be refined by adopting a smaller valence pion mass.

\subsection{Intrinsic soft function}

\begin{figure*}[th!]
    \centering
    \includegraphics[width=.32\linewidth]{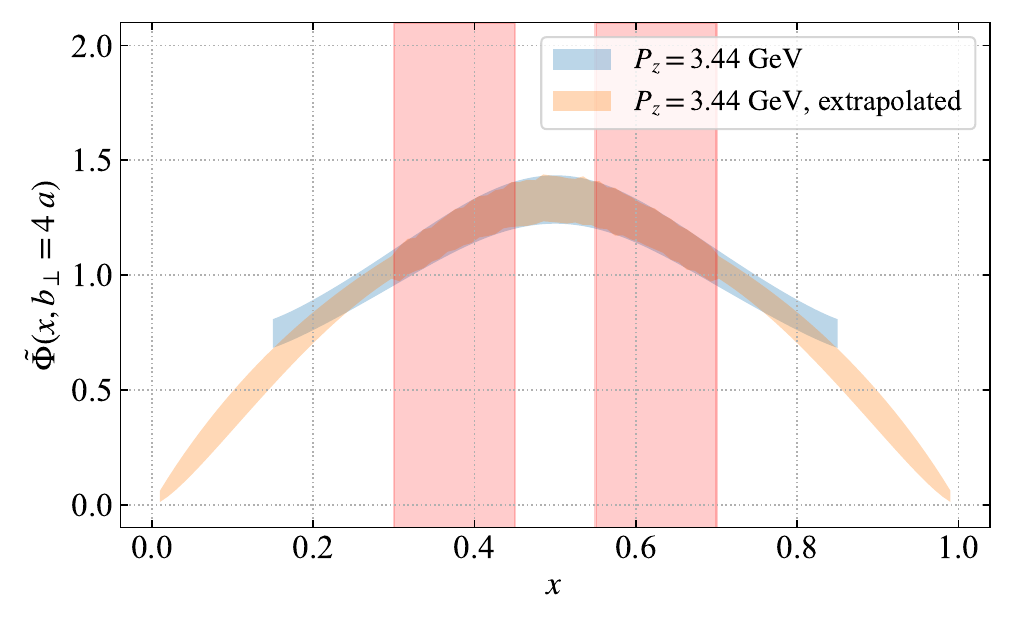}
    \includegraphics[width=.32\linewidth]{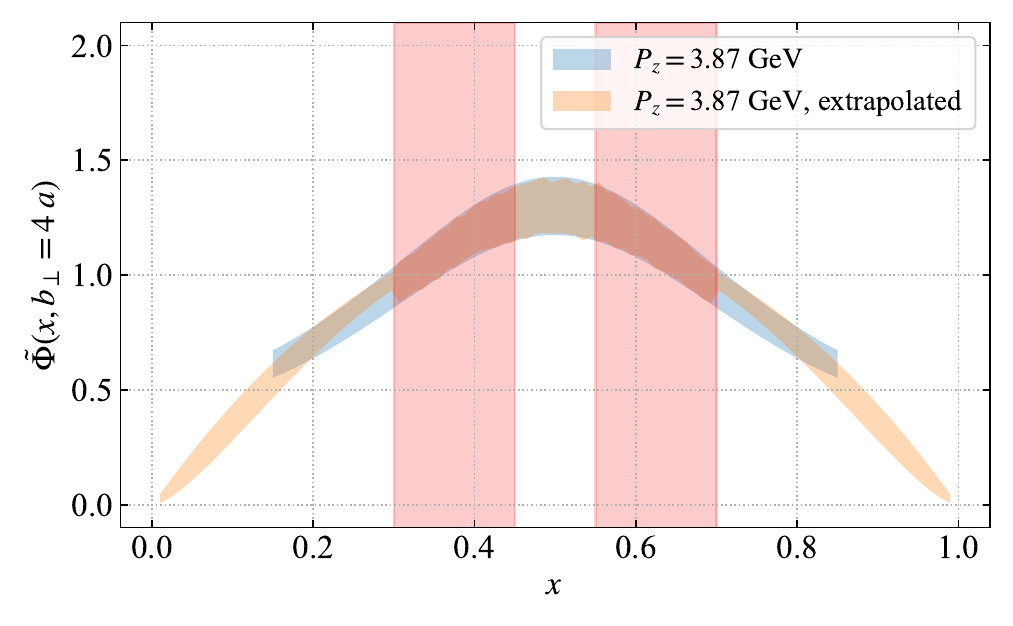}
    \includegraphics[width=.32\linewidth]{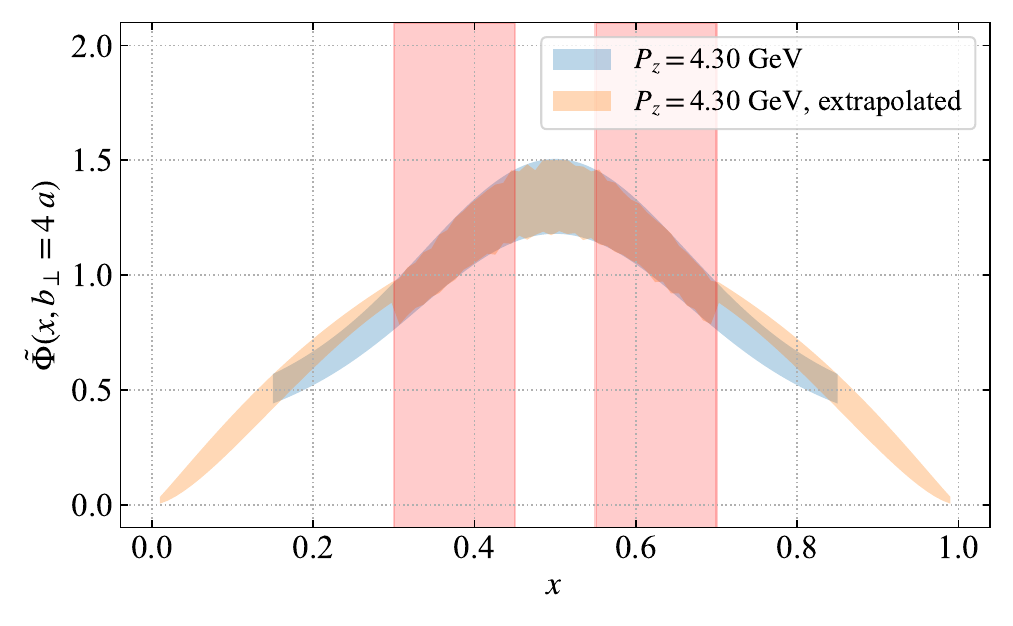}
    \caption{
    Extrapolations of reduced quasi-TMDWF using Eq.~\eqref{eq:reduce_wf_extra}. From left to right, the three panels correspond to hadron momenta $P^z=3.44$, 3.87 and 4.30 GeV, respectively. All of three cases are evolved to $P_{\rm FF} = 2.58$ GeV using the rapidity evolution of quasi-TMDWF. The shaded red bands indicate the regions constitute the input for the extrapolation fit.
    \label{fig:reduced_qtmdwf_extra}}
\end{figure*}

\begin{figure}[th!]
    \centering
    \includegraphics[width=.9\linewidth]{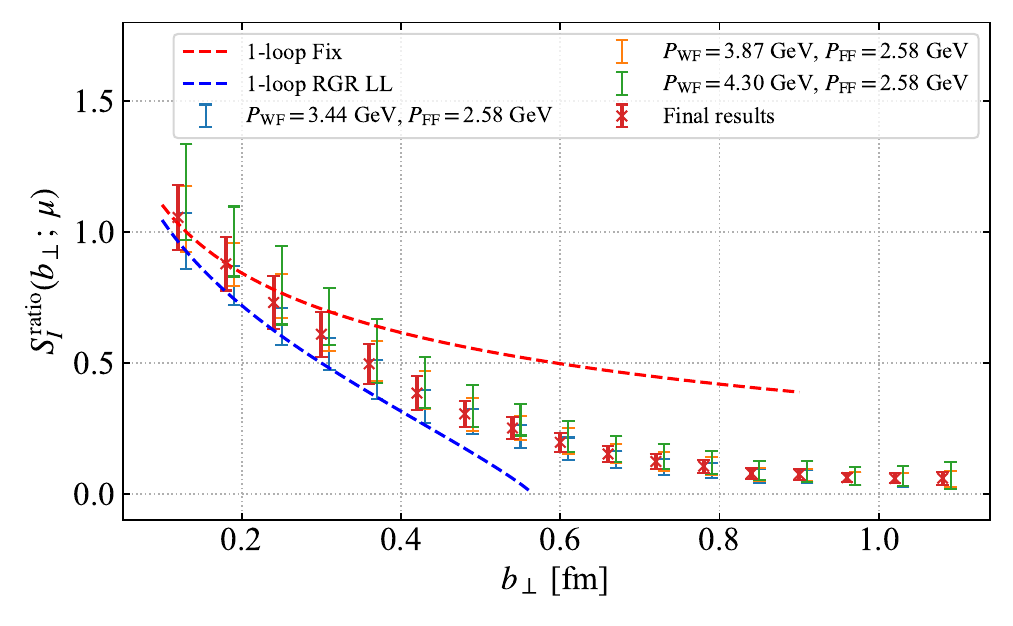}
    \caption{Intrinsic soft function in CG at $\mu = 2$ GeV in the ratio scheme, for different hadron momentum pairs of the quasi-TMDWF, $P_\mathrm{WF}$,  and form factor, $P_\mathrm{FF}$. The final results are shown as red points with error bars (see text for details). The corresponding one-loop fixed-order perturbative results are the red dashed line, and the one-loop RG-resummed (RGR) perturbative results are the blue dashed line. 
    \label{fig:softf}}
\end{figure}

The intrinsic soft function can be extracted from the analysis of the form factor with a large momentum transfer, aided by the quasi-TMDWF, as discussed in \sec{softfunction}. The details of extracting the bare quasi-TMDWF and the form factor can be found in Apps.~\ref{app:tmdwf} and \ref{app:ff}.

To extract the intrinsic soft function using Eq.~\eqref{eq:soft_func_calc}, we integrate the quasi-TMDWF over $ x_1 $ and $ x_2 $. However, near the endpoint regions ($ x \approx 0 $ and $ x \approx 1 $), quasi-distributions suffer from significant power corrections. To mitigate this issue, we extrapolate the reduced quasi-TMDWF $ \tilde{\Phi} $ using the functional form  
\begin{align}
    \tilde{\Phi}(x) = c x^n (1-x)^n ~,
    \label{eq:reduce_wf_extra}
\end{align}
where $ c $ and $ n $ are fitting parameters. A comparison of the reduced quasi-TMDWF before and after extrapolation is shown in Fig.~\ref{fig:reduced_qtmdwf_extra}, where the three panels correspond to hadron momenta $P^z=3.44$, 3.87 and 4.30 GeV, respectively. All three cases are evolved to $P_{\rm FF} = 2.58$ GeV using the rapidity evolution of quasi-TMDWF. The red shaded bands indicate the regions that constitute the input for the extrapolation fit, with the selected fit range of $ x \in [0.3, 0.45]$ and $x \in [0.55, 0.7]$, where the results of three different momenta show consistency. The extrapolation form derived from the fitting process is applied to the endpoint regions with $x<0.3$ or $x>0.7$. Fig.~\ref{fig:reduced_qtmdwf_extra} shows that the extrapolated results outside the fit range are in alignment with the observed trends. Furthermore, since the Sudakov kernel diverges near the endpoints ($x \approx 0$) after the RG resummation, the integral in Eq.~\eqref{eq:soft_func_calc} is restricted to the range $ x_1, x_2 \in [0.05, 0.95] $. This cutoff has a small impact on the results of the intrinsic soft function, since the integrand in the denominator of Eq.~\eqref{eq:soft_func_calc} converges near the endpoint regions.

To assess systematic effects of power corrections, we performed integration with three different momentum pairs, formed by the momentum of the form factor $P_{\rm FF} = 2.58$ GeV and three momenta of the quasi-TMDWF $P_{\rm WF} = 3.44$, 3.87 and 4.30 GeV. In order to bridge the gap between $P_{\rm FF}$ and $P_{\rm WF}$, the quasi-TMDWF is evolved to $P_{\rm FF}$ by solving the rapidity evolution equation. The extracted intrinsic soft function, according to Eq.~\eqref{eq:soft_func_calc}, is shown in \fig{softf}, where comparisons across different momentum pairs indicate that power corrections have only a minor impact. Note that the superscript ``ratio" of $S_I^{\rm ratio}$ indicates the ratio scheme in the renormalization of the quasi-TMDWF.

For the final results of $ S_I^{\rm ratio}$, the central values and statistical uncertainties are determined from a correlated average over the three momentum pairs formed by $P_{\rm FF}$ and $P_{\rm WF}$. The systematic uncertainties are determined by the spread of the mean values between these three momentum pairs, quantified as half the absolute difference between the maximum and minimum values. The extracted intrinsic soft function at $\mu = 2$ GeV, shown as red points with error bars in \fig{softf}, is compared to perturbative predictions of fixed order and RG resummed (RGR) at leading logarithmic (LL) precision. Further details on RG resummation can be found in App.~\ref{app:rg_resum}. In particular, at small $ b_\perp $, our lattice results exhibit reasonable agreement with perturbative predictions, highlighting the robustness of our approach. In the large $ b_\perp $ region, where perturbation theory becomes unreliable, our lattice results are expected to provide a more accurate description of the intrinsic soft function, offering valuable non-perturbative insights.

\subsection{Pion TMDWF in $x$ space}
\begin{figure}[th!]
    \centering
    \includegraphics[width=.9\linewidth]{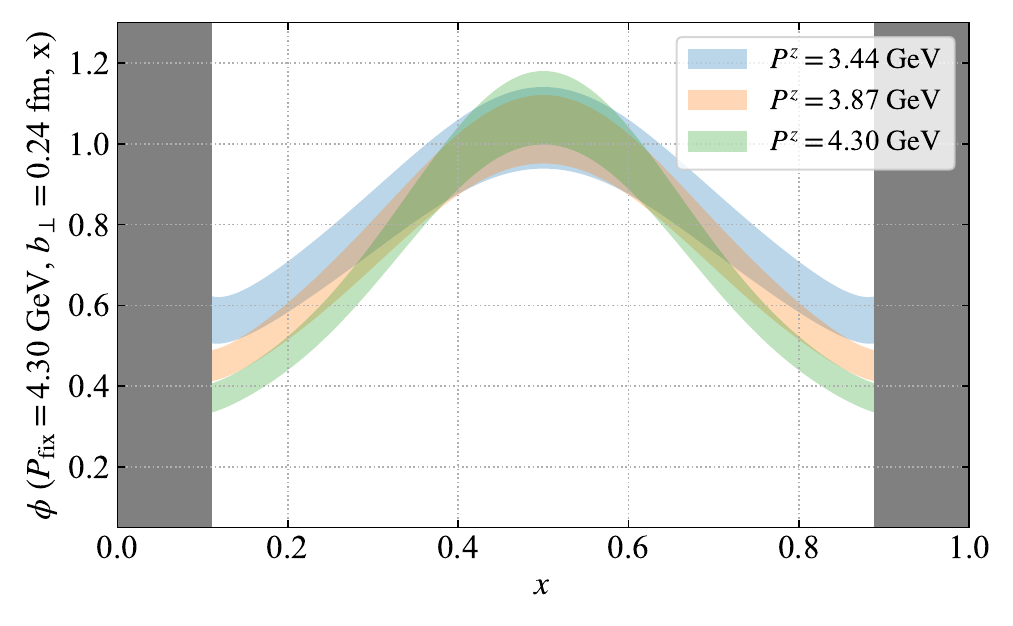}
    \includegraphics[width=.9\linewidth]{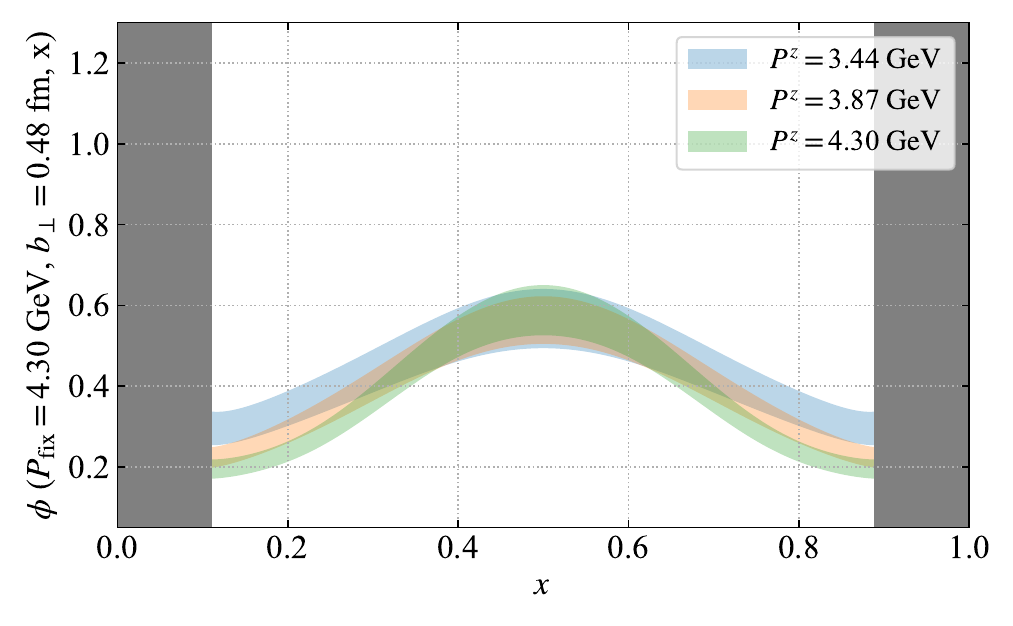}
    \caption{The light-cone pion TMDWF at $y_n = 0$, $P_\mathrm{fix} = 4.30$ GeV and $\mu = 2$ GeV of different hadron momenta, $P^z$, as the functions of momentum fraction, $x$, and for two transverse separations $b_\perp = 4a$ (upper panel) and $b_\perp = 8a$ (lower panel). The shaded gray bands ($x < 0.11$ and $x > 0.89$) indicate the endpoint regions where the estimated combined systematics are larger than $30\%$. The detailed estimation of the systematics is explained in App.~\ref{app:power_correction}.}
    \label{fig:tmdwf_xdep}
\end{figure}

By integrating the CG kernel, intrinsic soft function, and the renormalized quasi-TMDWF, the light-cone TMDWF can be extracted by employing the factorization formula found in Eq.~\eqref{eq:factorization_tmdwf_short}. Choosing scales $y_n=0$, $\zeta = (2x P_\mathrm{fix})^2$, and $\bar{\zeta} = (2\bar{x} P_\mathrm{fix})^2$, the factorization formula can be rewritten in a more explicit form:
\begin{align}
\begin{aligned}
    &\sqrt{S_I\left(b_{\perp} ; \mu\right) } \cdot \tilde{\phi}_{\Gamma}\left(x, b_{\perp}, P^z; \mu\right) = \phi\left(x, b_{\perp}; \mu, \zeta, \bar{\zeta} \right) \\ 
    & \quad \quad \times  H_\phi \left(x, \bar{x}, P^z; \mu\right)  \exp \left[\ln \left( \frac{P^z}{P_\mathrm{fix}} \right) \gamma^{\overline{\mathrm{MS}}}\left(b_{\perp}; \mu\right)\right] \\
    & \quad \quad +\mathcal{O}\left(\frac{\Lambda_{\mathrm{QCD}}}{x P^z}, \frac{1}{b_{\perp}\left(x P^z\right)}, \frac{\Lambda_{\mathrm{QCD}}}{\bar{x} P^z}, \frac{1}{b_{\perp}\left(\bar{x} P^z\right)}\right) ~,
    \label{eq:factorization_tmdwf}
\end{aligned}
\end{align}
where the CS scale is evolved from $\zeta_0 = (2x P^z)^2$ and $\bar{\zeta}_0$ to $\zeta = (2x P_\mathrm{fix})^2$ and $\bar{\zeta}$ using the CS kernel extracted from quasi-TMDWFs.

The final results for the TMDWF at $y_n = 0$, $P_\mathrm{fix} = 4.30$ GeV, and $\mu = 2$ GeV are shown in \fig{tmdwf_xdep} as a function of the momentum fraction $ x $ for three different hadron momenta. Two representative transverse separations, $ b_\perp = 4a $ (upper panel) and $ b_\perp = 8a $ (lower panel), are selected for illustration. As can be seen, the variation between different momenta remains mild in the moderate $ x $ region, demonstrating the validity of power expansion in large $ P^z $ within the quasi-TMD framework, where power corrections are small. The shaded gray bands ($x < 0.11$ and $x > 0.89$) indicate the endpoint regions where the estimated combined systematics are greater than $30\%$. The combined systematics include the variation of scales and power corrections, which are estimated by the variation of the initial scale $\mu_0$ of the RGR procedure by a factor of $\sqrt{2}$ and the spread in the central values of the TMDWFs with different momenta, respectively. More detailed discussion can be found in App.~\ref{app:power_correction}.

\subsection{Pion TMDPDF in $b_\perp$ and $k_\perp$ space}

\begin{figure}[th!]
    \centering
    \includegraphics[width=.9\linewidth]{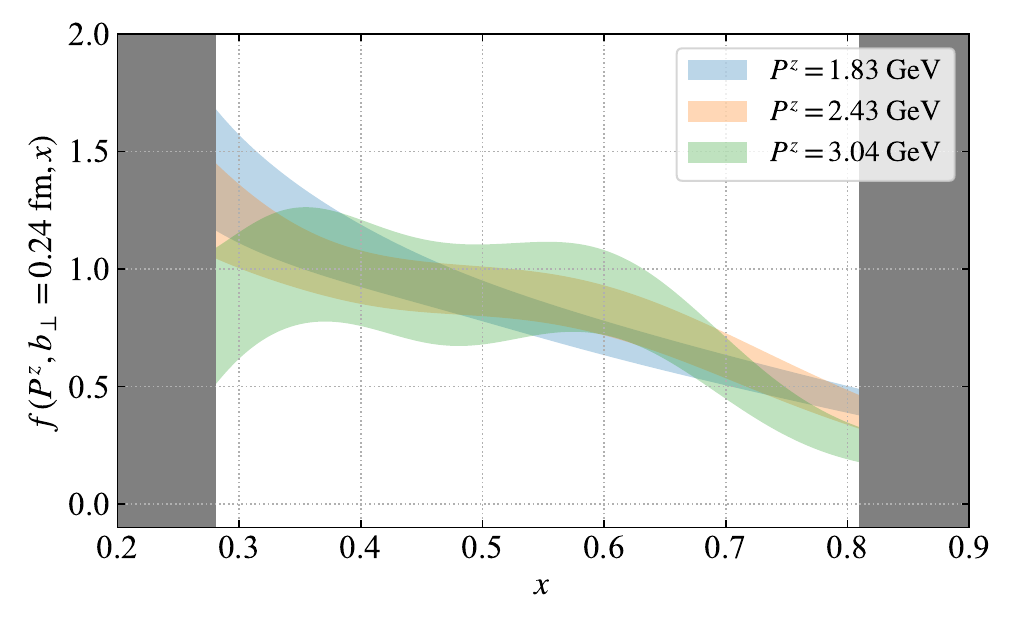}
    \includegraphics[width=.9\linewidth]{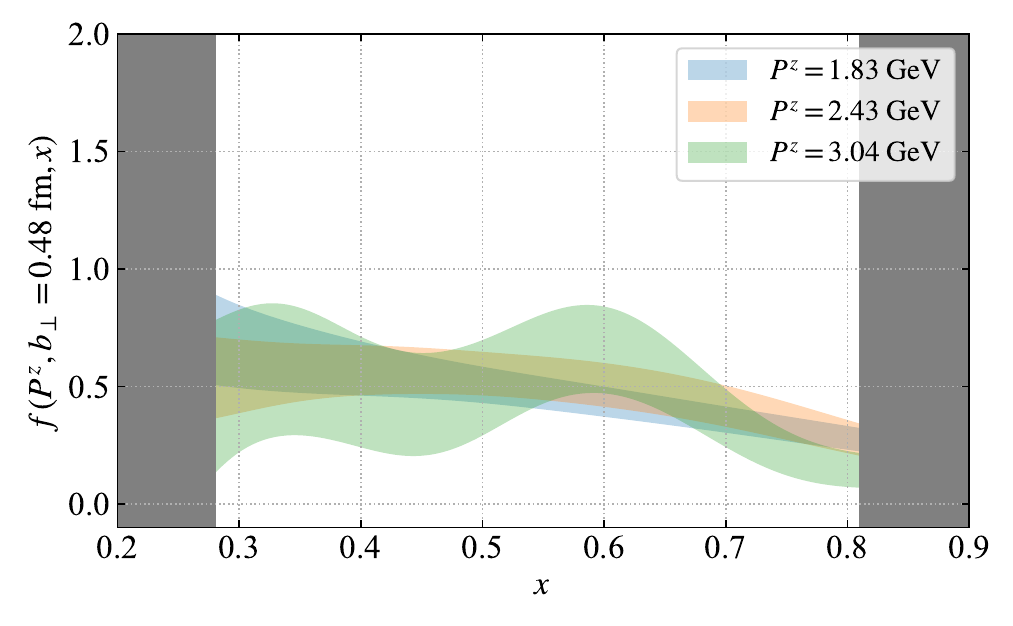}
    \caption{The unpolarized light-cone pion TMDPDF at $\zeta = \mu^2 = 4~\mathrm{GeV}^2$ of different hadron momenta are shown as the function of momentum fraction, $x$, and for transverse separations, $b_\perp = 4a$ (upper panel) and $b_\perp = 8a$ (lower panel). The shaded gray bands ($x < 0.28$ and $x > 0.81$) indicate the endpoint regions where the estimated combined systematics are larger than $30\%$. The detailed estimation of the systematics is explained in App.~\ref{app:power_correction}.
    \label{fig:tmdpdf_xdep}}
\end{figure}

\begin{figure}[th!]
    \centering
    \includegraphics[width=0.9\linewidth]{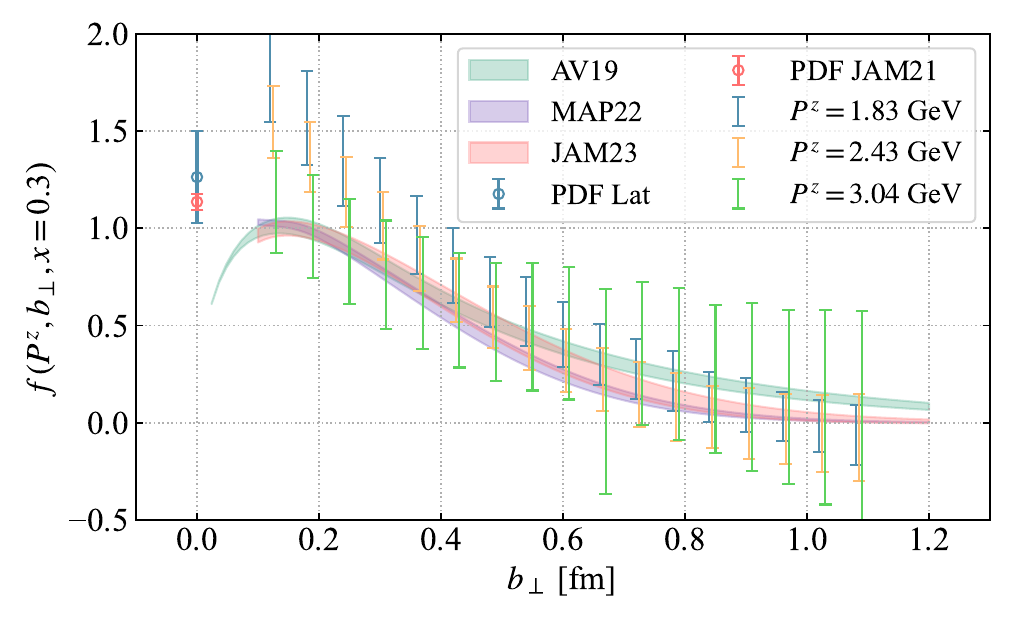}
    \includegraphics[width=0.9\linewidth]{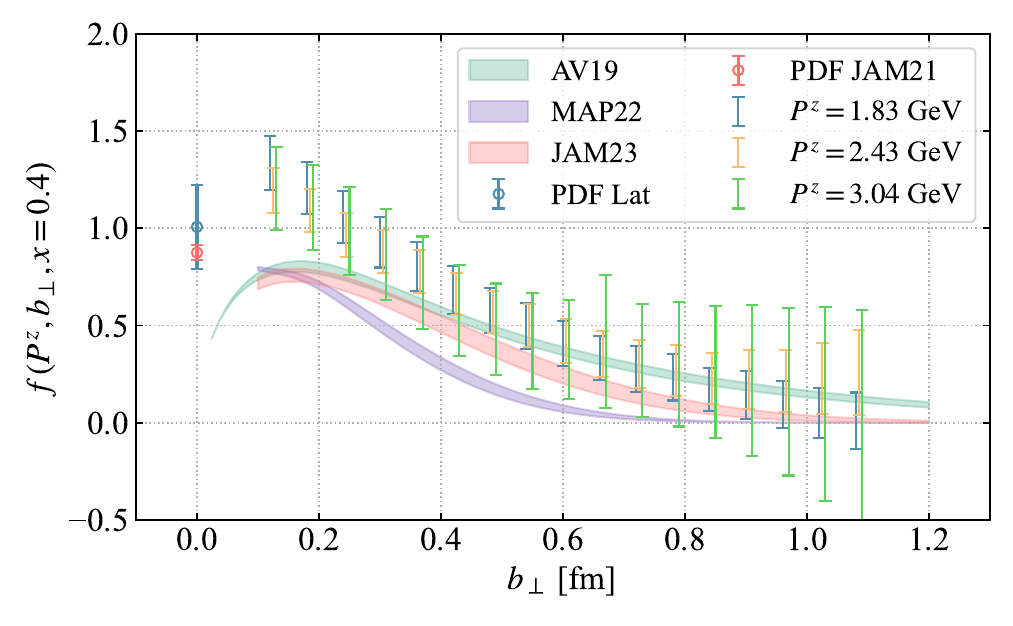}
    \includegraphics[width=0.9\linewidth]{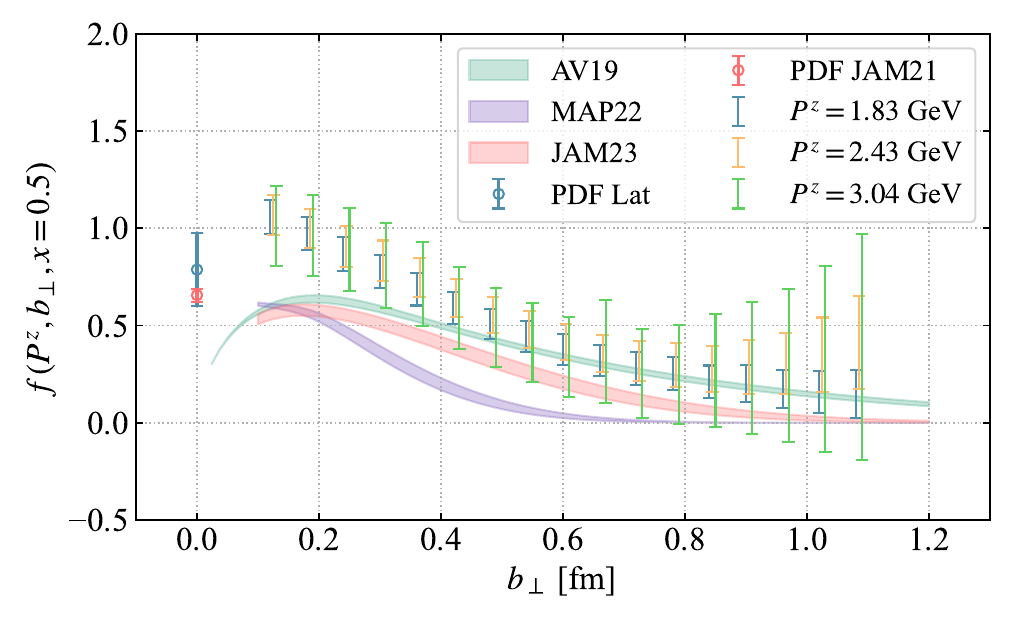}
    \caption{Unpolarized light-cone TMDPDF of pion at $x=0.3$, $x=0.4$ and $x=0.5$ in $b_\perp$ space up to $b_\perp \gtrsim 1$ fm, and for differnt hadron momenta, $P^z$. The lattice results are illustrated alongside the phenomenological results from 
    AV19~\cite{Vladimirov:2019bfa},
    MAP22~\cite{Cerutti:2022lmb} and JAM23~\cite{Barry:2023qqh}. In addition, the collinear PDF from lattice calculation in CG (PDF Lat)~\cite{Gao:2023lny} and global fit (PDF JAM21)~\cite{Barry:2021osv} are plotted for comparison.}
    \label{fig:tmdpdf_bdep}
\end{figure}

The combination of the CG kernel, intrinsic soft function, and renormalized quasi-TMD beam functions enables the extraction of the light-cone TMDPDF using the factorization formula in Eq.~\eqref{eq:factorization_tmdpdf}. The final results for the TMDPDF at $\zeta = \mu^2 = 4~\mathrm{GeV}^2$ are shown in \fig{tmdpdf_xdep} as a function of the momentum fraction $ x $ for three different hadron momenta. Two representative transverse separations, $ b_\perp = 4a $ (upper panel) and $ b_\perp = 8a $ (lower panel), are selected for illustration. As can be seen, the variation between different momenta remains mild in the moderate $ x $ region, demonstrating the validity of power expansion in large $ P^z $ within the quasi-TMD framework, where power corrections are small. The shaded gray bands ($x < 0.28$ and $x > 0.81$) indicate the endpoint regions where the estimated combined systematics are greater than $30\%$, with details provided in App.~\ref{app:power_correction}.

In \fig{tmdpdf_bdep}, we examine the $ b_\perp $ dependence of the TMDPDF at $ x = 0.3 $, $ x = 0.4 $, and $ x = 0.5 $. The results for different hadron momenta exhibit good consistency, which is improved as the $x$ value approaches $x=0.5$, thereby confirming our expectation of the behavior of power corrections. For comparison, we include recent global fits of the pion TMDPDF from AV19~\cite{Vladimirov:2019bfa}, MAP22~\cite{Cerutti:2022lmb} and JAM23~\cite{Barry:2023qqh}. Although slight deviations in magnitude are observed, the lattice results exhibit a qualitative agreement with global fits, especially at the smaller $x$ values, where the experimental data give a better constraint.

Since the collinear PDF and the $ b_\perp \to 0 $ limit of the unpolarized TMDPDF differ only by a perturbative expansion in $ b_\perp $~\cite{Moos:2023yfa}, we also compare our results to collinear PDFs. Specifically, we include the collinear PDF extracted from lattice calculations in CG (PDF Lat)~\cite{Gao:2023lny} and the global fit result from JAM21 (PDF JAM21)~\cite{Barry:2021osv}. As shown in the plots, the collinear PDFs closely match the TMDPDF in the small $ b_\perp $ region, supporting the expected theoretical behavior.

\begin{figure*}[th!]
    \centering
    \includegraphics[width=0.32\linewidth]{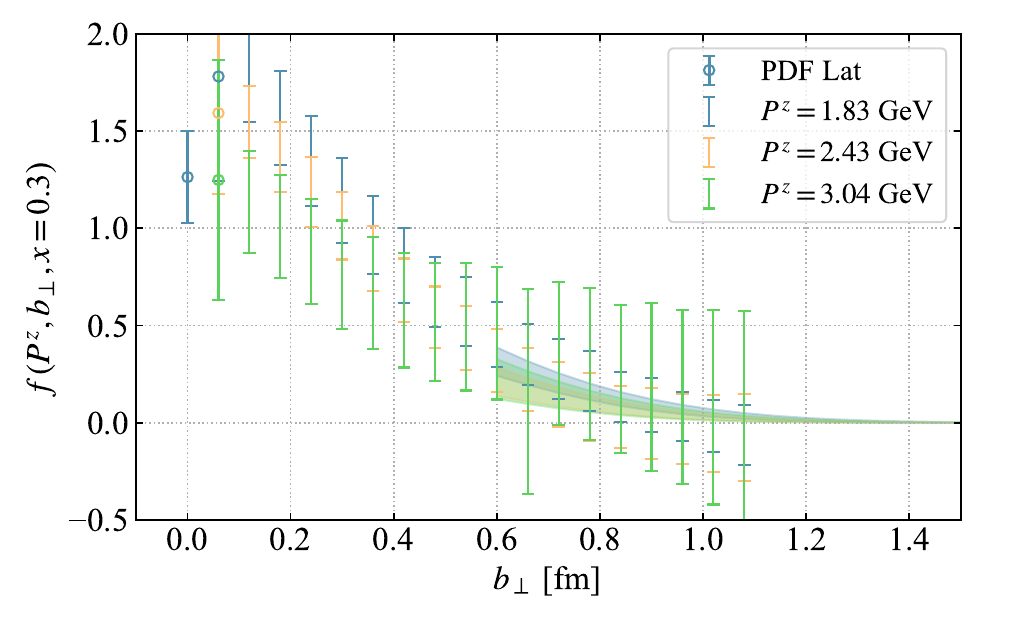}
    \includegraphics[width=0.32\linewidth]{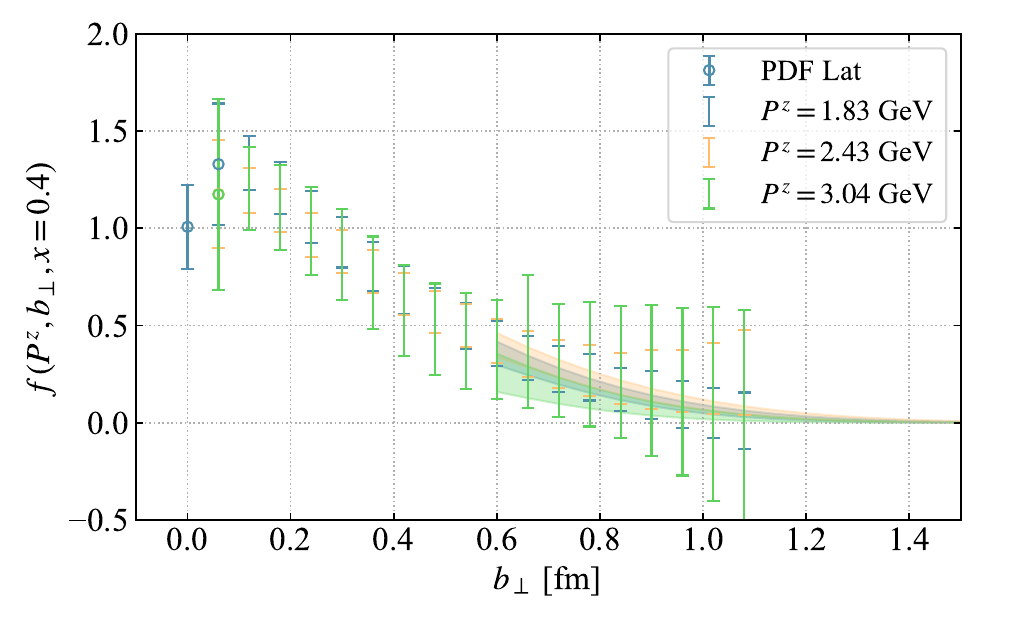}
    \includegraphics[width=0.32\linewidth]{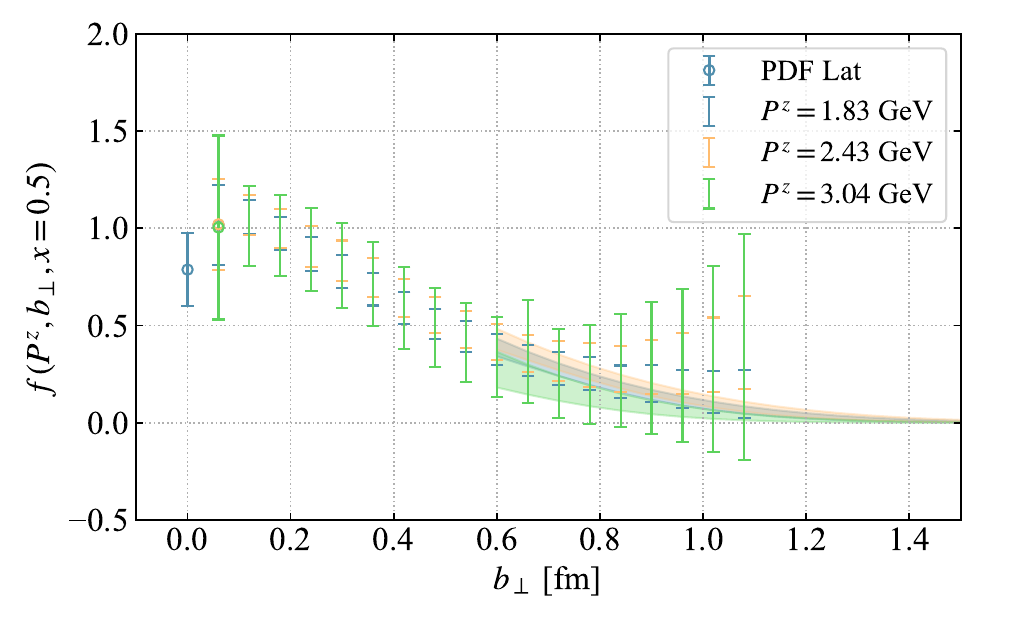}
    \caption{Unpolarized light-cone TMDPDF of pion at $x=0.3$, $x=0.4$ and $x=0.5$ in $b_\perp$ space after extrapolation to $b_\perp = 3$ fm. The collinear PDF from lattice calculation in CG (PDF Lat)~\cite{Gao:2023lny} is plotted at $b_\perp = $ 0 fm. The point at $b_\perp = 0.06$ fm is determined by cubic interpolation.}
    \label{fig:tmdpdf_bdep_extrapolated}
\end{figure*}

An important advantage of the CG method is the absence of linear divergence, which allows us to probe large $ b_\perp $ regions with a reasonable SNR. At $ b_\perp \gtrsim 1 $ fm, the TMDPDF smoothly decays to values consistent with zero, facilitating a feasible Fourier transform into the $ k_\perp $ space. In this study, we combined our lattice result for the collinear pion PDF (PDF Lat) with the pion TMDPDF to interpolate the distribution at short distances around $ b_\perp \approx a $, and subsequently applied a Gaussian form to extrapolate the large $b_\perp$ regime up to $3$ fm, the extrapolation form is
\begin{align}
    f (b_\perp) = A e^{- m \cdot b_\perp^2} ~,
    \label{eq:b_extrapolation_form}
\end{align}
where $A$ and $m$ are fit parameters. This extrapolation form is inspired by the confinement in the transverse plane, while the model dependence of the results will be investigated in future works with improved data precision. The extrapolated unpolarized light-cone TMDPDF of three momenta are plotted in Fig.~\ref{fig:tmdpdf_bdep_extrapolated}.

\begin{figure}[th!]
    \centering
    \includegraphics[width=0.9\linewidth]{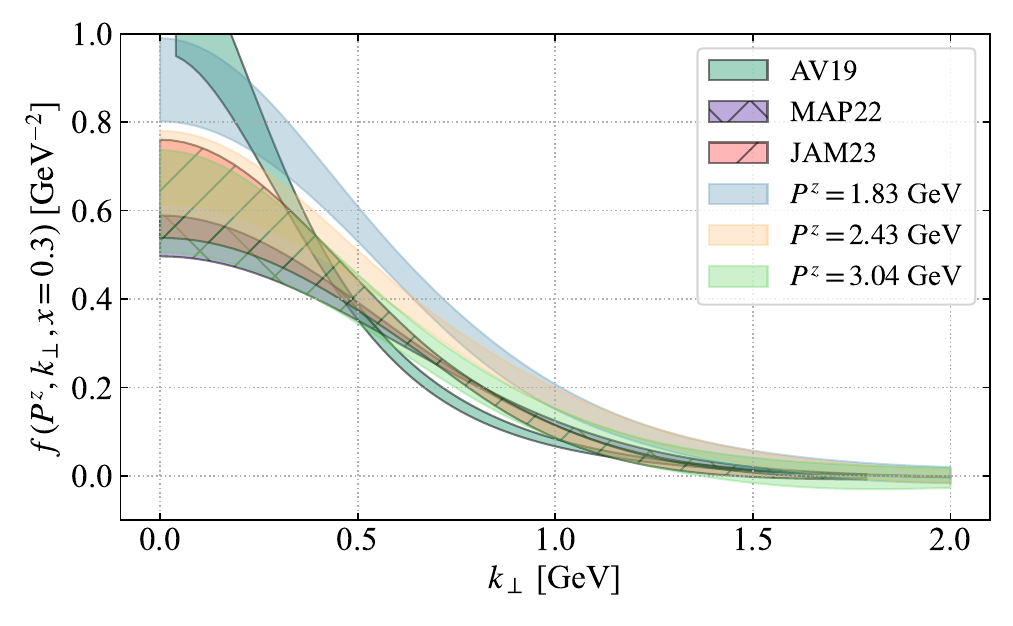}
    \includegraphics[width=0.9\linewidth]{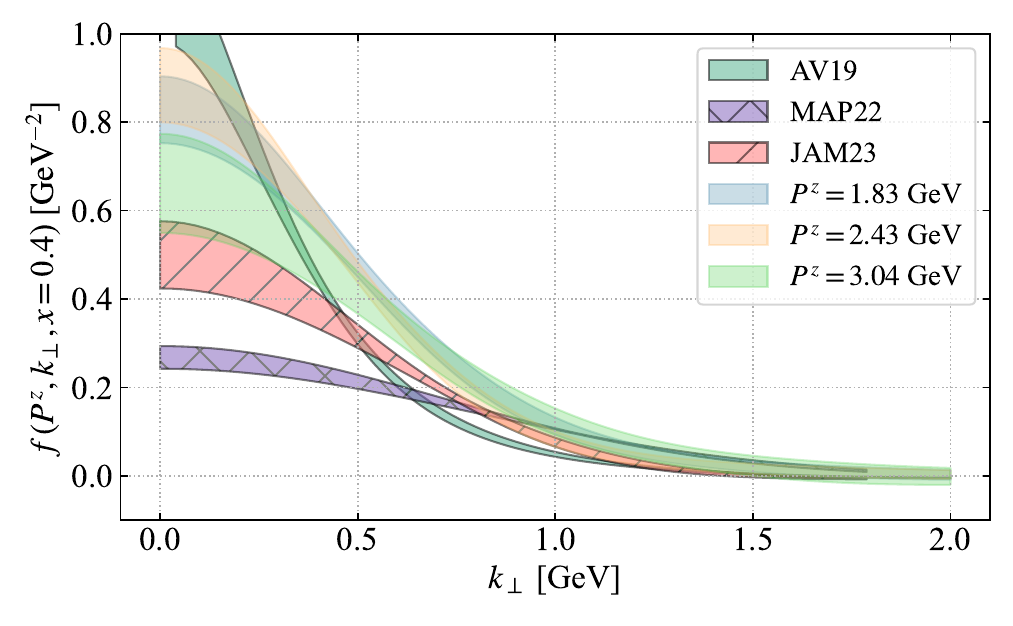}
    \includegraphics[width=0.9\linewidth]{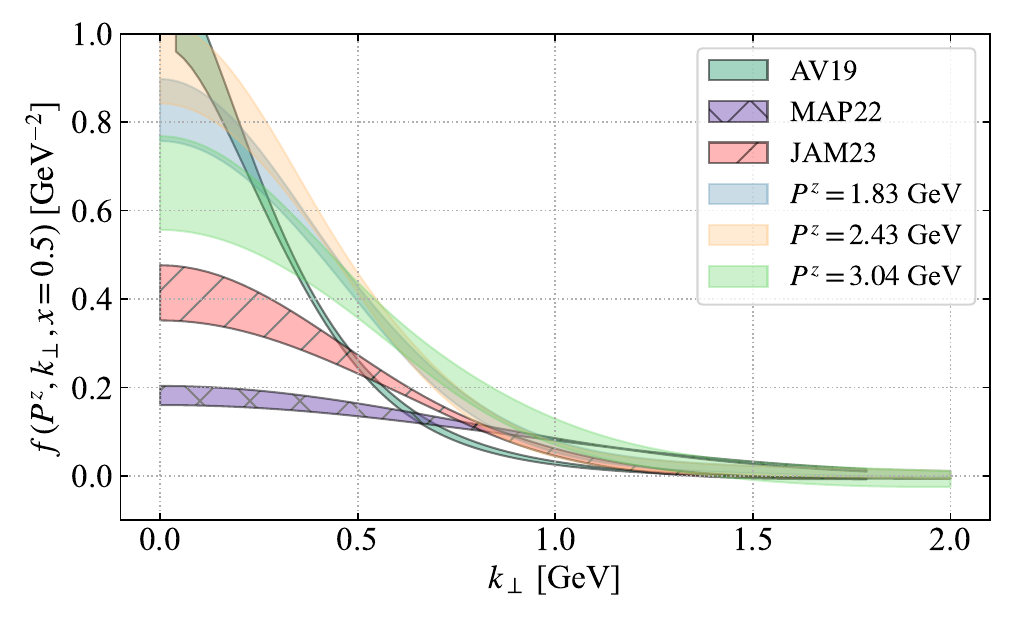}
    \caption{Unpolarized light-cone TMDPDF of pion at $x=0.3$, $x=0.4$ and $x=0.5$ in $k_\perp$ space. The lattice results are illustrated alongside the phenomenological results from 
    AV19~\cite{Vladimirov:2019bfa},
    MAP22~\cite{Cerutti:2022lmb} and JAM23~\cite{Barry:2023qqh}.} 
    \label{fig:tmdpdf_kdep}
\end{figure}

After extrapolation, we performed a discretized Fourier transform, with the results presented in \fig{tmdpdf_kdep}. Although the extrapolation in $b_\perp$ introduces minor model dependencies, the resulting distributions exhibit qualitative consistency with global fits, especially at the smaller $x$ values, where the experimental data give a better constraint. This demonstrates the potential of lattice QCD to explore the transverse momentum structure of hadrons from first principles. In future studies with higher precision, the extrapolation modeling uncertainties, as well as other systematic effects, will be investigated more comprehensively to further improve the robustness of such calculations.

\section{Conclusion}\label{sec:conclusion}

In this work, we have presented the first lattice QCD calculation of the pion unpolarized valence-quark TMDPDF within the framework of LaMET. The calculation was performed on a 2+1 flavor QCD ensemble with a lattice spacing of $a = 0.06$~fm and a pion mass of 300~MeV. By leveraging high statistics, off-axis momenta, a well-tuned boosted Gaussian smeared pion source, and the novel CG approach, we attained significant hadron momenta reaching 3~GeV. We computed the quasi-TMD beam function, extracted the CS kernel from quasi-TMDWF, and derived the intrinsic soft function. To enhance perturbative accuracy, we performed RGR of the Sudakov kernel at NLL order throughout our analysis. These advancements collectively enabled the determination of both the pion TMDWF and TMDPDF from the first principles.

Our results of the CS kernel and intrinsic soft function demonstrate consistency with perturbative calculations in the small-$b_\perp$ region, where perturbative methods remain valid. Additionally, the extracted CS kernel agrees with both previous lattice QCD studies and phenomenological fits of the experimental data.

By combining the renormalized quasi-TMDPDF, the CS kernel, and the intrinsic soft function, we obtained the pion valence TMDPDF across a range of $b_\perp$. Our results extend beyond $b_\perp\gtrsim1$ fm, allowing for a feasible Fourier transform into momentum space. The extracted TMDPDFs in the $b_\perp$ and $k_\perp$ space exhibit qualitative agreement with phenomenological fits. This agreement demonstrates the potential of lattice QCD to validate and complement global analyses.

The outcome of this study highlights the efficacy of the CG quasi-TMD approach in probing the transverse momentum structure of hadrons. Our results provide a valuable first-principles complement to global fits, especially in the non-perturbative region where experimental data remain scarce. Future improvements, such as the incorporation of finer lattice spacings, larger momenta, and improved control over systematic uncertainties, will further enhance the precision of lattice QCD determinations of TMDPDFs. Additionally, extending this methodology to other hadrons, including the nucleon, will provide deeper insights into the partonic structure of QCD bound states.

\begin{acknowledgments}

We thank Min-Huan Chu and Hai-Tao Shu for sharing the pion form factor data. We thank Lorenzo Rossi for sharing the pion TMD results of MAP22 and Patrick C. Barry for sharing the pion TMD results of JAM23. We thank Matteo Cerutti for sharing the CS kernel results of MAPNN25, Congyue Zhang for sharing the CS kernel results of EEC24, and Aleksandra Lelek for sharing the CS kernel results of PB24. We thank Yushan Su for a valuable discussion on the RG resummation.

This material is based upon work supported by the U.S. Department of Energy, Office of Science, Office of Nuclear Physics through Contract No.~DE-SC0012704, No.~DE-AC02-06CH11357, within the framework of Scientific Discovery through Advanced Computing (SciDAC) award Fundamental Nuclear Physics at the Exascale and Beyond, and under the umbrella of the Quark-Gluon Tomography (QGT) Topical Collaboration with Award DE-SC0023646. The work of YZ is also partially supported by the Laboratory Directed Research and Development (LDRD) funding from Argonne National Laboratory, provided by the Director, Office of Science, of the U.S. Department of Energy under Contract No.~DE-AC02-06CH11357.

This research used awards of computer time provided by the INCITE program at Argonne Leadership Computing Facility, a DOE Office of Science User Facility operated under Contract DE-AC02-06CH11357, the ALCC program at the Oak Ridge Leadership Computing Facility, which is a DOE Office of Science User Facility supported under Contract DE-AC05-00OR22725, and the National Energy Research Scientific Computing Center, a DOE Office of Science User Facility supported by the Office of Science of the U.S. Department of Energy under Contract DE-AC02-05CH11231 using NERSC awards NP-ERCAP0028137 and NP-ERCAP0032114. Computations for this work were carried out in part on facilities of the USQCD Collaboration, which is funded by the Office of Science of the U.S. Department of Energy. We gratefully acknowledge the computing resources provided on Swing, a high-performance computing cluster operated by the Laboratory Computing Resource Center at Argonne National Laboratory. This research also used resources of the Argonne Leadership Computing Facility, a U.S. Department of Energy (DOE) Office of Science user facility at Argonne National Laboratory and is based on research supported by the U.S. DOE Office of Science-Advanced Scientific Computing Research Program, under Contract No. DE-AC02-06CH11357.

The measurement of the correlators was carried out with the \texttt{Qlua} software suite~\cite{qlua}, which utilized the multigrid solver in \texttt{QUDA}~\cite{Clark:2009wm,Babich:2011np}. Data analysis was performed using the Python package LaMETLat, which is specifically implemented to support the analysis of Lattice QCD data within the LaMET framework. The package is available at \href{https://github.com/Greyyy-HJC/LaMETLat}{https://github.com/Greyyy-HJC/LaMETLat} and is expected to undergo further development for prospective research applications.

\end{acknowledgments}

\appendix

\section{Renormalization group resummation}
\label{app:rg_resum}

\subsection{RGR for TMD hard kernel}

As mentioned in Sec.~\ref{sec:framework}, the quasi-TMD beam function and quasi-TMDWF can be matched to their light-cone counterparts using the TMD matching coefficient $C_{\mathrm{TMD}}$. In CG, the matching coefficient is calculated up to NLO fixed-order perturbation theory as ~\cite{Zhao:2023ptv}
\begin{align}
    C_{\rm TMD} \left(x P^z, \mu \right) = 1 + \frac{\alpha_s C_F}{4 \pi} \left[ - \frac{L_p^2}{2} - 3 L_p - 12 + \frac{7 \pi^2}{12} \right] ~,
    \label{eq:wilson_fixed}
\end{align}
in which $L_p = \ln\frac{\mu^2}{(2 x P^z)^2}$ and $C_F = 4/3$. The RG evolution (RGE) of the matching coefficient is given in the literature~\cite{Stewart:2010qs} as
\begin{align}
    \frac{d}{d\ln \mu} \ln C_{\rm TMD} = \gamma_H = - \Gamma_{\rm cusp} L_p + \gamma_C ~.
\end{align}
Combining with the fixed-order results in Eq.~\eqref{eq:wilson_fixed}, we can solve to get the anomalous dimensions up to NLO
\begin{align}
    \Gamma_{\rm cusp}^{(1)} = 2 C_F \cdot \frac{\alpha_s (\mu)}{4 \pi} ~,
\end{align}
and
\begin{align}
    \gamma_C^{(1)} = (-6 C_F) \cdot \frac{\alpha_s (\mu)}{4 \pi} ~.
\end{align}
Employing the defined QCD beta function $\beta$, one can incorporate the evolution of the coupling constant as described in $d \ln \mu = \frac{d \alpha_s}{\beta}$, and subsequently perform the integration as
\begin{align}
\begin{aligned}
    &\ln C_{\rm TMD} (\mu) - \ln C_{\rm TMD} (\mu_0) \\
    &\quad = \int_{\alpha_s(\mu_0)}^{\alpha_s(\mu)} \frac{d \alpha_s}{\beta (\alpha_s)} \left[ - 2 \Gamma_{\rm cusp} \int_{\alpha_s(\mu_0)}^{\alpha_s} \frac{d \alpha^\prime_s}{\beta (\alpha^\prime_s)} + \gamma_C (\alpha_s) \right] ~,
\end{aligned}
\end{align}
where we can choose the physical scale $\mu_0 = 2xP^z$. Due to the double logarithmic term $L^2_p$ in Eq.~\eqref{eq:wilson_fixed}, the next-to-leading logarithmic (NLL) result needs the tree-level $C_{\rm TMD} (\mu_0)$, the one-loop $\gamma_C$ and the two-loop $\Gamma_{\rm cusp}$.

\subsection{RGR for Sudakov kernel}

The NLO fixed-order Sudakov kernel with gamma structure in the insertion current taking the form $\Gamma = \gamma_\perp$ or $\Gamma = \gamma_5 \gamma_\perp$ is given in Refs.~\cite{Manohar:2003vb,Collins:2017oxh,Deng:2022gzi}
\begin{align}
    C_{\rm Sud} (\mu) = 1 + \frac{\alpha_s C_F}{4 \pi} \left[ - \ln^2 \frac{\mu^2}{Q^2} - 3 \ln \frac{\mu^2}{Q^2} - 8 + \frac{\pi^2}{6} \right] ~,
    \label{eq:sudakov_fix}
\end{align}
where $Q^2 = -q^2 = - (p_1 - p_2)^2$. The NLO RGE of Sudakov kernel can be derived as 
\begin{align}
    \frac{d}{d \ln \mu} \ln C_{\rm Sud}(\mu) = \gamma_1 (\mu) ~,
\end{align}
with the anomalous dimension
\begin{align}
    \gamma_1^{(1)} (\mu) = - \frac{\alpha_s(\mu) C_F}{4\pi} \left[ 4 \ln \frac{\mu^2}{Q^2} + 6 \right] ~.
\end{align}
Similarly, incorporating the evolution of coupling constant via $d \ln \mu = \frac{d \alpha_s}{\beta}$, one can do the integral as
\begin{align}
\begin{aligned}
    &\ln C_{\rm Sud} (\mu) - \ln C_{\rm Sud} (\mu_0) \\
    &\quad = \int_{\alpha_s(\mu_0)}^{\alpha_s(\mu)} \frac{d \alpha_s}{\beta (\alpha_s)} \left[ - \frac{\alpha_s C_F}{4\pi} \left( 8  \int_{\alpha_s(\mu_0)}^{\alpha_s} \frac{d \alpha^\prime_s}{\beta (\alpha^\prime_s) } + 6 \right) \right] ~,
\end{aligned}
\end{align}
where we can choose the physical scale $\mu_0 = Q$. Due to the double logarithmic term in Eq.~\eqref{eq:sudakov_fix}, the NLL result needs the tree-level $C_{\rm Sud} (\mu_0)$ and the one-loop $\gamma_1$.

\subsection{RGR for intrinsic soft function}

In $\overline{\rm MS}$ scheme, the NLO fixed-order intrinsic soft function in CG is~\cite{Zhao:2023ptv}
\begin{align}
    S_I (b_\perp, \mu) = 1 - \frac{\alpha_s C_F}{\pi} L_b ~,
\end{align}
where $L_b = \ln \frac{\mu^2 b_\perp^2 e^{2\gamma_E}}{4}$. The RGE can be derived as
\begin{align}
    \frac{d}{d\ln \mu} \ln S_I (b_\perp, \mu) = -2 \frac{\alpha_s (\mu) C_F}{\pi} ~.
\end{align}
Similarly, incorporating the evolution of coupling constant via $d \ln \mu = \frac{d \alpha_s}{\beta}$, one can do the integral as
\begin{align}
\begin{aligned}
    &\ln S_I (b_\perp, \mu) - \ln S_I (b_\perp, \mu_0) = -2 \frac{C_F}{\pi} \int_{\alpha_s(\mu_0)}^{\alpha_s(\mu)} \frac{d \alpha_s}{\beta} \alpha_s ~.
\end{aligned}
\end{align}
The leading logarithmic (LL) result is given with the combination of the tree-level $S_I (b_\perp, \mu_0)$ and the one-loop anomalous dimension.

\section{Scheme conversion}
\label{app:scheme_conversion}

As discussed in Sec.~\ref{sec:framework}, using different renormalization schemes for quasi-TMDWF, one can get the intrinsic soft function in the corresponding renormalization schemes. This section derives the scheme conversion between the $\overline{\mathrm{MS}}$ scheme and the ratio scheme as defined in renormalization Eqs.~\eqref{eq:renorm_qtmdpdf} and \eqref{eq:renorm_qtmdwf}.

Using the ratio scheme, the renormalized quasi-TMDWF is defined as
\begin{widetext}
\begin{align}
    \tilde{\varphi}^{\text {ratio }}\left(z, b_{\perp},  P^z; \mu \right)=\frac{\tilde{\varphi}^0\left(z, b_{\perp},  P^z; a\right)}{\tilde{\varphi}^0\left(z=0, b_{\perp},  P^z =0; a\right)}=\frac{\tilde{\varphi}^{\overline{\rm M S}}\left(z, b_{\perp},  P^z; \mu\right)}{\tilde{\varphi}^{\overline{\rm M S}}\left(z=0, b_{\perp},  P^z =0; \mu\right)} ~,
\end{align}
\begin{align}
    \tilde{\phi}^{\text {ratio }}(x, b_\perp, P^z; \mu)= P^z \int \frac{dz}{2\pi} e^{i z (xP^z)} \tilde{\varphi}^{\text {ratio }}(z, b_\perp, P^z; \mu)=\int \frac{d \lambda}{2\pi} e^{i \lambda x} \frac{\tilde{\varphi}^{\overline{\rm M S}}\left(z, b_{\perp},  P^z; \mu\right)}{\tilde{\varphi}^{\overline{\rm M S}}\left(z=0, b_{\perp},  P^z =0; \mu\right)} ~,
\end{align} 
\end{widetext}
in which the denominator can be approximately replaced by the short distance factorization coefficient $C_0$ in CG, then we got
\begin{align}
    \tilde{\phi}^{\text {ratio }}(x, b_\perp, P^z; \mu)= \frac{\tilde{\phi}^{\overline{\rm MS}}(x, b_\perp, P^z; \mu)}{C_0 (b_\perp; \mu)} ~.
\end{align}
Combining with the Eq.~\eqref{eq:soft_func_calc}, we have the scheme conversion of intrinsic soft function as
\begin{align}
    S_I^{\rm ratio}(b_\perp; \mu) = S_I^{\overline{\rm MS}}(b_\perp; \mu) \cdot C^2_0 (b_\perp; \mu) ~.
\end{align}
The NLO fixed-order result of $C_0$ can be found in Ref.~\cite{Gao:2023lny} as
\begin{align}
    C_0\left(b_{\perp}; \mu\right)=1+\frac{\alpha_s C_F}{2 \pi}\left(\frac{1}{2}-\frac{L_b}{2}\right) ~.
\end{align}
One can do the RGR in the similar way as in App.~\ref{app:rg_resum} to get the LL result of $C_0$ as
\begin{align}
    C_0\left(b_\perp; \mu\right)= \exp \left(\frac{C_F}{\beta_0} \ln \frac{\alpha_s(\mu)}{\alpha_s\left(2 e^{-\gamma_E} / b_{\perp}\right)}\right) ~.
\end{align}

\section{Bare matrix elements of quasi-TMDWF}
\label{app:tmdwf}

\begin{figure}[th!]
    \centering
    \includegraphics[width=.9\linewidth]{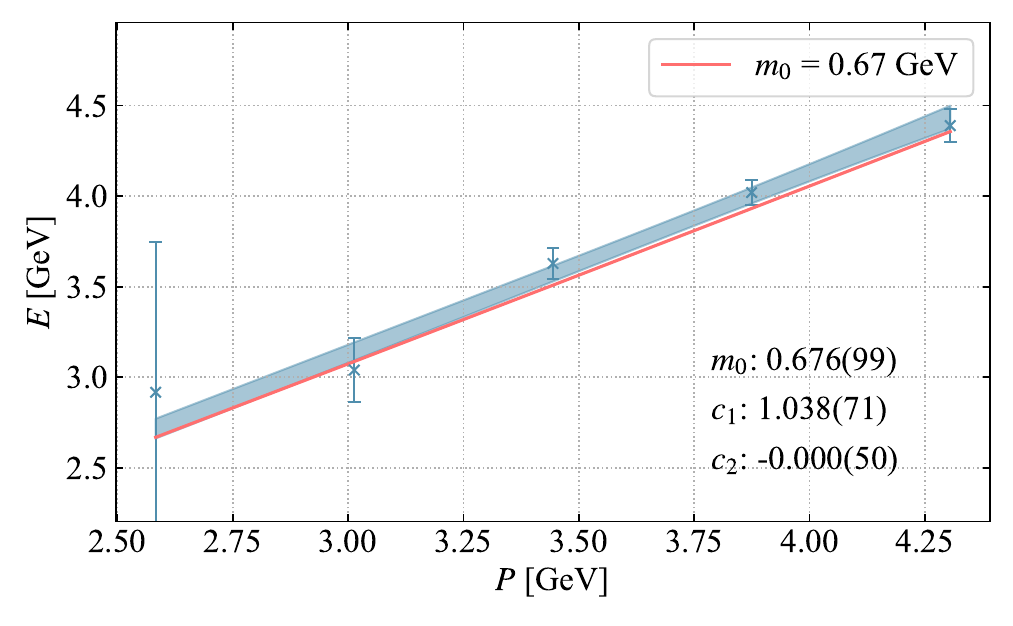}
    \caption{The ground-state energies $E_0$ extracted from two-state fit of the two-point functions. The red line represents the exact dispersion relations with the valence pion mass $m_0 = 670$ MeV.
    \label{fig:disp_TMDWF}}
\end{figure}

\begin{figure*}
    \centering
    \includegraphics[width=.32\linewidth]{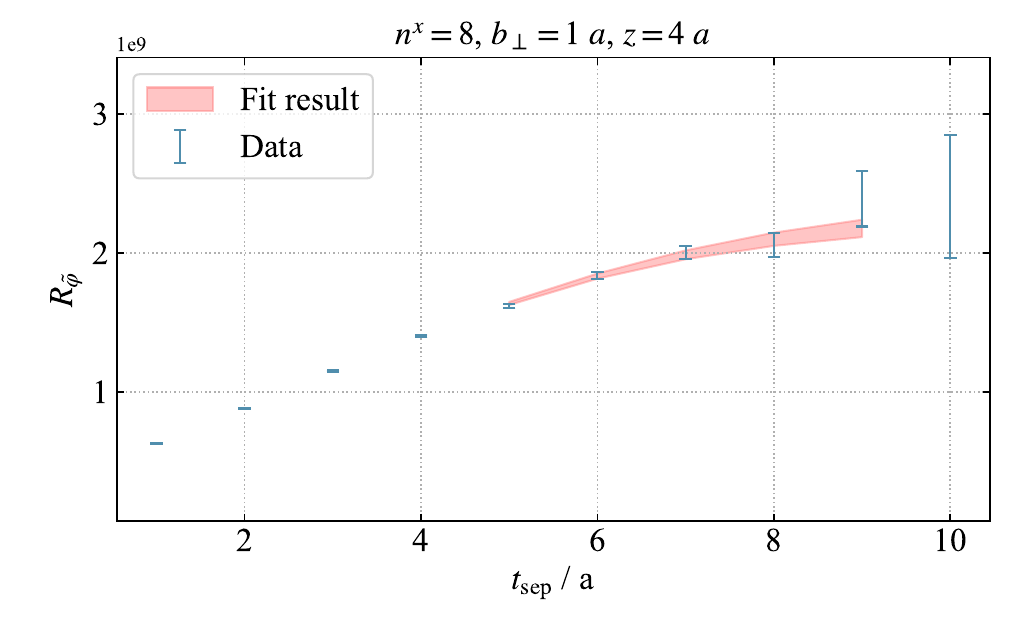}
    \includegraphics[width=.32\linewidth]{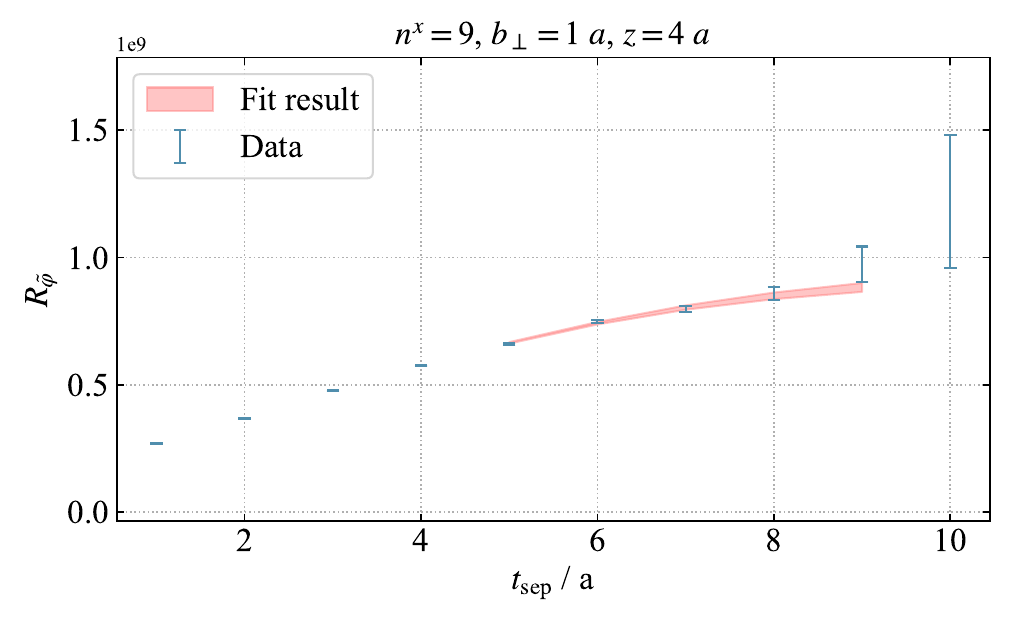}
    \includegraphics[width=.32\linewidth]{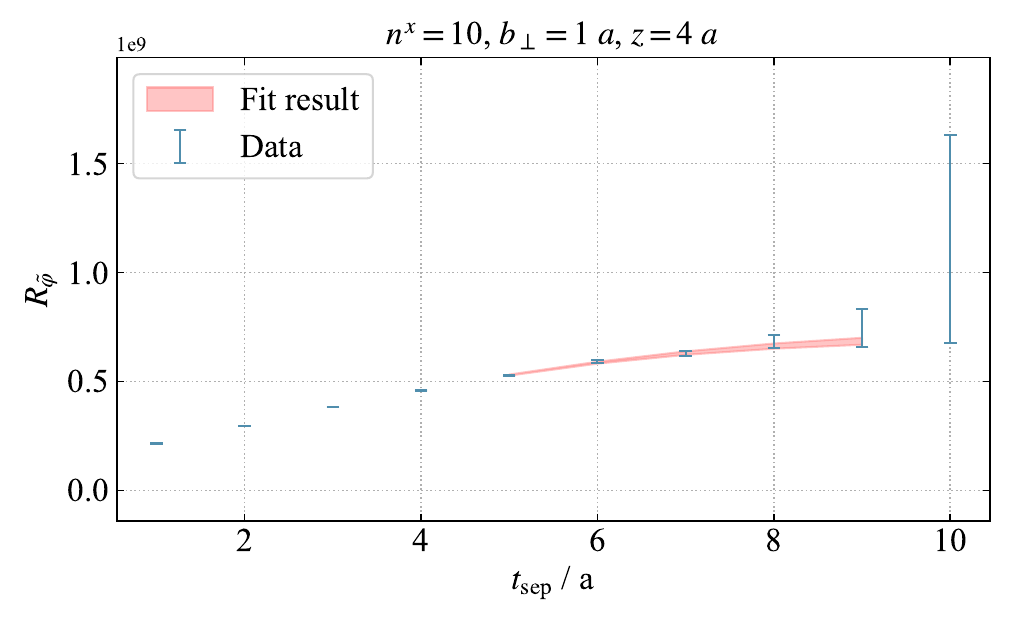}
    \includegraphics[width=.32\linewidth]{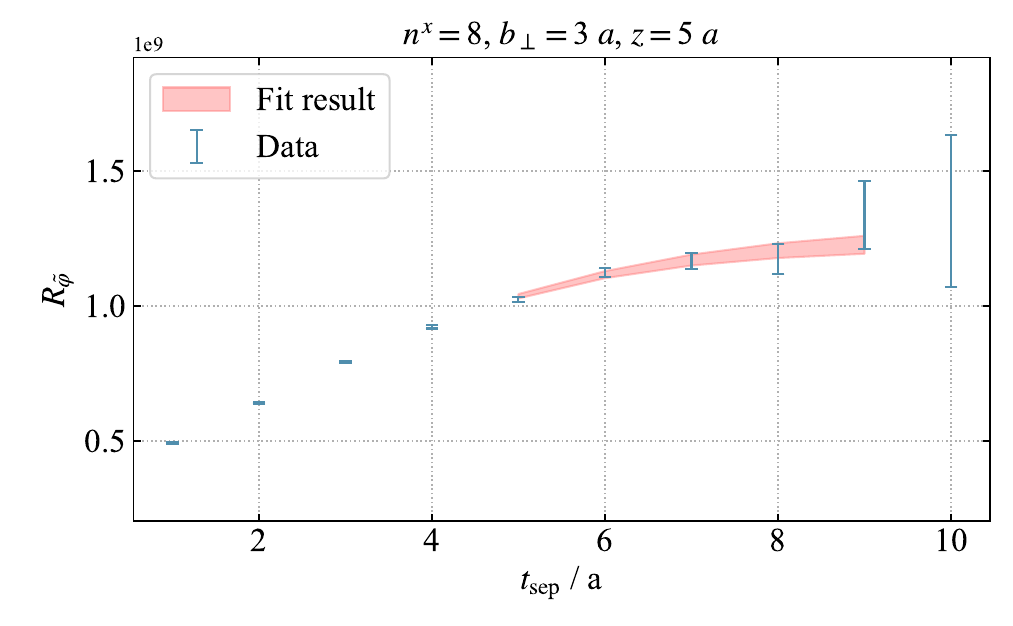}
    \includegraphics[width=.32\linewidth]{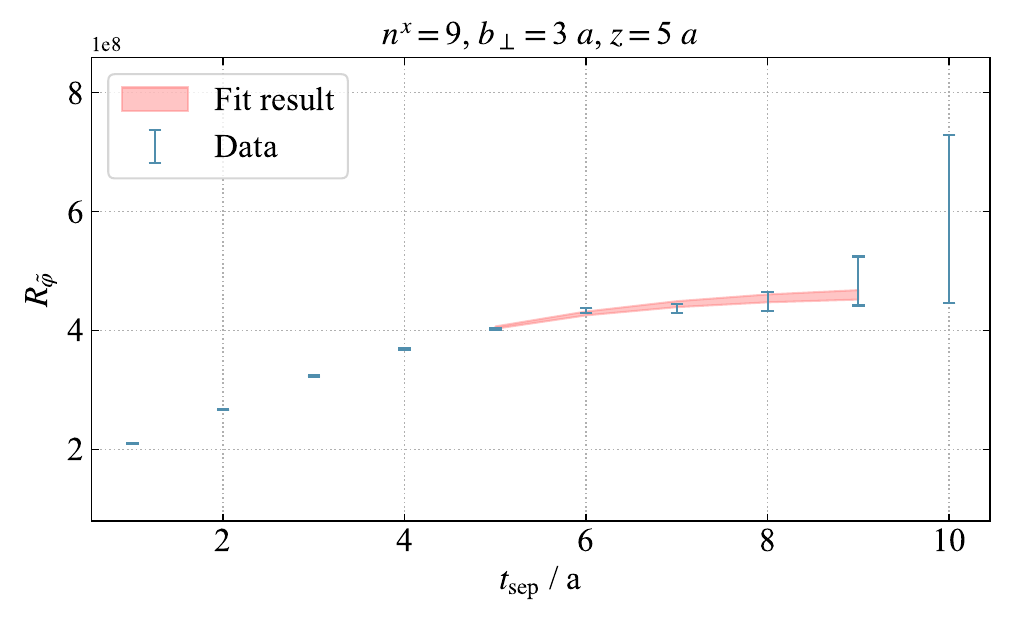}
    \includegraphics[width=.32\linewidth]{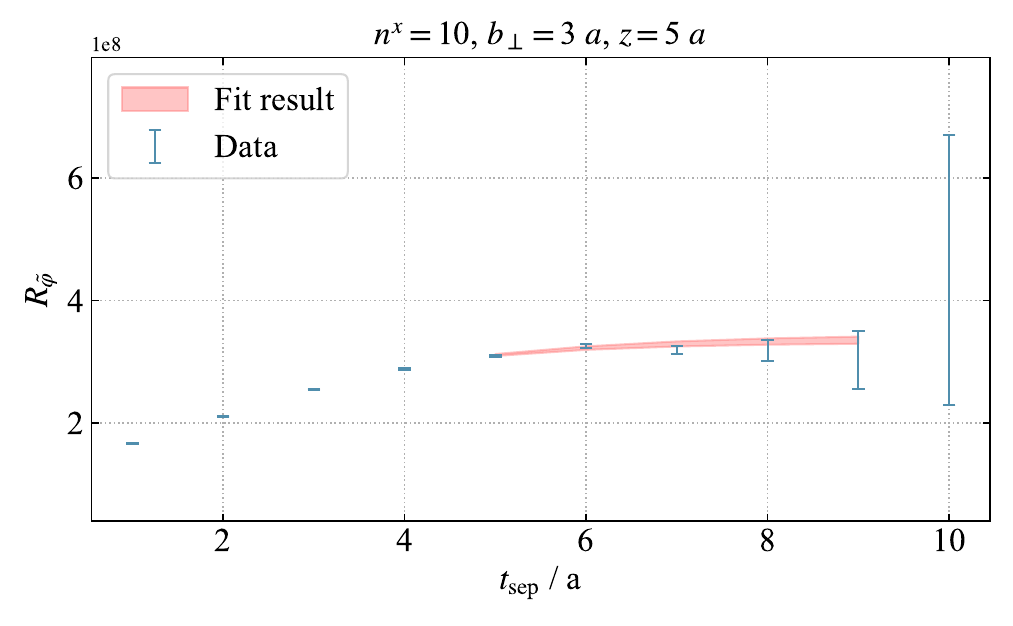}
    \caption{From left to right, the ratios of quasi-TMDWF to two-point functions $R_{\tilde{\varphi}}$ with hadron momentum $P^z=3.44$, 3.87 and 4.30 GeV are shown as functions of $t_{\rm sep}$. The upper and lower panels are for the cases with with $(b_\perp, z) = (1, 4)~a$ and $(b_\perp, z) = (3, 5)~a$, repectively. The colored bands are joint fit results.
    \label{fig:joint_fit_qtmdwf}}
\end{figure*}

\begin{figure*}[th!]
    \centering
    \includegraphics[width=.4\linewidth]{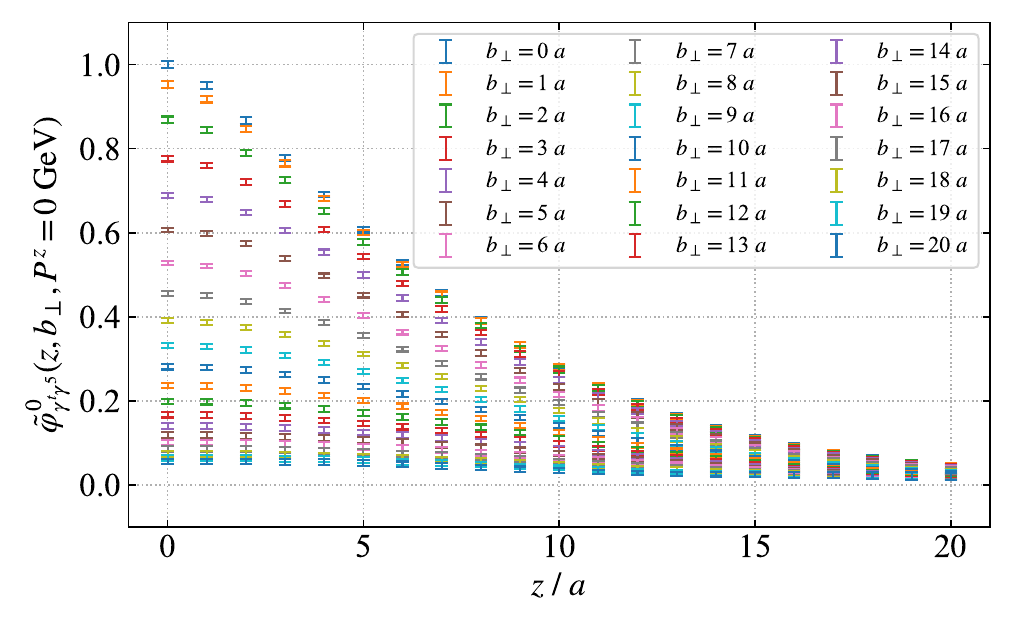}
    \includegraphics[width=.4\linewidth]{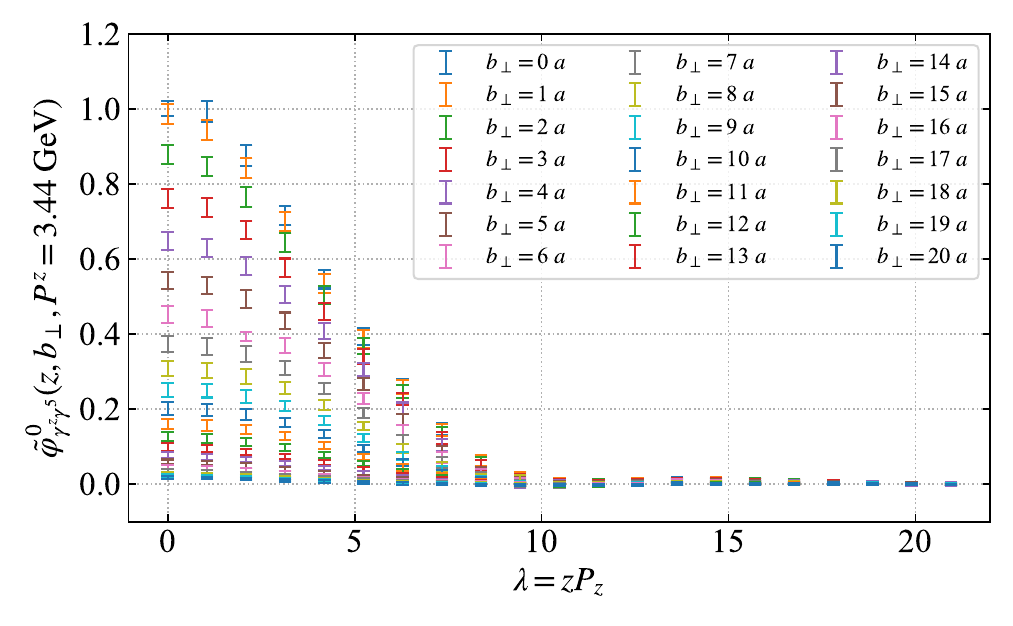}
    \includegraphics[width=.4\linewidth]{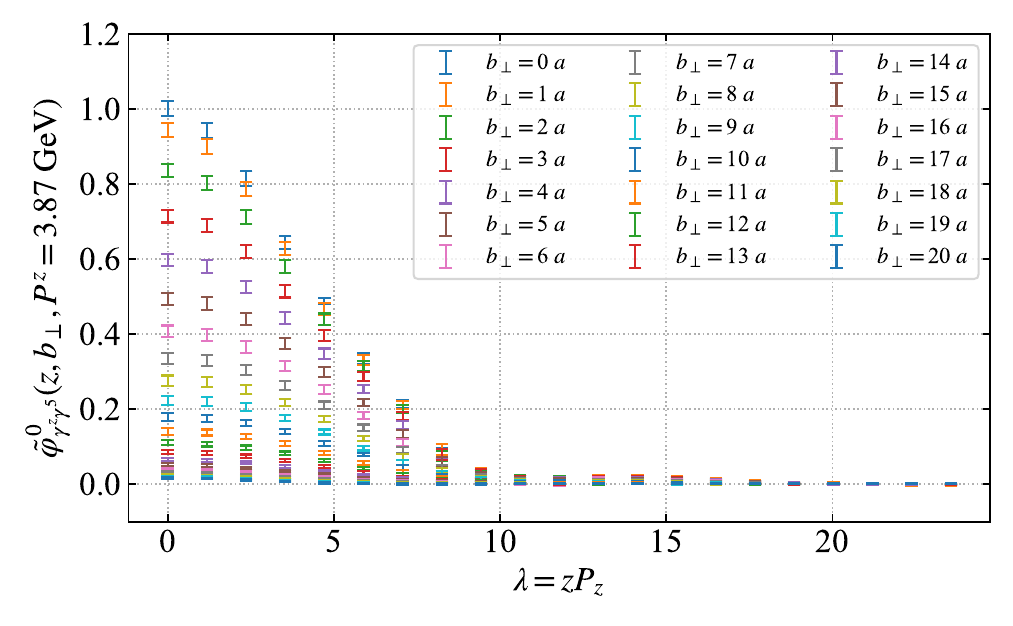}
    \includegraphics[width=.4\linewidth]{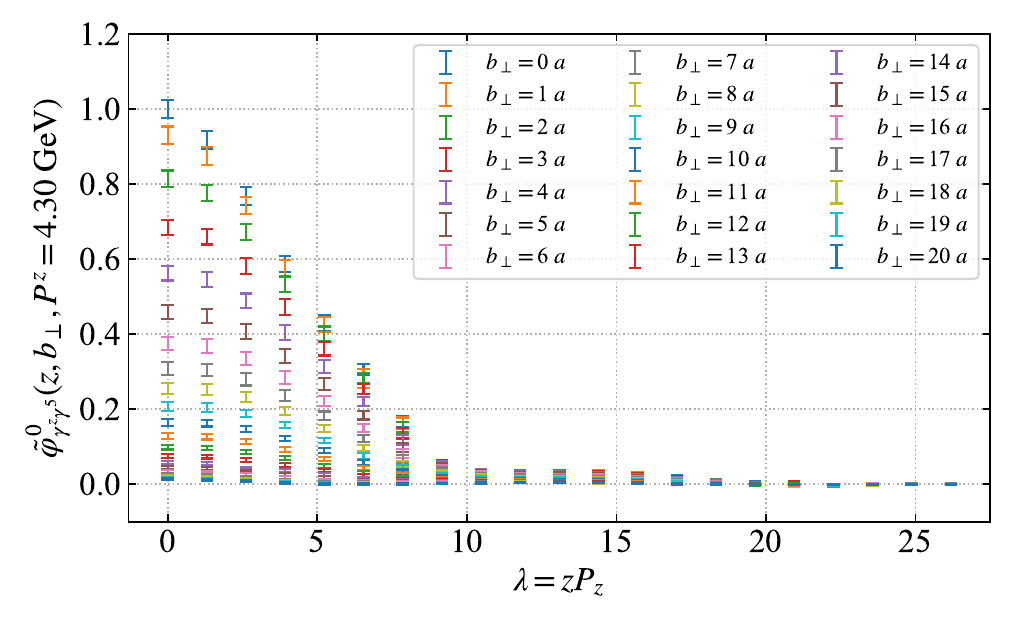}
    \caption{Bare matrix elements of quasi-TMDWF in the coordinate space. The non-zero-momentum correlators extracted from two-state joint fits of $C_{\text{2pt}}$ and $C_{\tilde{\varphi}}$, while the zero-momentum correlators are extracted from one-state fit of $C_{\tilde{\varphi}}$. All bare matrix elements are normalized using the mean value of the local matrix element.
    \label{fig:bare_zdep_qtmdwf}}
\end{figure*}

\begin{figure*}[th!]
    \centering
    \includegraphics[width=.32\linewidth]{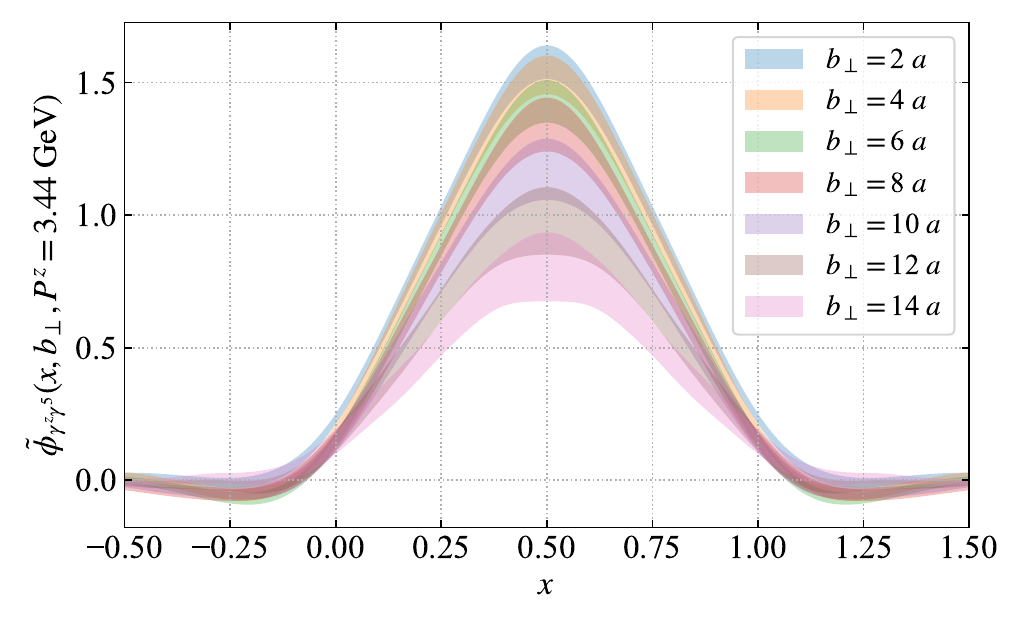}
    \includegraphics[width=.32\linewidth]{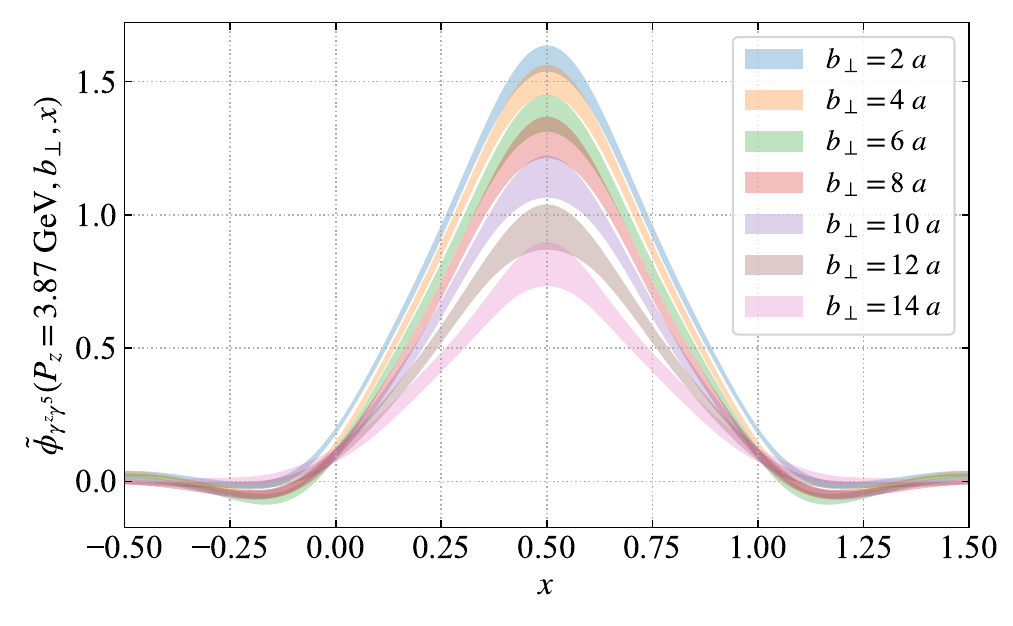}
    \includegraphics[width=.32\linewidth]{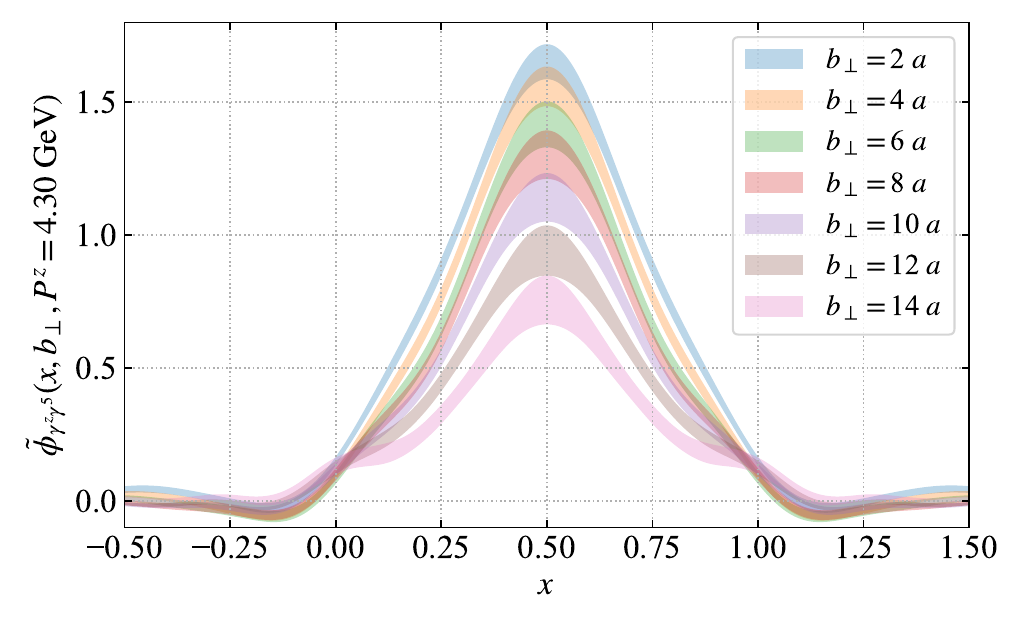}
    \caption{Quasi-TMDWF of pion in the momentum space. From left to right, the hadron momenta are $P^z$ = 3.44, 3.87 and 4.30 GeV. The small bumps in the right panel are caused by the small non-zero tails in the coordinate space, primarily manifest at large $b_\perp$.}
    \label{fig:renorm_xdep_qtmdwf}
\end{figure*}

The quasi-TMDWF is extracted from the non-local two-point correlator
\begin{align}
\begin{aligned}
C_{\tilde{\varphi}}\left(t_{\rm sep} \right)&= \left\langle O_\Gamma^{\rm CG} \pi_s^\dagger(\vec{P},0) \right\rangle \\
&=\left\langle \bar{q}(z/2, b_\perp, t_{\rm sep}) \Gamma q (-z/2, 0, t_{\rm sep}) \pi_s^\dagger(\vec{P},0)\right\rangle ~,
\end{aligned}
\end{align}  
which can be expressed in terms of a spectral decomposition as  
\begin{align}
\begin{aligned}
    &C_{\tilde{\varphi}}\left(t_{\rm sep} \right)\\
    &\quad=\sum_{n=0}^{N_{\mathrm{s}}-1} \frac{z_n}{2 E_n} \langle \Omega | O^{\rm CG}_{\Gamma} | n \rangle \left(e^{-E_n t_{\rm sep}}+e^{-E_n\left(L_t-t_{\rm sep}\right)}\right) ~,
    \label{eq:qtmdwf_decompose}
\end{aligned}
\end{align}
where $\ket{\Omega}$ is the interactive vacuum state, $\ket{n}$ is the $n$-th energy eigenstate. When $\Gamma = \gamma^z \gamma^5$, the matrix element of the ground state is $\langle \Omega | O^{\rm CG}_{\gamma^z \gamma^5} | 0 \rangle = i f_\pi P^z \tilde{\varphi}^0_{\gamma^z \gamma^5}$, where $f_\pi$ is the bare pion decay constant, and $\tilde{\varphi}^0_\Gamma$ is the bare matrix element of quasi-TMDWF in the coordinate space. When $\Gamma = \gamma^t \gamma^5$, the matrix element of the ground state is $\langle \Omega | O^{\rm CG}_{\gamma^t \gamma^5} | 0 \rangle = f_\pi E_0 \tilde{\varphi}^0_{\gamma^t \gamma^5}$, where $E_0 = \sqrt{m_\pi^2 + (P^z)^2}$.

In order to determine the energy spectrum and facilitate the extraction of the quasi-TMDWF, the local two-point functions of the pion are additionally computed with a valence pion mass of $m_\pi = 670$ MeV for the corresponding hadron momenta. The ground-state energies from the two-state fit of the two-point functions are illustrated in Fig.~\ref{fig:disp_TMDWF}, in comparison with the exact dispersion relations with $m_0 = 670$ MeV. Good consistency is found up to the hadron momentum $P^z= 4.30$ GeV.

For correlators with nonzero momenta, we employ the two-state joint fit of $C_{\text{2pt}}$ and $C_{\tilde{\varphi}}$, while for zero-momentum correlators, the one-state fit is adopted to fit $C_{\tilde{\varphi}}$. Some of the two-state fits are selected as examples in Fig.~\ref{fig:joint_fit_qtmdwf}. As a means of illustration, both data and fit results are plotted as the ratio of $C_{\tilde{\varphi}}$ and $C_{\text{2pt}}$, which is defined as
\begin{align}
    R_{\tilde{\varphi}}\left(t_{\rm sep} \right) = \frac{C_{\tilde{\varphi}}\left(t_{\rm sep} \right)}{C_{\text{2pt}}\left(t_{\rm sep} \right)} ~.
\end{align}
This ratio will converge to $\langle \Omega | O^{\rm CG}_{\Gamma} | 0 \rangle / z_0$ in the $t_{\rm sep} \to \infty$ limit. More details on the ground-state fit can be found in App.~\ref{app:gsfit}.

The bare matrix element of quasi-TMDWF in the coordinate space can be found in Fig.~\ref{fig:bare_zdep_qtmdwf}, different momenta are normalized using the mean value of the local matrix element at $(z, b_\perp) = (0, 0)$, so that some artifacts such as discretization effects can be canceled out. Due to the good convergence in the large-$\lambda$ region, the renormalized matrix elements $\varphi_\Gamma$ can be Fourier transformed directly to the momentum space.

Fig.~\ref{fig:renorm_xdep_qtmdwf} illustrates the quasi-TMDWF in momentum space, where it can be seen that the dependence on momentum is not significant, implying that the CS kernel is not a large quantity. The small bumps in the right panel are caused by the small non-zero tails in the coordinate space, primarily manifest at large $b_\perp$.

\section{Bare matrix elements of form factor}
\label{app:ff}

\begin{figure}[th!]
    \centering
    \includegraphics[width=.9\linewidth]{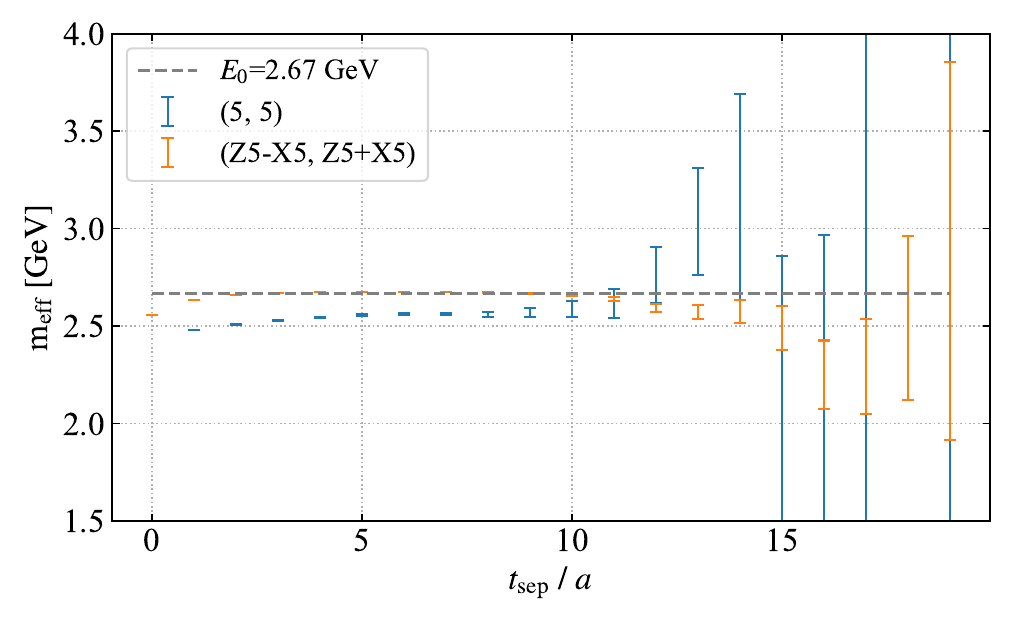}
    \caption{The effective mass of two-point function, where $(5, 5)$ means using $\gamma^5$ for both source and sink, and $(Z5-X5, Z5+X5)$ means using $\gamma^z\gamma^5 - \gamma^x\gamma^5$ and $\gamma^z\gamma^5 + \gamma^x\gamma^5$ for source and sink, respectively. The latter choice of the interpolating operator was proposed in Ref.~\cite{Zhang:2025hyo} to enhance the SNR and suppress excited-state contamination when the hadron momentum is large.
    \label{fig:ff_meff}}
\end{figure}

\begin{figure}[th!]
    \centering
    \includegraphics[width=.9\linewidth]{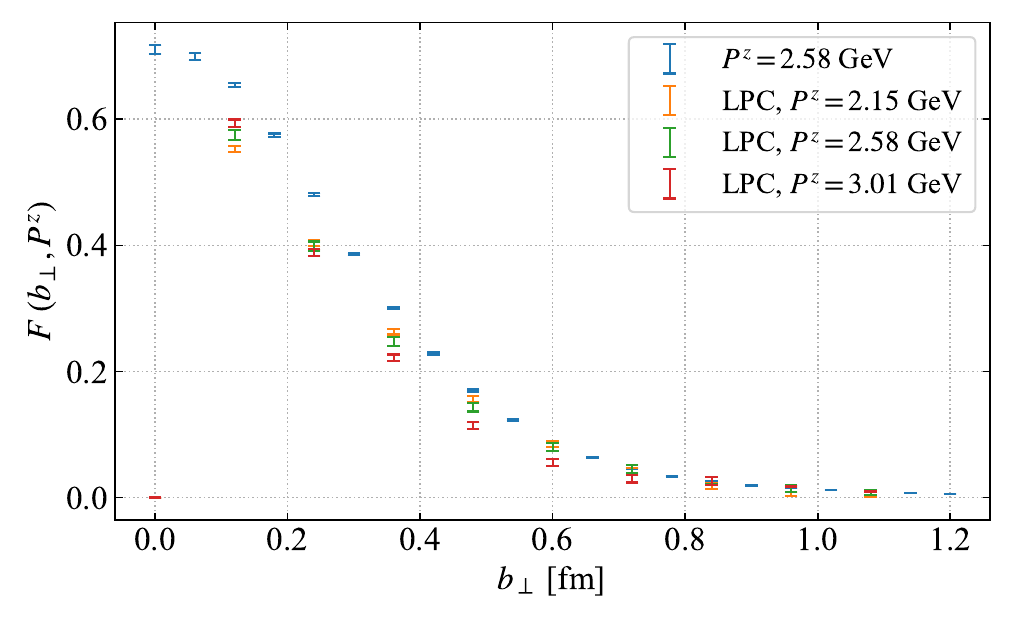}
    \caption{The final results of form factor $F$ is given by averaging over form factors of four gamma structures following the Fierz rearrangement~\cite{Li:2021wvl}. The form factor results calculated on MILC ensembles with $2+1+1$-flavor with HISQ action~\cite{Follana:2006rc} in Ref.~\cite{LatticePartonLPC:2023pdv} are plotted in the figure for comparison.
    \label{fig:ff_comparison}}
\end{figure}

The bare pion form factor is extracted from the ratio defined in Eq.~\eqref{eq:ff_ratio}. When the hadron momentum is large, the leading-twist contribution in the denominator of Eq.~\eqref{eq:ff_ratio} comes from the two-point functions with structures $\gamma^t \gamma^5$ and $\gamma^z \gamma^5$. According to the suggestion in Ref.~\cite{Zhang:2025hyo}, we chose the two-point function with structure $\gamma^z \gamma^5 + \gamma^x \gamma^5$, which has the same leading-twist contribution as that of $\gamma^z \gamma^5$, to enhance the SNR and suppress excited-state contamination when the hadron momentum is large. Compared to the two-point functions with $\gamma^5$ for both source and sink, the effective mass from the new interpolating operator is plotted in Fig.~\ref{fig:ff_meff}. It can be seen that the new interpolating operator leads to better SNR and faster convergence to the ground state. The difference between the two-point function of $\gamma^t \gamma^5$ and $\gamma^z \gamma^5$ is estimated by a factor of $P^t / P^z$ in the denominator of Eq.~\eqref{eq:ff_ratio}. Thus, the ratio in Eq.~\eqref{eq:ff_ratio} can be expressed in terms of a spectral decomposition as
\begin{align}
\begin{aligned}
    &R_F\left(t_{\mathrm{sep}}, \tau\right)= \frac{-4 N_c}{1 + (P^t / P^z)^2} \frac{C_{\mathrm{F}}\left(t_{\mathrm{sep}}, \tau\right)}{\left|C_{2 \mathrm{pt}}\left(t_{\mathrm{sep}} / 2\right)\right|^2}\\
    &\quad \quad =\frac{-4 N_c}{1 + (P^t / P^z)^2}\frac{\Sigma_{n, m} \frac{z_n F_{n m} z_m^{*}}{4 E_n E_m} \cdot e^{-E_n\left(t_{\mathrm{sep}}-\tau\right)} e^{-E_m \tau}}{\left|\Sigma_n \frac{z_n^*}{2 E_n} A_n \cdot e^{-E_n t_{\mathrm{sep}} / 2}\right|^2} ~.
\end{aligned}
\end{align}
Here $z_n = \langle n | \pi^\dagger | \Omega \rangle$ is the same overlap amplitude defined in Sec.~\ref{sec:quasi-tmd}, $F_{nm} = \left\langle n\right| \bar{q}\left(b_{\perp}\right) \Gamma q\left(b_{\perp}\right) \bar{q}(0) \Gamma q(0)\left|m\right\rangle$ is the matrix element, and $A_n = \left\langle \Omega \right| \bar{q}(0) (\gamma^z \gamma^5 + \gamma^x \gamma^5) q(0)\left|n \right\rangle$ is the amplitude with a local insertion current. When both the $t_{\rm sep}$ and $\tau$ are large enough, we can keep the first two energy states and approximate the ratio as
\begin{align}
\begin{aligned}
    &R_F\left(t_{\mathrm{sep}}, \tau\right) \approx \frac{-4 N_c}{1 + (P^t / P^z)^2} \\
    &\quad \quad \times \frac{F_{00}}{|A_0|^2}  \frac{1 + c_1 \cdot \left( e^{-\Delta E \cdot (t_{\rm sep} - \tau)} + e^{-\Delta E \cdot \tau} \right)}{1 + c_2 \cdot e^{-\Delta E \cdot t_{\rm sep} / 2}} ~,
\end{aligned}
\end{align}
where the amplitude can be expanded as $A_0 = i f_\pi P^z$. According to Eq.~\eqref{eq:ff_def}, the ground state matrix element of the form factor correlator is $F_{00} = F^0 f_\pi^2 \left( (P^z)^2 + (P^t)^2 \right) / (-4 N_c)$, where $F^0$ is the bare form factor. In the limit $t_{\rm sep} \to \infty$, the ratio $R_F\left(t_{\mathrm{sep}}, t\right)$ asymptotically converges to the bare form factor $F^0$. Therefore, we employed the two-state fit of $R_F$ to extract the bare form factor $F^0$, with more details provided in App.~\ref{app:gsfit}. Note that the fit results of $F^0$ need to be multiplied by an extra volume factor $\mathrm{Vol} = L_t^3$ due to the lattice setup of wall source.

To suppress the higher-twist effects, we applied the Fierz rearrangement~\cite{Li:2021wvl} to combine different $\Gamma$ structures in Eq.~\eqref{eq:ff_def}. Ignoring the renormalization factor $Z_A / Z_V$, the final results of the form factor is given as
\begin{align}
\begin{aligned}
    &F(b_\perp, P^z) =\\
    &\quad \frac{1}{4} \left[ F(b_\perp, P^z, \Gamma = \gamma^x \gamma^5) + F(b_\perp, P^z, \Gamma = \gamma^y \gamma^5) \right. \\
    &\quad \quad \left. + F(b_\perp, P^z, \Gamma = \gamma^x ) + F(b_\perp, P^z, \Gamma = \gamma^y ) \right] ~,
\end{aligned}
\end{align}
which is plotted in Fig.~\ref{fig:ff_comparison}. The form factor results calculated on MILC ensembles with lattice spacing $0.12$ fm and HISQ action~\cite{Follana:2006rc} in Ref.~\cite{LatticePartonLPC:2023pdv} are plotted in the figure for comparison. The small deviation can be caused by the systematics of the lattice in different ensembles, such as discretization effects, as well as the different choices of gamma structure for the two-point function.

\section{Ground state fit}
\label{app:gsfit}

\begin{figure*}[th!]
    \centering
    \includegraphics[width=0.4\linewidth]{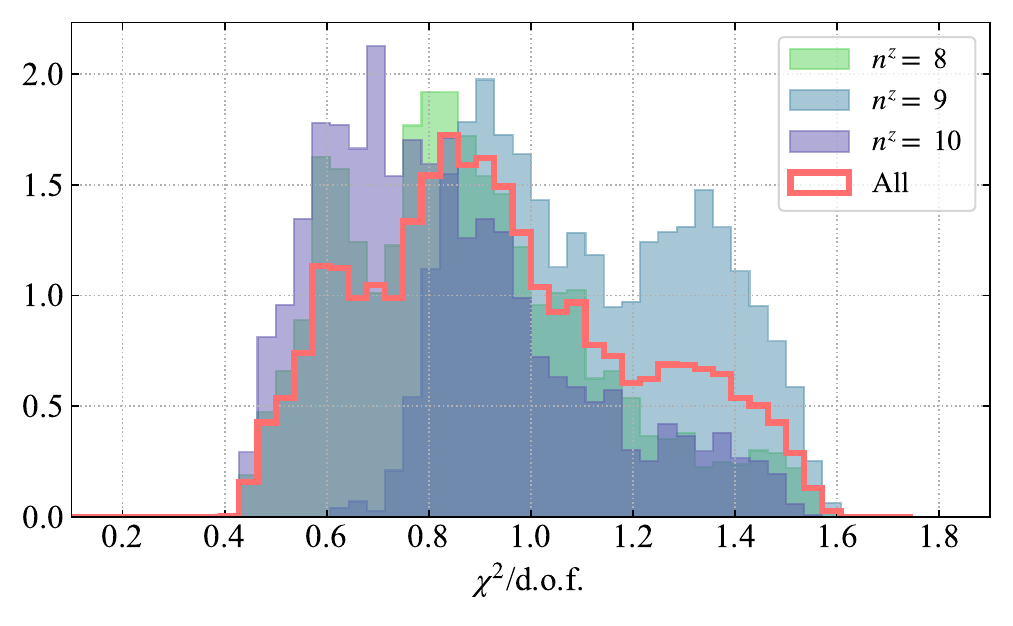}
    \includegraphics[width=0.4\linewidth]{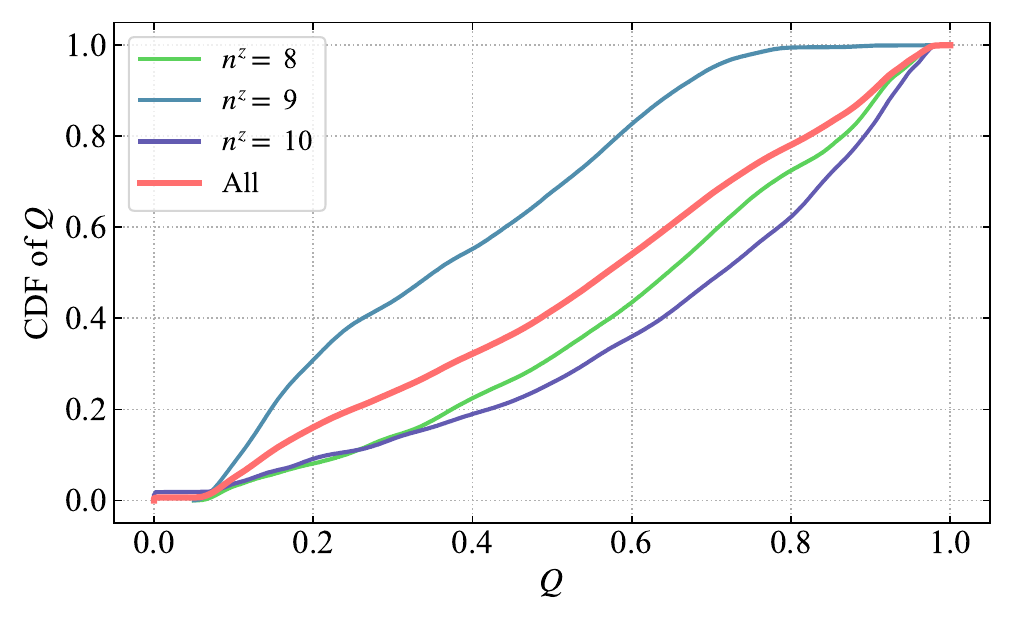}
    \caption{The evaluation of joint fits of quasi-TMDWF matrix elements. The left panel is the density distribution of $\chi^2 / \rm{d.o.f.}$ and the right panel is the cumulative distribution function (CDF) of p-value.}
    \label{fig:gsfit_stat_qtmdwf}
\end{figure*}
\begin{figure*}[th!]
    \centering
    \includegraphics[width=0.4\linewidth]{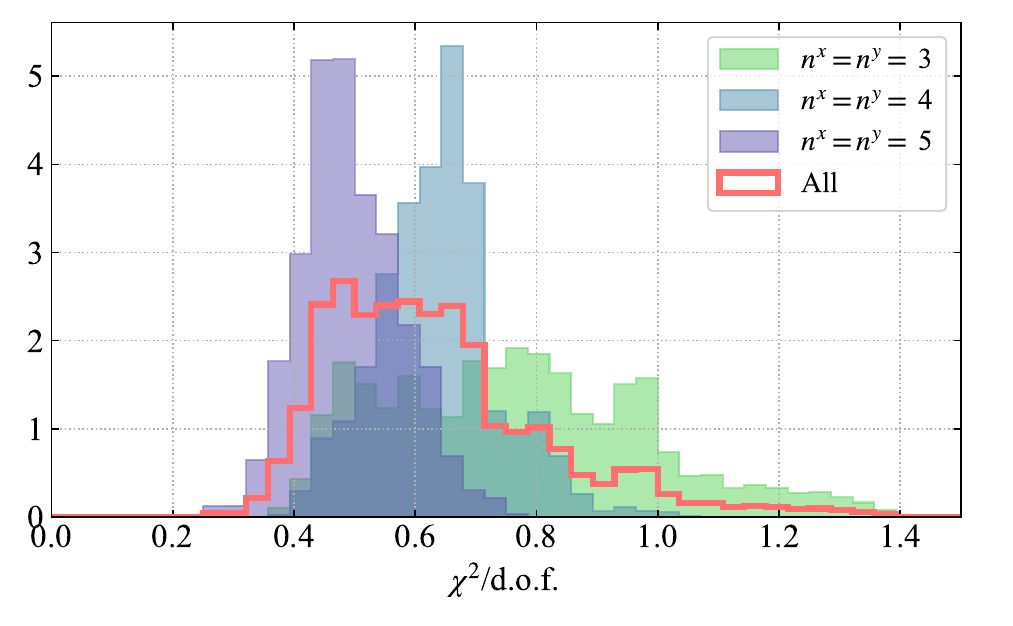}
    \includegraphics[width=0.4\linewidth]{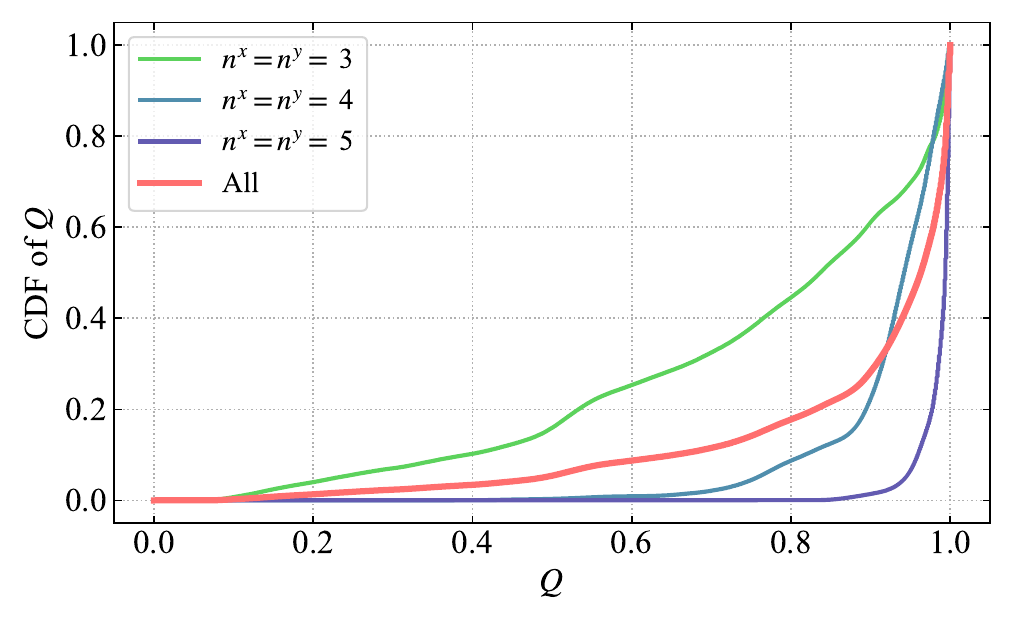}
    \caption{The evaluation of chained fits on $R_{\tilde{h}}$ of quasi-TMD matrix elements. The left panel is the density distribution of $\chi^2 / \rm{d.o.f.}$ and the right panel is the cumulative distribution function (CDF) of p-value.}
    \label{fig:gsfit_stat_qtmdpdf}
\end{figure*}

We have performed a fully correlated Bayesian analysis of the two-point and three-point correlation functions to extract the bare matrix element $\tilde{h}^{0}_{\Gamma}(x, b_\perp, P^z; \mu)$ defined in Eq.~\eqref{eq:qtmdpdf_def_2}, $F^0\left(b_{\perp}, P_1, P_2, \Gamma, \Gamma^\prime \right)$ defined in Eq.~\eqref{eq:ff_def} and $\tilde{\varphi}^{0}_{\Gamma}(x, b_\perp, P^z; \mu)$ defined in Eq.~\eqref{eq:qtmdwf_def_1}. The correlation functions across different data sets are taken into account by performing a Bayesian least-squares fit on each sample of Jackknife resampling. The parameter settings for the ground state fit are collected in \tb{gsfit}.
\begin{table}[th!]
\renewcommand{\arraystretch}{1.2} 
    \centering
    \begin{tabular}{|c|c|c|c|c|}
    \hline\hline
    Fit & Parts & $N_s$ & $t_{\rm sep}$ range & $\tau$ range \\
    \hline
    $\tilde{\varphi}^{0}_{\gamma^t \gamma^5}$ & $C_{\tilde{\varphi}}(t_{\rm sep})$ & 1 & $t_{\rm sep} \in [12, 15]$ & / \\
    \hline
    $F^0$ & $R_{F}(t_{\rm sep}, \tau)$ & 2 & $t_{\rm sep} \in \{ 6, 8, 10, 12 \}$ & $\tau \in [3, t_{\rm sep} - 3]$ \\
    \hline
    \multirow{2}{*}{$\tilde{\varphi}^{0}_{\gamma^z \gamma^5}$} & $C_{\rm 2pt}(t_{\rm sep})$ & 2 & $t_{\rm sep} \in [3, 10]$ & / \\
    \cline{2-5}
     & $C_{\tilde{\varphi}}(t_{\rm sep})$ & 2 & $t_{\rm sep} \in [5, 9]$ & / \\
    \hline
    \multirow{2}{*}{$\tilde{h}^{0}_{\gamma^t}$} & $C_{\rm 2pt}(t_{\rm sep})$ & 2 & $t_{\rm sep} \in [3, 12]$ & / \\
    \cline{2-5}
     & $R_{\tilde{h}}(t_{\rm sep}, \tau)$ & 2 & $t_{\rm sep} \in \{ 6, 8, 10, 12 \}$ & $\tau \in [3, t_{\rm sep} - 3]$ \\
    \hline\hline
    \end{tabular}
    \caption{Collection of ground state fit settings. $N_s$ = 1 means that only one ground state is included in the fit functions. The $t_{\rm sep}$ range of $C_{\tilde{\varphi}}$ varies in some sets of $(P^z, b_\perp)$ due to the different behaviors of excited states.}
    \label{tb:gsfit}
\end{table}

For the convenience of the readers, the fit functions corresponding to each part of Table~\ref{tb:gsfit} are listed below. The function $C_{\tilde{\varphi}}$ in the ground state fit of zero-momentum $\tilde{\varphi}^{0}_{\gamma^t \gamma^5}$ is
\begin{align}
    C_{\tilde{\varphi}}\left(t_{\rm sep} \right)= \tilde{\varphi}^{0}_{\gamma^t \gamma^5} \left(e^{-E_0 t_{\rm sep}}+e^{-E_0\left(L_t-t_{\rm sep}\right)}\right) ~,
\end{align}
where the fit results of $\tilde{\varphi}^{0}_{\gamma^t \gamma^5}$ are normalized using the mean value of the local matrix element at $z = b_\perp = 0$. The function $R_{F}(t_{\rm sep}, \tau)$ in the ground state fit of bare form factor $F^0$ is
\begin{align}
    R_{F}(t_{\rm sep}, \tau) = F^0 \frac{1 + c_1 \cdot \left( e^{-\Delta E \cdot (t_{\rm sep} - \tau)} + e^{-\Delta E \cdot \tau} \right)}{1 + c_2 \cdot e^{-\Delta E \cdot t_{\rm sep} / 2}} ~,
\end{align}
where the fit results of $F^0$ need to be multiplied by an extra volume factor $V = L_t^3$ due to the lattice setup of wall source. The fit function of $C_{\rm 2pt}(t_{\rm sep})$ in the ground state fit of $\tilde{\varphi}^{0}_{\gamma^z \gamma^5}$ is
\begin{align}
    C_{\text{2pt}}\left(t_{\rm sep}\right) = \sum_{n=0}^{1} \frac{\left|z_n\right|^2}{2 E_n} \left(e^{-E_n t_{\rm sep}}+e^{-E_n\left(L_t-t_{\rm sep}\right)}\right) ~,
\end{align}
and the function $C_{\tilde{\varphi}}(t_{\rm sep})$ in the ground state fit of $\tilde{\varphi}^{0}_{\gamma^z \gamma^5}$ is
\begin{align}
\begin{aligned}
    C_{\tilde{\varphi}}(t_{\rm sep}) &= \frac{z_0}{2 E_0} E_0\tilde{\varphi}^{0}_{\gamma^z \gamma^5} \left(e^{-E_0 t_{\rm sep}}+e^{-E_0\left(L_t-t_{\rm sep}\right)}\right) \\
    &+ \frac{z_1}{2 E_1} O_{01} \left(e^{-E_1 t_{\rm sep}}+e^{-E_1\left(L_t-t_{\rm sep}\right)}\right) ~,
\end{aligned}
\end{align}
where the fit results of $\tilde{\varphi}^{0}_{\gamma^z \gamma^5}$ are normalized using the mean value of the local matrix element at $z = b_\perp = 0$. The function $C_{\rm 2pt}(t_{\rm sep})$ in the ground state fit of $\tilde{h}^{0}_{\gamma^t}$ is
\begin{align}
    C_{\text{2pt}}\left(t_{\rm sep}\right) = \sum_{n=0}^{1} \left|z_n\right|^2 \left(e^{-E_n t_{\rm sep}}+e^{-E_n\left(L_t-t_{\rm sep}\right)}\right) ~,
\end{align}
and function $R_{\tilde{h}}(t_{\rm sep}, \tau)$ in the ground state fit of $\tilde{h}^{0}_{\gamma^t}$ is 
\begin{align}
\begin{aligned}
    R_{\tilde{h}}\left(t_{\mathrm{sep}}, \tau \right) &= \frac{|z_0|^2 \tilde{h}^{0}_{\gamma^t} e^{-E_0 t_{\mathrm{sep}}} }{\sum_{n=0}^{1} |z_n|^2 \left(e^{-E_n t_{\rm sep}}+e^{-E_n\left(L_t-t_{\rm sep}\right)}\right)} \\
    &+ \frac{z_0 z_1 O_{01} e^{-E_0\left(t_{\mathrm{sep}}- \tau \right)} e^{-E_1 \tau}}{\sum_{n=0}^{1} |z_n|^2 \left(e^{-E_n t_{\rm sep}}+e^{-E_n\left(L_t-t_{\rm sep}\right)}\right)}\\
    &+ \frac{z_1 z_0 O_{10} e^{-E_1\left(t_{\mathrm{sep}}- \tau \right)} e^{-E_0 \tau}}{\sum_{n=0}^{1} |z_n|^2 \left(e^{-E_n t_{\rm sep}}+e^{-E_n\left(L_t-t_{\rm sep}\right)}\right)} \\
    &+ \frac{|z_1|^2 O_{11} e^{-E_1 t_{\mathrm{sep}}}}{\sum_{n=0}^{1} |z_n|^2 \left(e^{-E_n t_{\rm sep}}+e^{-E_n\left(L_t-t_{\rm sep}\right)}\right)} ~,
\end{aligned}
\end{align}
which has a different form from Eqs.~\eqref{eq:qtmdpdf_3pt} and \eqref{eq:qtmdpdf_ratio} due to the cancellation of energy factors in the numerator and the denominator. In the fit functions, all overlap factors $z_n$ are set as real numbers, while any potential complex phases are absorbed into the matrix elements $O_{01}$ and $O_{10}$.

To evaluate the overall quality of the ground state fits, the density distribution of $\chi^2 / \rm{d.o.f.}$ and the cumulative distribution function (CDF) of the p-value are plotted in Figs.~\ref{fig:gsfit_stat_qtmdwf} and \ref{fig:gsfit_stat_qtmdpdf}, corresponding to the matrix elements of the quasi-TMDWF and the quasi-TMD beam function, respectively. The figures reveal that $\chi^2/\rm{d.o.f.}$ is predominantly distributed within the expected range, with p-values exceeding $0.05$ for the majority of fits, thereby suggesting a high level of overall quality. In addition, as examples, the fit results of the quasi-TMD and quasi-TMDWF are plotted alongside the data points in Figs.~\ref{fig:ratio_fit_qtmdpdf} and \ref{fig:joint_fit_qtmdwf}, respectively.

\section{Stability of extrapolation}
\label{app:extrapolation}

\begin{figure*}[th!]
    \centering
    \includegraphics[width=.4\linewidth]{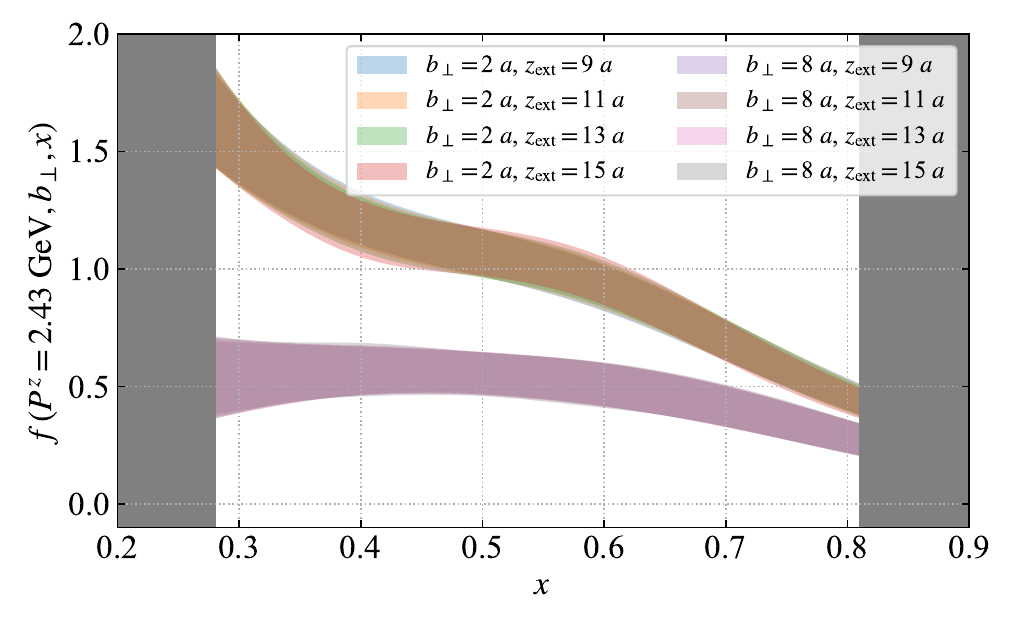}
    \includegraphics[width=.4\linewidth]{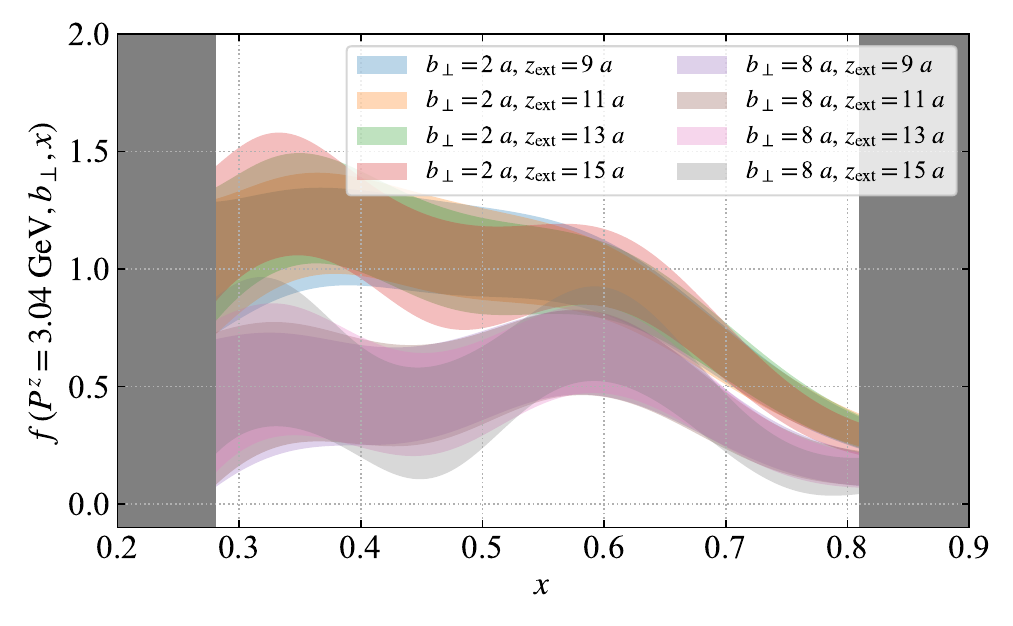}
    \caption{The unpolarized light-cone pion TMDPDF of different $z_{\rm ext}$ are ploted in comparison as the function of momentum fraction $x$. Two hadron momenta $P^z = 2.43$ GeV (left panel) and $P^z = 3.04$ GeV (right panel) are selected as examples. The variations among different $z_{\rm ext}$ are mild in the moderate region near $x=0.5$. The shaded gray bands ($x < 0.28$ and $x > 0.81$) indicates the endpoint regions where the estimated combined systematics are larger than $30\%$. The detailed estimation of the systematics is explained in App.~\ref{app:power_correction}.
    \label{fig:tmdpdf_xdep_zext}}
\end{figure*}

As mentioned in Sec.~\ref{sec:quasi-tmd}, the light-cone TMDPDF in momentum space within the moderate $x$ region is insensitive to extrapolation strategies. The unpolarized light-cone pion TMDPDF with different $z_{\rm ext}$ are plotted in Fig.~\ref{fig:tmdpdf_xdep_zext} for comparison.

It shows that the variations of the TMDPDF among the different $z_{\rm ext}$ are mild in the moderate region near $x=0.5$, which shows the stability of the extrapolation.

\section{Estimation of systematics}
\label{app:power_correction}

\begin{figure*}[th!]
    \centering
    \includegraphics[width=0.4\linewidth]{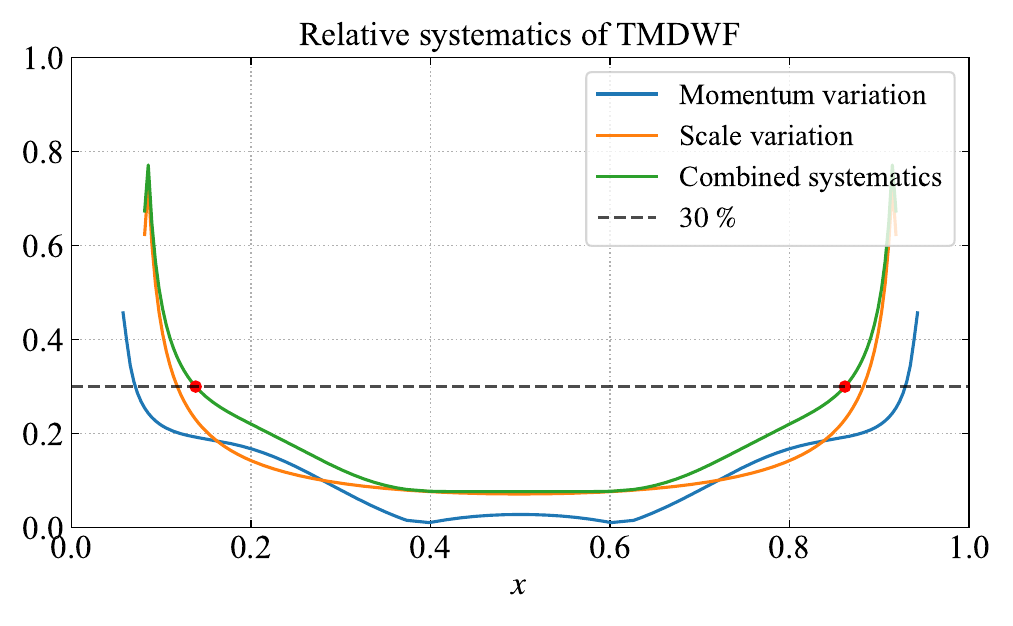}
    \includegraphics[width=0.4\linewidth]{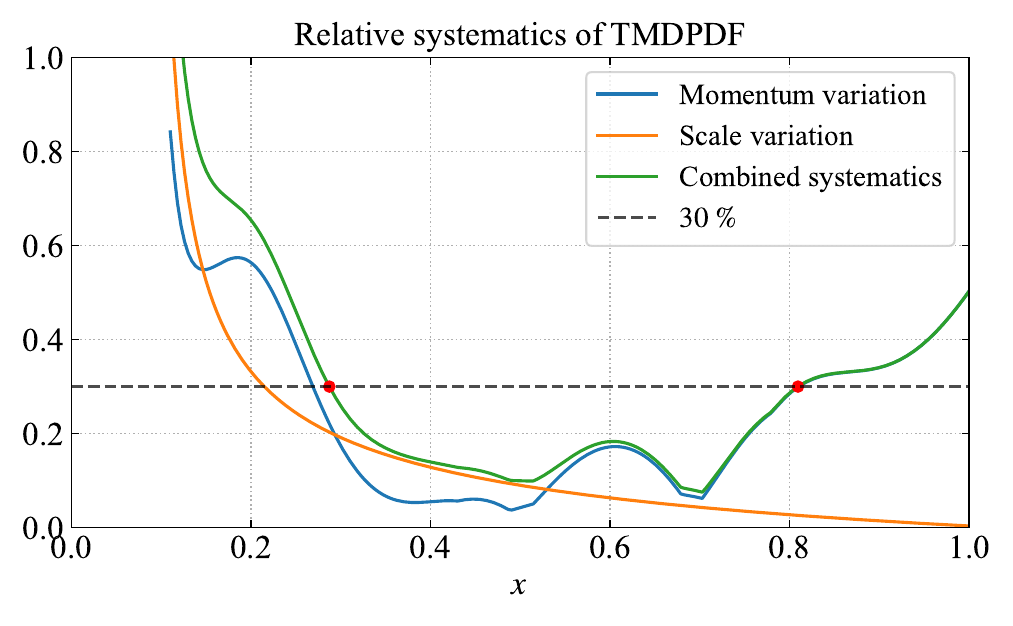}
    \caption{The estimation of the relative systematics of TMDWF (left panel) and TMDPDF (right panel). The scale variation is estimated by varying the initial scale $\mu_0$ of the RGR procedure by a factor of $\sqrt{2}$. The momentum variation is quantified as the ratio of the spread in the central values of the light-cone distributions, evaluated at three different momenta, to their mean. The combined systematics are estimated by the root-sum-square of two variations. A threshold of $30\%$ on the combined systematics is used to define the reliable region, yielding $x \in [0.11, 0.89]$ for the TMDWF and $x \in [0.28, 0.81]$ for the TMDPDF. These regions are indicated by the red markers in the figure.}
    \label{fig:sys_analysis}
\end{figure*}

As mentioned in Sec.~\ref{sec:framework}, the results of TMDWF and TMDPDF are only reliable in the moderate $x$ region due to various systematics such as uncontrolled power corrections in the endpoint regions. In this section, we will give an approximate estimation of the systematics based on the results of light-cone distributions, with the objective of delineating the reliable range within the $x$-space.

The estimation of the systematics is separated into two parts, the systematics of varying the initial scale of RGR, and the variation of light-cone distributions calculated at different pion momenta. Taking TMDWF and TMDPDF at $b_\perp = 6~a$ as representatives, the relative systematics are plotted in Fig.~\ref{fig:sys_analysis}. The systematics of scale variation is estimated by varying the initial scale $\mu_0$ of the RGR procedure by a factor of $\sqrt{2}$. Note that the systematics of scale variation is dependent on the hadron momentum. For estimation purposes, the maximum hadron momentum is selected as the representative. The momentum variation is quantified as the ratio of the spread in the central values of the light-cone distributions, evaluated at three different momenta, to their mean. These two sources of systematic uncertainty are regarded as independent, so they are combined using the root-sum-square method to provide an estimation of the combined systematics. A threshold of $30\%$ on the combined systematics is used to define the reliable region, yielding $x \in [0.11, 0.89]$ for the TMDWF and $x \in [0.28, 0.81]$ for the TMDPDF.

\FloatBarrier 

\bibliography{revised}

\end{document}